\documentclass{aa}  
\usepackage{graphicx}       
\usepackage{natbib}
\usepackage{longtable}
\usepackage{supertabular}

\newcommand{\AGN}{{AGN}}
\newcommand{\BCG}{{BCG}}
\newcommand{\BCDG}{{BCDG}}

\newcommand{\FIR}{\emph{FIR}}
\newcommand{\FUV}{\emph{FUV}}
\newcommand{\HI}{\textsc{H\,i}}
\newcommand{\Ha} {H$\alpha$}

\newcommand{\HII}{\textsc{H\,ii\ }}

\newcommand{\HIPASS}{{HIPASS}}

\newcommand{\IMF}{{IMF}}

\newcommand{\IRAS}{{IRAS}}
\newcommand{\LCBG}{{LCBG}}
\newcommand{\LIRG}{{LIRG}}

\newcommand{\Lo} {$L_{\odot}$}
\newcommand{\Mo} {$M_{\odot}$}
\newcommand{\Myr}{Myr}          % DICHOSOS Myr

\newcommand{\NIR} {\emph{NIR}}

\newcommand{\SFR} {{\sc Sfr}}
\newcommand{\TDG} {{TDG}}

\newcommand{\VLT}{{VLT}}
\newcommand{\WRBUMP} {WR bump}

\newcommand{\ergcms}{erg\,cm$^{-2}$\,s$^{-1}$}
\newcommand{\ergs}{erg\,s$^{-1}$}
\newcommand{\Moy}{$M_{\odot}$\,yr$^{-1}$}

\newcommand{\MHi}{$M_{\rm H\,I}$}
\newcommand{\Mdyn}{$M_{\rm dyn}$}
\newcommand{\Mkep}{$M_{\rm Kep}$}
\newcommand{\Mdust}{$M_{\rm dust}$}

\newcommand{\MHii}{$M_{\rm H\,II}$}
\newcommand{\MHil}{$M_{\rm H\, I}/L_B$}
\newcommand{\Mdynl}{$M_{\rm dyn}/L_B$}
\newcommand{\Mdustl}{$M_{\rm dust}/L_B$}

\newcommand{\MHiil}{$M_{\rm H\, II}/L_B$}
\newcommand{\Mstar}{$M_{\star}$}
\newcommand{\Mstarl}{$M_{\star}/L_B$}

\newcommand{\Mol}{$M_{\odot}/L_B$}

\newcommand{\Mbar}{$M_{\rm bar}$}

\newcommand{\kms}{km\,s$^{-1}$}

\newcommand{\tableline}{\hline}

\newcommand{\HeI}{He\,{\sc i}}

\newcommand{\WHb}{$W$(H$\beta$)}

\newcommand{\Te}{$T_{\rm e}$}

\newcommand{\CHb}{$c$(H$\beta$)}

\newcommand{\abox}{12+log(O/H)}

\newcommand{\nodata}{...}
\begin{document}
   \title{Massive star formation in Wolf-Rayet galaxies}

%\thanks{Based on observations made with NOT (Nordic Optical Telescope), INT (Isaac Newton 
%Telescope) and WHT (William Herschel Telescope) operated on the island 
%of La Palma jointly by Denmark, Finland, Iceland, Norway and Sweden (NOT) or 
%the Isaac Newton Group (INT, WHT) in the Spanish Observatorio del Roque de Los 
%Muchachos of the Instituto de Astrof\'\i sica de Canarias. 
%Based on observations made at the Centro Astron\'omico Hispano Alem\'an (CAHA) at Calar Alto, 
%operated by the Max-Planck Institut f\"ur Astronomie and 
%the Instituto de Astrof\'{\i}sica de Andaluc\'{\i}a (CSIC).}}

  \subtitle{V. Star formation rates, masses and the importance of galaxy interactions }

   \author{\'Angel R. L\'opez-S\'anchez
          \inst{1,2}
		  %\and
		  %C\'esar Esteban\inst{2,3}
          %\and
          %Jorge Garc\'{\i}a-Rojas\inst{1}
          }

   \offprints{\'Angel R. L\'opez-S\'anchez, \email{Angel.Lopez-Sanchez@csiro.au}}

\institute{CSIRO\,Astronomy\,and\,Space\,Science\,/\,Australia\,Telescope\,National\,Facility,\,PO-BOX\,76,\,Epping,\,NSW\,1710,\,Australia \and Instituto de  
Astrof{\'\i}sica de Canarias, C/ V\'{\i}a L\'actea S/N, E-38200, La Laguna, Tenerife, Spain}
%\and Departamento de Astrof\'{\i}sica de la Universidad de La Laguna, E-38071, La Laguna, Tenerife, Spain}

       %       \email{wuchterl@amok.ast.univie.ac.at}
       %  \and
       %      University of Alexandria, Department of Geography, ...\\
       %      \email{c.ptolemy@hipparch.uheaven.space}
       %      \thanks{The university of heaven temporarily does not
       %              accept e-mails}
       %      }

   \date{Received Feb 20, 2010; Accepted May 3, 2010}

% \abstract{}{}{}{}{} 
% 5 {} token are mandatory
 
  \abstract
   {}
{We have performed a comprehensive analysis of a sample of 20 starburst galaxies that show the presence of a substantial population of very young  
massive stars, most of them classified as Wolf-Rayet galaxies.}
  {In this paper, the last of the series, we analyze the global properties of our galaxy sample using multiwavelength data extracted from our own  
observations (H$\alpha$ fluxes, $B$ and  $H$-band magnitudes) and from the literature, that include X-ray, \FUV, \FIR, and radio (both \HI\ spectral  
line %at 21~cm 
and 1.4~GHz radio-continuum) measurements.}
{The agreement between our \Ha-based star-formation rates (\SFR) and those provided by indicators at other wavelengths is remarkable, but we consider  
that the new \Ha-based calibration provided by Calzetti et al. (2007) should be preferred over older calibrations. The \FUV-based \SFR\ provides a  
powerful tool to analyze the star-formation activity in both global and local scales independently to the \Ha\ emission. We provide empirical  
relationships between the ionized gas mass, neutral gas mass, dust mass, stellar mass, and dynamical mass with the $B$-luminosity. Although all mass  
estimations increase with increasing luminosity, we find important deviations to the general trend in some objects, that seem to be consequence of  
their particular evolutionary histories. The analysis of the mass-to-light ratios give similar results. We investigate the mass-metallicity relations  
and conclude that both the nature and the star-formation history are needed to understand the relationships between both properties. The majority of  
the galaxies follow a Schmidt-Kennicutt scaling law of star-formation that agrees with that reported in individual star-forming regions within M~51  
but not with that found in normal spiral galaxies. Dwarf galaxies seem to be forming stars more efficiently than the outskirts of spiral galaxies. We  
found a relation between the reddening coefficient 
%(derived from our spectra) 
and the warm dust mass 
%(derived from the \FIR\ flux), 
indicating that the extinction is mainly internal to the galaxies. The comparison with the closed-box model also indicates that environment effects  
play and important role in their evolution.} % of the galaxies. }
  {Considering all multi-wavelength data, we found that 17 up to 20 galaxies are clearly interacting or merging with low-luminosity dwarf objects or  
\HI\ clouds. The remaining three galaxies (Mkn~5, SBS~1054+364, and SBS~1415+437) show considerable divergences of some properties when comparing with  
similar objects. Many of the interacting/merging features are only detected when deep optical spectroscopy and a detailed multi-wavelength analysis,  
including \HI\ observations, are obtained. We conclude that interactions do play a fundamental role in the triggering mechanism of the strong  
star-formation activity observed in dwarf starburst galaxies.}

\titlerunning{Massive star formation in Wolf-Rayet galaxies V: Multiwavelength global analysis}

\authorrunning{L\'opez-S\'anchez}

   % --- galaxies: individual: ******}
   \maketitle
%
%________________________________________________________________

%\input{MSFinWRG_V_datos_05feb10}

\section{Introduction}

%Interacting galaxies have elevated emission in practically all wavelenghts:  
%soft X-ray emission (e.g., Read \& Ponman 1998), UV emission (Petrosian et al. 1978), 
%\Ha\ emission (e.g., Balzano 1983; Kennicutt et al. 1987; Bushouse 1987; 
%Barton et al. 2000), IR emission (e.g., Lonsdale et al. 1984; Young et al. 1986; 
%Heckman et al. 1986; Solomon \& Sage 1988) and radio continuum emission 
%(e.g., Hummel 1981; Condon et al. 1982; Hummel et al. 1990).

\subsection{Galaxy interactions and starburst activity}

Since the discovery of the starburst galaxies \citep{SS70}, many studies have tried to understand the processes that trigger the strong  
star-formation activity in these objects. 
%In particular, dwarf starburst galaxies have been always a challenge, 
%as the wave-density theory (REF) cannot be applied because of the low mass of these objects. 
%Several alternative mechanisms have been proposed to explain starbursts in dwarf galaxies. 
%Some of them favored internal processes, such as the Stochastic Self-Propagating 
%Star Formation (Gerola, Seiden, \& Schulman, 1980) model, that assumes statistical 
%fluctuations of the star formation rate (\SFR), or the idea of the cyclic gas re-processing 
%%(expulsion and later accretion) 
%of the interstellar medium (ISM) in dwarf objects (Davies \& Phillipps 1988), 
%for example, gas compression by shocks due to the mass lost by galactic winds followed by the 
%%subsequent 
%cooling of the ISM \citep{Th91,Hi00}. However, in many cases these 
%hypotheses fail to explain some observational characteristics (i.e., Cair\'os ****)
%%Thereby, this theory predicts an intermittent behavior of the bursts 
%%in dwarf galaxies, with long inactivity periods \cite{Th91,Hi00}. 
%Hovever, there are increasing evidences that indicate that interactions 
%and mergers do trigger the starburst activity in dwarf galaxies.
%%However, others authors have proposed galaxy interactions as the massive star 
%%formation triggering mechanism in dwarf galaxies \citep{S88,ME00**}. 
The hypothesis that gravitational interaction (not necessary merging) of galaxies enhances star formation or leads to 
starburst activity was made soon after the recognition of the starburst phenomenon. 
%The first evidence that linked galaxy interaction and star formation activity 
%was reported by Morgan (1958), who noted \emph{hotspots} in galaxies. 
%In the following year, Vorontson-Velyaminov (1959) published the first catalogue 
%of interacting galaxies, that was followed by the famous \emph{Atlas of Peculiar 
%Galaxies} developed by Arp (1966). 
%Analyzing the Vorontson-Velyaminov (1959) and Arp's catalogues, 
%S\'ersic \& Pastoriza (1967) found that the majority of galaxies 
%harboring hotspots show extreme blue colors in their centers. 
\citet{Larson78} %Larson \& Tinsley (1978)
%suggested that tidal forces in interacting galaxies could trigger star formation activity  
%Larson \& Tinsley's (1998) classical study of a normal and peculiar (Arp) sample of galaxies 
did a study of normal and peculiar \citep{Arp66} sample of galaxies and
demonstrated that recent ($\leq 10^8$ yr) star-formation is more likely to occur in interacting than in non-interacting galaxies.
%Infrared observations confirmed the existence of very intense starbursts 
%in major disk-disk mergers (e.g., Joseph \& Wright 1985).
%Hence, mergers have been linked to extreme infrared luminosities seen in 
%LIRG (e.g., Soifer et al. 1984; Armus et al. 1987; 
%Sanders \& Mirabel 1996; Genzel et al. 1998), 
%increasing the fraction of interaction galaxies and 
%(in the most extreme cases) of mergers with 
%the far-infrared luminosity (Sanders 1997)
%the latter being a direct measure of the SFR. 
%Indeed, about 10\% of galaxies with \FIR\ luminosities less than 10$^{11}$ \Lo\ 
%are in interacting/merging systems, but about 10$^{12}$ \Lo\ this fraction 
%becomes almost 100\% (Sanders, 1997; Colina *****).
%This is, however, an upper limit to the occurrence of starbursts since 
%an unknown fraction of ultraluminous galaxies is powered by an active 
%galactic nucleus (AGN) rather than a starburst (Schaerer 1999).
Since then, numerous %morphological 
studies of individual galaxies have revealed the fossil remnants of interaction/merger activity, increasing the evidences that interactions and  
mergers trigger star-formation phenomena in spiral galaxies \citep{Koribalski96,K98,Nikolic04}.
%(Koribalski 1996; Kennicutt 1998; Nikolic et al. 2004).
Infrared observations confirmed the existence of very intense starbursts in major disk-disk mergers (e.g., Joseph \& Wright 1985; Solomon \& Sage  
1988; Sanders \& Mirabel 1996; Genzel et al. 1998; Arribas et al. 2004). %OK ref
Actually, almost 100\%\ of galaxies with far-infrared (\FIR) luminosities of about 10$^{12}$~\Lo\ 
are in interacting/merging systems \citep{Sanders97}. Furthermore, %OK ref
analysis of large galaxy surveys (e.g., CfA2: Barton, Geller \& Kenyon 2000; 2dF: Lambas et al. 2003; SDSS: Nikolic et al. 2004) has provided new  
evidences for interaction-induced starburst activity. 

According to hierarchical clustering models of galaxy formation, larger galactic structures build up and grow through the accretion of dwarf galaxies 
(White \& Frenk, 1991; Kauffman \& White 1993; Springer et al. 2005). 
%Hierarchical formation models of galaxies (i.e. Kauffman \& White 1993; Springer et al. 2005) 
%predict that galaxy interactions should be more common at high redshifts.
Observations of local and distant luminous blue galaxies (LBG) and Lyman break galaxies seem to confirm that galaxy  interactions are more common at  
high redshifts.
(e.g., Guzman et al. 1997; Hopkins et al. 2002; Erb et al. 2003; Werk, Jangren \& Salzer 2004; Colina, Arribas \& Monreal-Ibero 2005; Overzier et al.  
2009; Cardamone et al. 2009) but many details are still unclear (i.e., Basu-Zych et al. 2009). 
Indeed, detailed studies of local interacting/merging galaxies 
%in the Local Universe 
provide vital clues to our knowledge in galaxy formation and evolution, 
as they constrain the properties of the hierarchical formation models. 
%as they induce profound morphological and kinematical galaxy transformations 
%and drive the final destiny of the galaxies. 

Recent observations also suggest that interactions and mergers between dwarf galaxies also trigger the star-formation activity
and play a fundamental role in the evolution of dwarf galaxies
%strongly suggest that interactions and mergers also play a 
%fundamental role in the evolution of the dwarf galaxies 
(i.e., M\'endez \& Esteban, 2000; \"Ostlin et al. 2001, 2004;  Bergvall \& \"Ostlin 2002; Johnson et al. 2004; Bravo-Alfaro et al. 2004, 2006;  
Cumming et al. 2008, Garc\'{\i}a-Lorenzo et al. 2008; L\'opez-S\'anchez \& Esteban, 2008, 2009; James et al. 2010). 
%being many of them Blue Compact Dwarf Galaxies (\BCDG s), 
%and hence interacting phenomena play a fundamental role in the evolution of dwarf galaxies.
Many of these studies have been done in Blue Compact Dwarf Galaxies (\BCDG s), that are low-luminosity, low-metallicity ($\sim$10 \% solar) galaxies  
showing compact and irregular morphologies and undergoing an intense and short-lived episode of star formation (i.e., Izotov \& Thuan 1999; Cair\'os  
et al. 2001a,b; Papaderos et al. 2006), on top of an old underlying population with ages of several Gyrs (Noeske et al. 2003, 2005; Amor\'{\i}n et  
al. 2007, 2009). Recent numerical simulations \citep{Bekki08} satisfactory explain the physical properties of \BCDG s\ as a consequence of the  
merging of two dwarf galaxies with larger fraction of gas and extended gas disks.
%However, interaction features in dwarf objects seem difficult to detect, in part 
%because of the lack of deep observations and detailed multi-wavelength analyses.

Actually, much of our knowledge in interacting galaxies has been provided via \HI\ observations.
Neutral hydrogen gas is the best tracer for galaxy-galaxy interactions because, such the
\HI\ distribution is usually several times larger than the optical extent, it is more easily disrupted by 
external forces (tidal interactions, gas infall, ram pressure stripping)
%tidal forces from nearby objects 
than the stellar disk \citep{BroeilsvanWoerden94,SalpeterHoffman1996}.
The distribution and kinematics of atomic gas within galaxies usually is more or less regular, but in many cases they revealed complex entities  
between galaxies such as tails, ripples and bridges, arcs, or independent \HI\ clumps that, in many cases, show little disturbance in their  
corresponding optical images (e.g., Schneider et al. 1989; Yun, Ho \& Lo 1994; Hibbard \& van Gorkom 1996; Verdes-Montenegro et al. 2001, 2002, 2005;  
Putman et al. 2003; %Iyer et al. 2004; 
Koribalski et al. 2003; 2004; 2005; Temporin et al. 2003, 2005; Emonts et al. 2006; Ekta et al. 2008; Koribalski \& L\'opez-S\'anchez 2009; English  
et al. 2010; see also \emph{The \ion{H}{i} Rogues Gallery}, Hibbard et al. 2001). 
Several interferometric \HI\ surveys, such as \emph{The \HI\ Nearby Galaxy Survey} (THINGS, Walter et al. 2008),  the \emph{Local Volume \HI\ Survey}   
(LVHIS, Koribalski 2008) or the \emph{Faint Irregular Galaxies GMRT Survey} (FIGGS, Begum et al. 2008), are nowadays providing accurate \HI\ and  
dynamical masses in hundreds of nearby galaxies, many of them being dwarf objects, as they account for $\sim$85\%\ of the known galaxies in the Local  
Volume \citep{Karachentsev04}.

\subsection{The interplay between gas and stars in galaxies}

However, to understand interaction processes in dwarf galaxies we first have to know how stars and gas interact in low-mass environments. 
Indeed, feedback from massive stars is the dominant process that affects the interstellar medium (ISM)
of these galaxies. Violent star-formation phenomena may disrupt the galaxy's gas and even expel it to the intergalactic medium, as some theoretical  
models predict \citep{MacLowFerrara99}. But alternative models (e.g., Silich \& Tenorio-Tagle 1998) and the available observations \citep{Bomans05}  
suggest that dwarf galaxies keep their processed material.  
Furthermore, the links between the observational characteristics (fluxes, colors, morphologies or sizes) and the underlying physical properties of  
the galaxies (stellar, dust, gas, baryonic, and dark matter content, chemical abundances, star formation rate, star formation history) are still not  
well known. 

For example, there are still many caveats in the understanding of the interplay between the star formation rate (\SFR) and the properties of the ISM.  
A very important step was achieved with the Schmidth-Kennicutt power-law relation \citep{Schmidt59,Schmidt63,K98} that correlates the average \SFR\  
per unit area and the mean surface density of the cold gas (atomic plus molecular). But tracers of star-formation, including optical colors and \Ha\  
flux (e.g., Larson \& Tinsley, 1978; Kennicutt 1998; Calzetti et al. 2007), \FIR\ flux \citep{K98,Heckman99}, radio-continuum flux \citep{Condon92},  
and far-ultraviolet (\FUV) flux \citep{K98,Salim07}, often yield to very different values of the \SFR. 
Although the density of atomic gas is known in some cases, no many direct measurements of the molecular gas are available, being specially rare in  
dwarf galaxies (i.e., Taylor et al. 1998; Barone et al. 2000; Braine et al. 2000, 2001, 2004).

On the other hand, the physics underlying the relationship between stellar mass (or luminosity) with the metallicity is still far from clear, besides  
the important observational  (e.g., Tremonti et al. 2004; van Zee \& Haynes, 2006; Kewley \& Ellison 2008) and theoretical (e.g., De Lucia et al.  
2004; Tissera et al. 2005; De Rossi et al. 2006; Dav\'e \& Oppenheimer 2007) efforts that aimed to explain it. Indeed, one of the main problems is to  
derive the real metallicity of the ionized gas, as empirical calibrations based on the direct estimation of the electron temperature (\Te) of the  
ionized gas and theoretical methods based on photoionization models provide very different oxygen abundances (e.g., Yin et al. 2007; Kewley \&  
Ellison 2008; Esteban et al. 2009; L\'opez-S\'anchez \& Esteban, 2010).  Finally, the present understanding of correlations between the \HI\ content,  
stellar populations and star formation in dwarf starburst galaxies is still at a preliminary stage because of the lack of detailed optical/\NIR\  
images and spectra and/or interferometric \HI\ maps in these systems.

%Interacting galaxies have elevated emission in practically all wavelenghts:  
%soft X-ray emission (e.g., Read \& Ponman 1998), UV emission (Petrosian et al. 1978), 
%\Ha\ emission (e.g., Balzano 1983; Kennicutt et al. 1987; Bushouse 1987; 
%Barton et al. 2000), IR emission (e.g., Lonsdale et al. 1984; Young et al. 1986; 
%Heckman et al. 1986; Solomon \& Sage 1988) and radio continuum emission 
%(e.g., Hummel 1981; Condon et al. 1982; Hummel et al. 1990).

%The estimated star formation rates in them is so high that the material available for 
%the creation of new stars would be exhausted very soon compared to the age of the universe. 
%This problem is more marked in dwarf galaxies. 
%However, star formation episodes show a large variability in duration, 
%ranging from 10$^7$--10$^8$ years \cite{RL85} to more than 10$^9$ yr \cite{HG85}. 
%In fact, the age of the bursts and the galaxy that hosts them is also 
%a controversial problem. The knowing of both the triggering mechanism and 
%the age of the starbursts would help to understand the evolution of the galaxies.

\subsection{About this paper series}

In our paper series, we have presented a detailed photometric and spectroscopic study of a sample of strong star-forming galaxies, many of them  
previously classified as dwarf galaxies. 
The majority of these objects are  
Wolf-Rayet (WR) galaxies, that are a very inhomogeneous class of star-forming objects 
which have in common that the ongoing or most recent star formation event has produced stars
sufficiently massive to evolve to the WR stage \citep{SCP99}.  The presence of WR features in the spectra
of a galaxy constrains the properties of the 
%provides useful information about its 
star-formation processes. 
As the first WR stars typically appear around 2 -- 3 Myr after the starburst is initiated and
disappear within some 5 Myr \citep{MeynetMaeder05}, 
their detection informs about both the youth and strength of
the burst, offering the opportunity to study an approximately coeval
sample of very young starbursts \citep{SV98}. 

Our main aim is the study of the formation of massive stars in starburst galaxies and the role that interaction with or between dwarf galaxies and/or  
low surface brightness objects have in its triggering mechanism. 
In Paper~I \citep{LSE08} we exposed the motivation of this work, compiled the list of the analyzed 
WR galaxies (Table~1 of Paper~I) and presented the results of optical/\NIR\ broad-band and \Ha\ photometry. In Paper~II \citep{LSE09} we presented  
the results of the analysis of intermediate-resolution long slit spectroscopy of 16 objects of our sample of WR galaxies --the results for the other  
4 objects have been published separately. In Paper~II, we also specified the oxygen abundances of the ionized gas (they were computed following the  
direct \Te\ method in the majority of the cases ) and analyzed the kinematics of the ionized gas.  
%In many cases, two or more slit positions have been used 
%in order to analyze the most interesting zones, knots or morphological 
%structures belonging to each galaxy or even surrounding objects. 
In Paper~III \citep{LSE10a}, we studied the O and WR stellar populations within these galaxies, and compared with theoretical evolutionary synthesis  
models. In Paper~IV \citep{LSE10b}, we analyzed globally the optical/\NIR\ properties of the galaxies, concluding that such detailed analyses are  
fundamental to understand the star-formation histories of the galaxies. In this paper, the last of the series, we perform a comprehensive  
multiwavelength analysis considering all the optical and \NIR\ data but also including radio, \FIR, \FUV\ and X-ray data available in the  
literature.

The selection criteria of the galaxy sample were the following. We used the most recent catalogue of Wolf-Rayet galaxies \citep*{SCP99}, which contains a very inhomogeneous group of starbursting objects, to make a list of dwarf objects that could be observed from the Northern Hemisphere. Hence, we did not consider either spirals galaxies or giant \HII regions within them, and considered only dwarf objects, such as apparently isolated \BCDG s and dwarf irregular galaxies that had peculiar morphologies in previous, shallower imaging. We also chose two galaxies belonging to the Schaerer et al. (1999) catalogue that were classified as \emph{suspected} WR galaxies (Mkn 1087 and Tol 9), to confirm the presence of massive stars within them (see Papers~II and III). The galaxy IRAS 08339+6517 was also included because previous multiwavelength results suggested that the WR stars could still be present in its youngest star-forming bursts (see L\'opez-S\'anchez et al. 2006). With this, we got a list of $\sim$40 systems to observe and analyze using the telescopes available at Roque de los Muchachos (La Palma, Spain) and Calar Alto (Almer\'{\i}a, Spain) observatories. We added the southern galaxy NGC~5253, for which we obtained  deep echelle spectrophoto\-metry using 8.2m \VLT, because of the very intriguing properties it possesses (see L\'opez-S\'anchez et al. 2007, 2010). The final sample of 20 galaxies was created considering those galaxies for which we obtained optical/\NIR\ broad-band and \Ha\ images plus the deep optical spectroscopy during our observation runs.  We already have all these data for other $\sim15$ galaxies, the analysis of these systems will be presented in the future elsewhere, but its preliminary results seem to agree with the main results reported in this paper. Hence, our galaxy sample is not complete, but we consider it represents quite well dwarf galaxies experiencing a very strong star-formation burst. Indeed, this was the main bias introduced when choosing the galaxy sample, such as we focused only in galaxies in which WR stars are detected. It would be very interesting to extend this analysis to similar star-forming galaxies that do not show WR features, as the sample of \BCDG s\  analyzed by \citet{GildePaz03}.

The structure of this paper is the following. In Sect.~2 we describe the details of the radio, \FIR, \FUV, and X-ray data extracted from the  
literature, providing some very useful relations. Sect.~3 analyzes the star-formation activity in our sample galaxies considering all  
multi-wavelength calibrators to the \SFR. 
%We also provide new relationships between the $U$, $B$ and X-ray luminosities and the \SFR. 
We check if our sample galaxies follow the radio/\FIR\ correlation in Sect.~4. Next, Sect.~5 compiles, analyses and compares all mass estimations  
derived in this work. Several mass-metallicity relations are investigated in Sect.~6. We study whether our galaxies satisfy the Schmidt-Kennicutt  
relation in Sect.~7. Section~8 analyzes and compares several mass-to-light ratios.  The dust properties within our starburst galaxies are  
investigated in Sect.~9. We compare the predictions of the closed-box model with our observational data in Sect.~10. Finally, Sect~11 compiles a  
quantitative analysis of the interaction features considering all available multi-wavelength data. The conclusions reached in our analysis are  
compiled in Sect.~12. The Appendix describes the main results found in each of the analyzed WR galaxies. 

Hence, this is essentially an observational work. Each system has been carefully analyzed considering all available data (those specifically obtained  
for this work and those compiled from literature) with the final aim to understand its chemical and dynamical evolution, its stellar, dust, gas, and  
dark matter content, the relative importance of its stellar populations (WR, young, intermediate-age and old stars) and its star formation  
properties. Our data support the hypothesis that interactions between galaxies and dwarf or low surface-brightness objects (that can not be detected  
using less detailed and less deeper observations) have a considerable importance in the triggering mechanism of massive star formation activity in  
this kind and young starbursts. 
We have produced the most complete, detailed, and exhaustive data set of this kind of galaxies, so far, involving multi-wavelength data and a careful  
analysis of each individual object following the same procedures and equations.

%This result can allow to get a better understanding, for example, about the dwarf 
%galaxies density and its clustering in the Local Universe, as well as in the role 
%that galactic interactions have in the morphological evolution of dwarf galaxies.

\begin{table*}[t!]
\centering
  \caption{\footnotesize{Radio data compiled from the literature for our WR galaxy sample. We include the flux of the 21 cm \HI\ emission line,  
$F_{\rm H\, I}$, its equivalent width, $W_{\rm H\, I}$, and the radio-continuum flux at 1.4 GHz, $S_{\rm 1.4\, GHz}$.}}
  \label{datos_radio}
  \tiny
  \begin{tabular}{l ccc cc}
  \noalign{\smallskip}
    \tableline
	\noalign{\smallskip}
 Galaxy  &  $F_{\rm H\, I}$  & $W_{\rm H\, I}$ & Ref.  &  $S_{\rm 1.4\, GHz}$  &  Ref.     \\
         & [Jy km s$^{-1}$]  & [km s$^{-1}$]   &      &     [mJy]         &          \\
    \tableline
    \noalign{\smallskip} 
%HCG 31$^a$     &  21.75        &   228             & VM04    &   27.4$\pm$3.8   & VM05   \\
HCG 31 AC      &   5.15        &  169.2A+190.6C    & VM05    &   22$\pm$3       & VM05 \\
%Mkn 1089 (HCG 31 AC) & \nodata &     \nodata       & \nodata &   31.7$\pm$1.7   & Co98 \\
HCG 31 B       &   2.74        &    85.8           & VM05    &   2.1$\pm$0.3    & VM05 \\
HCG 31 F       &   0.866       &    74.6           & VM05    &   \nodata        & \nodata \\ 
HCG 31 G       &   2.74        &    84.9           & VM05    &   3.3$\pm$0.5    & VM05 \\
%Mkn 1090 (HCG 31 G) &  \nodata &     \nodata       & \nodata &   5.5$\pm$0.5    & Co98 \\
%\noalign{\smallskip} 
Mkn 1087       &   5.38        &    270     & GG81 &   12.1$\pm$0.6     & Co98 \\
Haro 15        &  3.11$\pm$1.01&    220     & GG81 &  17.8$\pm$1.0  & Co98\\
               %&  9.3$\pm$3.4  &86.3$\pm$7.2& Pa03 &  17.8$\pm$1.0  & Co98 \\
Mkn 1199       &   1.78$\pm$0.67 &    170     & DC04 &   36.2$\pm$1.2     & Co98\\
Mkn 5          &  2.12$\pm$0.27&22.4$\pm$4.9& Pa03 &  $<$2.8     & HSLD02 \\
IRAS 08208+2816& \nodata       & \nodata & \nodata &   15.2$\pm$0.6  & Co98 \\
IRAS 08339+6517& 3.68$\pm$0.46 &  $\sim$300 & Ca04 &  33.56$^a$ & Co90  \\
%Pox 4         &  0.98$^c$     &  65$^c$    & \HIPASS &   4.2$\pm$0.5  & Co98\\
%POX 4          &  2.30        &  130$^c$    & Ott &   4.2$\pm$0.5  & Co98\\
POX 4$^b$        &  4.31         &  130    & LS10b &   4.2$\pm$0.5  & Co98\\
UM 420     &    \nodata  & \nodata & \nodata     &   1.1$\pm$0.3  &  HSLD02 \\
%\noalign{\smallskip} 
%SBS 0926+606A  & 2.53$\pm$0.53 & 148$\pm$20 & T99  & 2.7$\pm$0.6   & HSLD02 \\
SBS 0926+606A$^c$  & 1.30$\pm$0.49 & 120$\pm$37 & P02  & 2.7$\pm$0.6   & HSLD02  \\
%SBS 0926+606B  & 1.88          & 61         & H05  & \nodata     & \nodata  \\
SBS 0926+606B$^c$  & 1.10$\pm$0.49 & 120$\pm$37 & P02  & \nodata     & \nodata  \\
%SBS 0926+606B  & 2.9           & 101        & H07  & \nodata     & \nodata  \\
%\noalign{\smallskip} 
SBS 0948+532   & \nodata  & \nodata & \nodata      & $<$0.9  &  HSLD02  \\
SBS 1054+365   & 4.03$\pm$0.39 & 117$\pm$11 & Z00   &   1.28$\pm$0.14 & BWH95 \\
%SBS 1054+365 Comp. & \nodata & \nodata & \nodata   & 1.28$\pm$0.14 & BWH95 \\
SBS 1211+540   & 0.71$\pm$0.12  & 47 & H05      & $<$0.9  &  HSLD02   \\
SBS 1319+579   & 8.4  & 134     & H07      & $<$2.9  &  HSLD02   \\
SBS 1415+437   & 4.73$\pm$0.32 &   66      &  H05  & $<$0.5  &  H05 \\
%\noalign{\smallskip} 
III Zw 107 &  4.48$\pm$0.79     &   200$\pm$25    & P03 & 8.0$\pm$0.5  & Co98+Y01 \\
Tol 9$^b$          &  5.02$\pm$0.40 &   185     &  LS10b & 19.2$\pm$0.7  & Co98  \\
Tol 1457-262$^d$  &  4.3    &    176       &  Kor06 & 38.9$\pm$1.8 & Co98+Y01 \\
%ESO 566-8$^{e,f}$  &   \nodata  & \nodata & CBG04     & 97.6$\pm$3.0  & Co98+Y01\\
Arp 252$^{d,e}$  &   \nodata  & \nodata & Kor06     & 97.6$\pm$3.0  & Co98+Y01\\

%NGC 5253   &  33.4$\pm$9.9  &  68.5$\pm$10.2  & P03 & 85.8$\pm$3.4 & Co98+Y01  \\
NGC 5253$^b$   &  43.1$\pm$2.6  &  106$\pm$6  &  LS10a & 87.1$\pm$3.5 & LS10a \\

	\noalign{\smallskip}    
  \tableline
  \end{tabular}
   \begin{flushleft}
 % $^a$ VM04 give separate values for members AC, B and G within HCG~31. \\
  $^a$ Co90 \citep{Condon90} gave radio-continuum values at 1.49 GHz. The value of $S_{\rm 1.4\, GHz}$ shown in the table was computed from $S_{\rm  
1.49\, GHz}$  using the \citep{CCB02} relation between both quantities, as it was explained in \citet{LSEGR06}. \\
  $^b$ A detailed analysis of the \HI\ gas within the galaxies POX~4, Tol~9 and NGC~5253 using the \emph{Australia Telescope Compact Array} (ATCA)  
will be soon presented elsewhere \citep{LS+10a,LS+10b}.\\
  $^c$ SBS~0926+606 was observed by \citet{HPKK07}, who gave a measurement of the \HI\ flux for both A and B galaxies. Only interferometric studies  
can disentangle the amount of neutral gas in the individual galaxies.  \\
       %  $^b$ Este valor puede estar sobrestimado porque Tol 9 pertenece al 
	   % grupo de galaxias Klemola 13. El gas at\'omico debe provenir especialmente 
	   % de la espiral cercana ESO 436-46 (ver Figura~\ref{tol9sds}). 
	   % Más detalles en \S\ref{tol9}. \\ 
 % $^c$ Barely detected in \HIPASS.\\ 
  $^d$ This galaxy was observed in \HI\ by \citet{Cassasola04} using a single-dish antenna, but it was not detected.\\
  $^e$ This galaxy is not detected in \HI\ in HIPASS (Koribalski 2006, priv. comm.).\\
  %$^e$ It is one of the galaxies of Arp 252, See \S~4.19 in Paper~I.\\
  \smallskip
  {\sc References}: 
%  BWH95: Becker, White \& Helfand (1995); 
  Ca04: \citet{CSK04}; %Cannon et al. (2004); 
  CBG04: \citet{Cassasola04}; %Casasola, Bettoni \& Galleta (2004) 
  Co90: \citet{Condon90}; %Condon et al. (1990); 
  Co98: \citet{Condon98}; %Condon et al. (1998); 
  DC04: \citet{Davoust04}; %Davoust \& Contini (2004); 
  GG81: \citet{GG81}; %Gordon \& Gottesman (1981); 
%  \HIPASS: on-line data; 
  HSLD02: \citet{Hopkins02}; 
  H05: \citet{HKP05} ;%Huchtmeier, Krishna \& Petrosian (2005); 
  H07: \citet{HPKK07}; %Huchtmeier et al. (2007);
  Kor06: Koribalski (2006), priv. comm.;
%  LS09: L\'opez-S\'anchez et al. (2009);
  LS10a: \citet{LS+10a};
  LS10b: \citet{LS+10b};
  Pa03: \citet{Paturel03}: HyperLEDA; 
  P02: \citet{Pustilnik02};  
%  T99: Thuan et al. (1999);  
  VM05: \citet{VM05}; 
  Y01: \citet{YRC01}; %Yun, Reddy \& Condon (2001); 
  Z00: \citet{Zasov00}.
\end{flushleft}
\end{table*}

\begin{table*}[t!]
\centering
  \caption{\footnotesize{\FIR\ and \FUV\ data for the WR galaxy sample analyzed in this work. \FIR\ data were provided by \IRAS, \FUV\ data were  
provided by GALEX.   
  %We have included the width, $\Delta \lambda$, of the four \FIR\ filters. 
  The value of the total \FIR\ flux was computed using Equation~\ref{fir}.}}
  \label{datos_fir}
  \tiny
  \begin{tabular}{l cccc c  c cc}
  \noalign{\smallskip}
    \tableline
	\noalign{\smallskip}
 Galaxy  &   $f_{\rm 12\,\mu m}$ &  $f_{\rm 25\,\mu m}$ &   $f_{\rm 60\,\mu m}$   &  $f_{\rm 100\,\mu m}$  &  $F_{FIR}$ & Reg. & $m_{FUV}$ &  
$f_{FUV}$  \\
         &  [Jy]   & [Jy]      &  [Jy]    & [Jy]    &   (a)  & (b) &[mag] & (c)  \\
    %\tableline
%	\noalign{\smallskip} 
%$\Delta \lambda$ [$\mu$m]       &  8.5--15  &  19--30   &   40--80  & 83--120 \\
    \noalign{\smallskip} 
	\tableline
	\noalign{\smallskip} 
HCG 31         &  0.110$\pm$0.020 & 0.580$\pm$0.040 &  3.92$\pm$0.31  & 5.84$\pm$0.47 & 2.01$\pm$0.16  &AC&14.97$\pm$0.06 & 7.54$\pm$0.44\\
               &                  &                 &                 &               &                &B& 16.80$\pm$0.09 & 3.32$\pm$0.26\\
               &                  &                 &                 &               &                &E& 18.01$\pm$0.08 & 0.456$\pm$0.035\\
               &                  &                 &                 &               &                &F& 18.24$\pm$0.11 & 0.460$\pm$0.047\\ 
               &                  &                 &                 &               &                &G& 15.93$\pm$0.06 & 3.11$\pm$0.18\\
               &                  &                 &                 &               &                &H& 20.63$\pm$0.40 & 0.0041$\pm$0.0014\\ 			   			     
Mkn 1087       &  0.103$\pm$0.029 & 0.414$\pm$0.058 &  3.03$\pm$0.33  & 4.44$\pm$0.40 & 1.54$\pm$0.16  &\nodata& \nodata&\nodata \\
Haro 15        &  0.118$\pm$0.034 & 0.297$\pm$0.089 &  1.36$\pm$0.12  & 1.97$\pm$0.20 & 0.690$\pm$0.064&\nodata& 15.12$\pm$0.07& 9.41$\pm$0.57\\
Mkn 1199       &  0.282$\pm$0.031 &  1.28$\pm$0.09  &  6.82$\pm$0.34  & 8.85$\pm$0.53 & 3.33$\pm$0.18  &\nodata& \nodata&\nodata \\
Mkn 5          &  $<$0.0503       &  $<$0.0533      &  0.21$\pm$0.04  & $<$0.8473     & $<$1.75        &\nodata& 17.76$\pm$0.07 & 1.59$\pm$0.10\\
IRAS 08208+2816&  0.126$\pm$0.029 & 0.278$\pm$0.067 &  1.15$\pm$0.09  & 1.70$\pm$0.17 & 0.588$\pm$0.051&\nodata& 16.59$\pm$0.07 & 3.75$\pm$0.23\\
IRAS 08339+6517&  0.250$\pm$0.025 &  1.13$\pm$0.02  & 5.81$\pm$0.04   & 6.48$\pm$0.09 & 2.71$\pm$0.02  & & 15.37$\pm$0.06 & 10.7$\pm$0.6\\
               &                  &                 &                 &               &                &c& 19.70$\pm$0.10 & 0.160$\pm$0.014\\
POX 4          &  $<$0.987        & 0.153$\pm$0.040 & 0.629$\pm$0.057 & $<$0.5798     & $<$0.278       &POX~4& 16.03$\pm$0.06 & 2.82$\pm$0.16\\
               &                  &                 &                 &               &                &Comp.& 19.06$\pm$0.12 & 0.269$\pm$0.030 \\
UM 420         &  \nodata         & $<$0.275           &    0.411:    &   0.613:      & 0.211:     &\nodata& 18.22$\pm$0.06 & 0.377$\pm$0.021\\
SBS 0926+606   &  $<$0.07553      &  $<$0.08818     & 0.269$\pm$0.046 & $<$0.5296     & $<$0.154       &A& 16.71$\pm$0.06 & 1.75$\pm$0.09\\
               &                  &                 &                 &               &                &B& 17.84$\pm$0.10 & 0.829$\pm$0.073\\ 	
SBS 0948+532   &  \nodata & \nodata& \nodata & \nodata                                & \nodata        &\nodata&  \nodata&\nodata \\
SBS 1054+365   &   $<$0.055     & $<$0.100  & 0.536$\pm$0.048 & 0.97$\pm$0.15         & 0.296$\pm$0.035&\nodata& 16.67$\pm$0.07 & 1.18$\pm$0.08 \\
SBS 1211+540   &  \nodata & \nodata& \nodata & \nodata                                & \nodata        &\nodata& 18.40$\pm$0.07 & 0.370$\pm$0.022\\
SBS 1319+579   &  \nodata & \nodata& 0.209: & 0.685:                                  & 0.154:         &\nodata& 17.00$\pm$0.06 & 0.864$\pm$0.048\\
               &                  &                 &                 &               &                & A     & 18.60$\pm$0.06 & 0.199$\pm$0.010\\           
SBS 1415+437   &  \nodata & \nodata& \nodata & \nodata                                & \nodata        &\nodata& 16.22$\pm$0.06 & 3.95$\pm$0.22\\
III Zw 107     &  $<$0.0968        & 0.336$\pm$0.050  & 1.37$\pm$0.20  & 1.72$\pm$0.31& 0.662$\pm$0.104&\nodata& \nodata&\nodata\\
Tol 9          &  0.111$\pm$0.030  & 0.465$\pm$0.051  & 2.71$\pm$0.22  & $<$5.516     & $<$1.58        &\nodata& 17.57$\pm$0.07& 4.22$\pm$0.22\\
Tol 1457-262   & $<$0.117          & 0.611$\pm$0.067  & 3.09$\pm$0.19  & 3.68$\pm$0.40& 1.47$\pm$0.11  &\nodata& 15.84$\pm$0.07& 6.95$\pm$0.47\\
               &                  &                 &                 &               &                &Obj~1& 16.15$\pm$0.06  & 5.26$\pm$0.29\\
               &                  &                 &                 &               &                &Obj~2& 17.41$\pm$0.13  & 1.64$\pm$0.18\\
               &                  &                 &                 &               &                &\#15& 19.71$\pm$0.21   & 0.199$\pm$0.037\\ 
Arp 252        &  0.188$\pm$0.023  & 0.994$\pm$0.050  & 3.91$\pm$0.20  & 4.11$\pm$0.25& 1.79$\pm$0.10  &A& 18.67$\pm$0.08 & 1.91$\pm$0.13\\
               &                   &                  &                &              &                &B& 18.84$\pm$0.07 & 0.441$\pm$0.027\\
NGC 5253       &  2.50$\pm$0.02    & 12.07$\pm$0.05   & 29.84$\pm$0.07 &30.08$\pm$0.21& 13.49$\pm$0.05 &\nodata& 12.81$\pm$0.06 & 123.0$\pm$6.5\\
	\noalign{\smallskip}   
  \tableline
  \end{tabular}
     \begin{flushleft}
	  (a) In units of $10^{-10}$ erg s$^{-1}$ cm$^{-2}$.\\ 
	  (b) Region within each system (HCG~31, IRAS~08339+6517, POX~4, SBS~0926+606, SBS~1319+579, Tol~1457-262, and Arp~252; see Paper~I for  
identification of the regions). The \FIR\ emission provided by \IRAS\ does not allow to distinguish between these regions, but \FUV\ data provided by  
GALEX does. In Arp~252, region~A is galaxy ESO~566-8 and region~B is galaxy ESO~566-7. \\
      (c) In units of $10^{-14}$ erg s$^{-1}$ cm$^{-2}$ \AA$^{-1}$.  
	 \end{flushleft}
\end{table*}

\section{Multi-wavelength data completeness}

We have performed an exhaustive literature search to complete the optical/\NIR\ observations of our WR galaxy sample with data from other wavelengths  
(radio, far-infrared, far-ultraviolet, and X-ray). Here we describe all these data and the useful properties we derive from them.

\subsection{Radio data}

\subsubsection{H\,I data at 21 cm}

%Neutral gas analysis have given fundamental clues about our knowledge of the galaxies. 
Observations in the hyperfine transition of the neutral hydrogen, \HI, with a rest frequency of 1420.405 MHz, have been key to understand the  
distribution and kinematics of the atomic gas within galaxies, included the Milky Way. Neutral gas observations are very important because they are  
used to determine both the neutral gas mass (\HI\ gas) and the dynamical mass (\Mdyn) of the systems. 
Single-dish \HI\ surveys, --e.g. Mathewson et al. 1992, the \ion{H}{i} 
\emph{Parkes Sky Survey} (\HIPASS, Barnes et al. 2001; Koribalski et al. 2004; 
Meyer et al. 2004), and the \emph{Arecibo Legacy Fast ALFA} survey (ALFALFA; Giovanelli et al. 2005)--, 
give spectra with detected \HI\ emission of thousands of galaxies. 
However, the best tool to analyze the neutral gas content in galaxies is via radio
interferometer observations (e.g., THINGS; LVHIS; FIGGS; \emph{The \ion{H}{i} Rogues Gallery}).
%(e.g., THINGS, Walter et al. ; LVHIS, Koribalski et al.). 
%The analysis of the kinematics of the neutral gas gives key clues 
%about the past and future dynamical evolution of the galaxies. 
Knowing the amount of available neutral gas, the timescale of the starbursts (i.e., the time in which the \ion{H}{i} cloud will be exhausted if the  
star formation activity continue at the current \SFR) can be calculated.  

%No obstante, se consigue información más detallada a partir de observaciones 
%interferométricas en la línea de \ion{H}{i} de galaxias externas, como las 
%que se ha hecho de galaxias irregulares y espirales con el interferómetro 
%Westerbork ({\sc Whisp}, Swaters y Balcells, 2002; Swaters et al. 2002 y
%referencias) o \emph{\ion{H}{i} Rogues Gallery} (Hibbard et al. 2001). 

Table~\ref{datos_radio} compiles all \HI\ 21 cm data found for our galaxy sample. The majority of the \HI\ data is provided by single-dish \HI\  
observations, but for some few cases (HGC~31 and IRAS~08339+6517) interferometric \HI\ maps are available.  Table~\ref{datos_radio} lists the \HI\  
flux density,   $f_{\rm H\, I}$ (in units of Jy km s$^{-1}$), and the \HI\ equivalent width, $W_{\rm H\, I}$ (in \kms). We note that for 3 galaxies  
(POX~4, Tol~9 and NGC~5253) we are using the data provided by our new interferometric maps obtained using the \emph{Australia Telescope Compact  
Array}. For these objects, we compile the integrated \HI\ flux and width; their detailed analysis will be soon presented elsewhere  
\citep{LS+10a,LS+10b}. 
The total \HI\ mass is computed applying
\begin{equation}
M_{\rm H\, I} = 2.356 \times 10^5 d^2 f_{\rm H\, I}
\end{equation}
%(Dahlem et al. 2005),
\citep{Roberts75,RH94}
%(Robert 1975; Robert \& Haynes 1994), 
where the distance to the galaxy, $d$, is expressed in Mpc and the result for the neutral gas mass is given in solar units. The dynamical mass of the  
system, \Mdyn, can be estimated from \HI\ radio observations considering the inclination-corrected maximum rotation velocity, $v_{max}^i$, that is  
obtained at radius $R_{max}$ and assuming a virial equilibrium,
\begin{equation}
M_{dyn} = 2.31 \times 10^5 R_{max} (v_{max}^i)^2,
\end{equation}
being the result in solar masses and assuming $v_{max}^i= \frac{W_{\rm H\, I}}{2 \sin i}$. The inclination angle, $i$, is defined as that found  
between the plane of the sky and the plane of the galaxy (hence, $i$=90$^{\circ}$ in an edge-on galaxy and $i$=0$^{\circ}$ in a face-on galaxy). We  
usually estimated this angle assuming that the elliptical shape of the galaxy is just a consequence of its orientation.  Note that the usual problem  
deriving virial masses is the unknowledge of the inclination angle, $i$, and sometimes also $R_{max}$, specially in galaxies showing disturbed  
morphologies.  We adopted the maximum radius observed in our deep optical images. Therefore, as the extension of the neutral gas is usually larger  
than the extension of the stellar component, our values of \Mdyn\ may be underestimated. The gas depletion timescale defined by \citet{SCM03} was  
computed using \MHi\ and the assumed \SFR\ derived for each galaxy (see below).

\subsubsection{Radio-continuum data}

For an individual star-forming galaxy, the \SFR\ is directly proportional to its radio luminosity (i.e., Condon 1992). Hence, the radio continuum  
flux is widely used as a dust-free indicator of the star formation rate. Nearly all of the radio-continuum luminosity from galaxies without a  
significant Active Galactic Nucleus (\AGN) can be traced to recently formed massive ($M\geq$ 8 \Mo) stars (Condon et al. 1992). The 10\% of the  
continuum emission at 1.4~GHz is due to free-free emission from extremely massive main-sequence stars (thermal emission) and almost 90\% is  
synchrotron radiation from relativistic electrons accelerated in the remnants of core-collapse supernovae (non-thermal emission). As the stars that  
contribute significantly to the radio emission have lifetimes $\tau\leq\ 3\ \times\ 10^7$ yr and the relativistic electrons have lifetimes $\tau\leq\  
10^8$ yr, the current radio luminosity is nearly proportional to the rate of massive star formation during the past $\tau\leq\ 10^8$ yr  
\citep{CCB02}:  
\begin{eqnarray}
%SFR_{1.4\,\rm{GHz}} (M\ > 5M_{\odot}) \sim 2.17 \times 10^{-22} L_{1.4\,\rm{GHz}}
\label{esfrghz} SFR_{1.4\,\rm{GHz}}\ (M\ > 5M_{\odot}) \sim 2.5 \times 10^{-22} L_{1.4\,\rm{GHz}},
\end{eqnarray}
where $L_{1.4\,\rm{GHz}}$ has units of W Hz$^{-1}$. 
%While this radio luminosity is reasonably consistent with other measures 
%of massive star formation (\Ha\ and \FIR\ luminosities), all of them are 
%insensitive to the formation rate of lower mass stars. By assuming a Salpeter 
%initial mass function, $\psi\propto M^{-2.35}$ over the mass range 0.1 and 
%100 \Mo, (Condon et al 2002) find by extrapolation:
%\begin{eqnarray}
%SFR_{1.4\,\rm{GHz}}\ (M\ > 0.1M_{\odot}) \sim 1.19 \times 10^{-21} L_{1.4\,\rm{GHz}}.
%\end{eqnarray}

%The radio continuum flux can be used as an extinction-free star formation indicator (Condon et al. 2002).  
Table~\ref{datos_radio} compiles all 1.4 GHz radio-continuum flux data available for our WR galaxy sample in the literature. The 1.4~GHz luminosity,  
$S_{1.4\,\rm{GHz}}$, can be computed using the expression given by \citet{YRC01}:
\begin{eqnarray}
\label{Lquince} \log  L_{1.4\,\rm{GHz}} = 20.07+2\log d +\log S_{1.4\,\rm{GHz}}, 
%= 1.197 \times 10^{20}\times D^2\times S_{1.4\,\rm{GHz}}.
\end{eqnarray}
where the result is given in units of W~Hz$^{-1}$, the distance $d$ is expressed in Mpc and $S_{1.4\,\rm{GHz}}$ is expressed in Jy.

%Radio emission from starbursts is a superposition of nonthermal synchroton 
%emission from supernovae and thermal emission from \HII regions (Condon 1992). 
Radio-continuum observations at several cm wavelengths are used to quantify the thermal and non-thermal contributions, and thereby distinguish older  
and supernova-rich regions from younger and mostly thermal areas (i.e., Deeg et al. 1993, Beck et al. 2000, Cannon et al. 2004, 2005). These  
observations also permit to detect extremely young, dense heavily embedded star clusters \citep{KobulnickyJohnson99,JohnsonKobulnicky03}. Although  
radio data at frequencies different of 1.4~GHz are not usually available for this kind of galaxies, we applied the equation provided by  
\citet{DPKC02},
\begin{eqnarray}
\label{ftermico} F_{1.4\,\rm{GHz\, thermal}} = 1.21 \times 10^{12} F_{H\alpha},
\end{eqnarray}
to obtain an estimation of the thermal emission at 1.4 GHz, $F_{1.4\,\rm{GHz\, thermal}}$, using the \Ha\ flux derived from our images (see Paper~I).  
In this equation, $F_{H\alpha}$ is in units of \mbox{erg\ cm$^{-2}$\ s$^{-1}$} and the result is given in mJy. The comparison between  
$F_{1.4\,\rm{GHz\, thermal}}$ and $F_{\rm 1.4\, GHz}$ allows the estimation of the non-thermal flux. Condon (1992) and \citet{Niklas97} indicated  
that the non-thermal component is more than 90\% of the total at this frequency. It is common to consider the non-thermal to thermal ratio, $R$;  
\citet{DPKC02} reported that the average value in starburst galaxies is $\log R= 1.3 \pm 0.4$.  Radio continuum and \FIR\ data help to discern  
between the normal or active (that is, a galaxy hosting an \AGN) nature.

\subsection{FIR data}

\begin{figure*}[t!]
\centering
\includegraphics[angle=0,width=\linewidth]{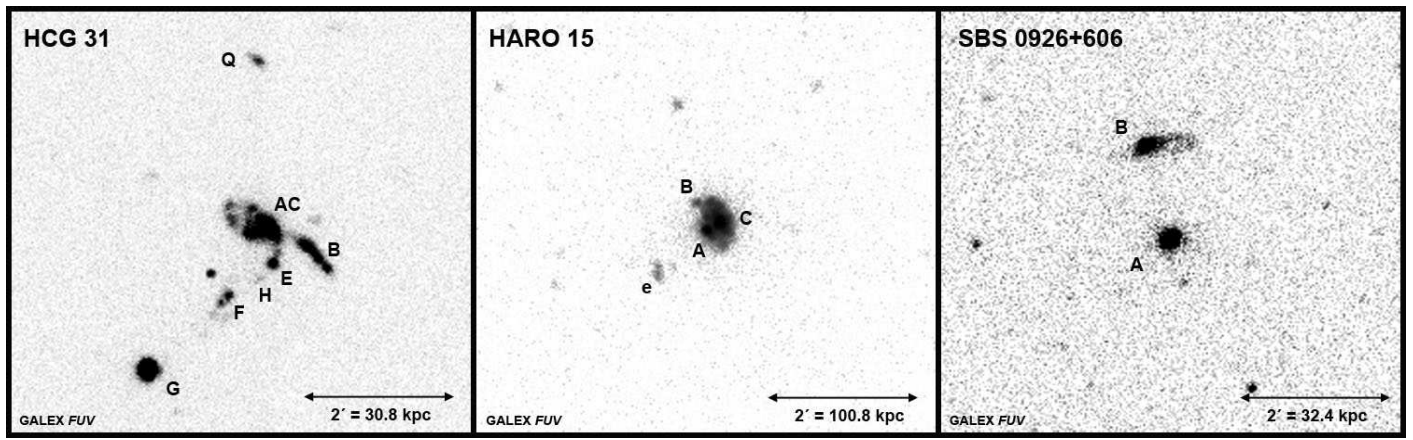}
\protect\caption[ ]{\footnotesize{Example of GALEX images, showing the \FUV\ emission in HCG~31, Haro~15 and SBS~0926+606. Regions within each object  
have been labeled following the notation given in Paper~I.}}
\label{galex}
\end{figure*}

Many of the problems found to derive the \SFR\ from optical data can be avoided by measuring the far-infrared (FIR) and sub-millimeter spectral  
energy distributions (SEDs). These are determined by the re-radiation as thermal continuum by the dust grains of stellar photospheric radiation  
absorbed in the visible and UV regions of the spectrum. Assuming that the dust completely surrounds the star forming regions, it acts as a bolometer  
reprocessing the luminosity produced by the stars. Therefore, the \SFR\ can be also computed
using theoretical stellar flux distributions and evolutionary models.
%, as STARBURST99 \citep{L99}, the \SFR\ can also be derived. 
\citet{K98} provides the following correlation between the \SFR\ (in units of \Moy) and the far-infrared flux:
\begin{eqnarray}
\label{esfrfir} SFR_{FIR}= 4.5 \times 10^{-44} L_{FIR},
\end{eqnarray}
where $L_{FIR}$ (given in units of \ergs) is obtained using the \FIR\ flux between 42.5 and 122.5 $\mu$m \citep{SM96},
\begin{eqnarray}
\label{fir}F_{FIR} = 1.26 \times 10^{-11} \big( 2.58 f_{60} + f_{100} \big),
\end{eqnarray}
being $f_{60}$ and $f_{100}$ the flux densities (in Jy) for 60$\mu$m and 100$\mu$m and the conventional expression between flux and luminosity, $L =  
4\pi d^2 F$. This relation can be applied only in starbursts with ages less than 10$^8$ yr, where the approximations assumed by \citet{K98} are  
valid. If the SFR value derived from $L_{FIR}$ agrees with that estimated from the \Ha\ luminosity we may consider that the correction by extinction  
done to derive the \Ha\ flux is correct. 

Assuming that all the UV and blue radiation from massive stars is absorbed by grains and is re-emitted as thermal radiation in the 40-120~$\mu$m  
band, Condon (1992) derives the following relation between \SFR\ and L$_{60\, \mu m}$ (in units of W\ Hz$^{-1}$):
\begin{eqnarray}
\label{esfrsesenta} SFR_{60\, \mu m}  \sim 1.96 \times 10^{-24} L_{60\, \mu m}, 
\end{eqnarray}
i.e.,
\begin{eqnarray}
SFR_{60\, \mu m}  \sim 2.346 \times 10^{-4} f_{60\, \mu m} d^2, 
\end{eqnarray}
being $f_{60\, \mu m}$ expressed in Jy and $d$ in Mpc.
Although some authors \citep{Lonsdale84} have argued that radiation from stars less massive than 5 \Mo\ will contribute significantly to FIR emission  
from galactic disks, this relation seems to give good values for the SFR. 

%Roussel et al. 2001 A\&A 372, 427 
\citet{RSVB01} provide an alternative \SFR\ calibration using the 
%the 7 $\mu$m  and 
15 $\mu$m luminosity:
%luminosities:
%\begin{eqnarray}
%SFR_{7\, \mu m}  \sim 2.4 \times 10^{-9} L_{7\, \mu m} (L_{\odot}), 
%\end{eqnarray}
\begin{eqnarray}
%\label{esfrquince} SFR_{15\, \mu m}  \sim 6.5 \times 10^{-9} L_{15\, \mu m},  
\label{esfrquince} SFR_{15\, \mu m}  \sim 3.66 \times 10^{-3}d^2 f_{15\, \mu m}, 
\end{eqnarray}
%being $L_{15\, \mu m}$ expressed in solar units. 
being $f_{15\, \mu m}$ the monochromatic flux at 15 $\mu$m and $d$ the distance in Mpc.
This formula is applicable only when the mid-infrared emission is dominated by unidentified infrared bands (UIBs) with a negligible very small grains  
(VSG) continuum, which is the case in disk galaxies, but is not always verified at 15 $\mu$m in galactic central regions. It may be assumed that  
$L_{15\, \mu m}\ \sim\ L_{12\, \mu m}$.

The warm dust mass can be estimated using the 60 and 100 $\mu$m fluxes and applying the relation given by \citet*{HSH95},
\begin{eqnarray}
\label{empolvo} M_{dust} = 4.78 d^2 f_{\rm 100\, \mu m}  \Bigg(\exp\Big[2.94\Big(\frac{f_{\rm 100\, \mu m}}{f_{\rm 60\, \mu m}}\Big)^{0.4}\Big]-1  
\Bigg),
\end{eqnarray}
where the distance is expressed in Mpc, the flux densities are in Jy and the result is given in \Mo.

We have used the far-infrared (\FIR) data provided by the \emph{Infrared Astronomical Satellite} (IRAS) to obtain the monochromatic fluxes at 12, 25,  
60 and 100 $\mu$m. These data are used to get an independent estimation of the \SFR\ and to derive the warm dust mass within every galaxy. We will  
also check if the galaxies follow the \FIR-radio relationship. Table~\ref{datos_fir} compiles all the \FIR\ data found for our sample of WR galaxies,  
three of them have no useful measurements at these frequencies. 
%The \FIR\ flux was derived from the 60 and 100 $\mu$m monochromatic fluxes applying
%\begin{eqnarray}
%\label{fir} F_{FIR} = 1.26 \times 10^{-11} \big( 2.58 f_{\rm 60\, \mu m} + f_{\rm 100\, \mu m} \big)
%\end{eqnarray}
%(Sanders \& Mirabel, 1996), being the monochromatic fluxes expressed in Jy and the result given in \mbox{erg s$^{-1}$ cm$^{-2}$}. 
%Se puede calcular la luminosidad total en infrarrojo aplicando:
%\begin{eqnarray}
%F_{IR} = 1.8 \times 10^{-11} \big( 13.48 f_{\rm 12\, \mu m} + 5.16 f_{\rm 25\, \mu m} + 2.58 f_{\rm 60\, \mu m} + f_{\rm 100\, \mu m} \big),
%\end{eqnarray}
%con las mismas unidades que antes. 

%\item {\bf Ultraviolet (\UV) observations}. In some cases, some \UV\ 
%imagery from the \emph{Hubble Space Telescope} (\HST) have been used 
%in order to know the distribution of the massive stellar clusters in 
%the galaxies. We have also checked all spectroscopical data in these 
%frequencies looking the \ion{He}{ii} $\lambda$1640 emission line, 
%also attributed to WR stars. 

\subsection{FUV data}

In the last years, the \emph{GALaxy Evolution eXplorer} (GALEX) satellite is providing astonishing ultraviolet ($UV$) images of galaxies, and revealing  
recent star-formation activity in their external regions (i.e., Gil de Paz, 2005, 2007; Thilker et al. 2005; Koribalski \& L\'opez-S\'anchez, 2009).  
%The GALEX full-of-view is $\sim$1.2$^{\circ}$ in diameter, 
%and the pixel size is 1.5 arcsec. 
The GALEX point spread function in the central 0.5$^{\circ}$ has a full width at half-maximum ($FWHM$) of $\sim$5 arcsec, matching quite nicely with  
the spatial resolution of our optical/\NIR\ images. We searched for GALEX observations of the galaxies that compose our sample in the $far-UV$-band  
(\FUV, 1350--1750 \AA),  all of them except four objects (Mkn~1087, Mkn~1199, SBS~0948+532 and III~Zw~107) have useful \FUV\ data. 
%that we compile in Table~\ref{datos_fir}. 

In general, the \FUV\ emission of our sample galaxies matches quite well with their optical emission. In many cases, \FUV\ emission is much more  
extended that the \Ha\ emission. Figure~\ref{galex} shows as examples the GALEX \FUV\ images of HCG~31, Haro~15, and SBS~0926+606. As we can see when  
comparing with our optical images (see Paper~I), the star-forming regions are clearly observed in \FUV. Just quick comments about three galaxies: 
\begin{itemize}
\item we note that the eastern tail of SBS~0926+606~B is quite bright in the \FUV\ image, 
suggesting a extended distributions of massive OB stars that we do not detect in our deep \Ha\ image (see Fig.~15 and Sec.~3.10.1 in Paper~I),
\item the star-forming galaxy \#15 in Tol~1457-262 is clearly detected in the \FUV\ emission, but the faint galaxy \#16 is not seen (Fig~31 and  
Sect.~3.18.1 in Paper~I), and
\item in Arp~252, \FUV\ emission is not only detected at the center of the galaxies (ESO~566-8 and ESO~566-7) but also throughout the tails and in  
tidal dwarf candidates \emph{c, e} and \emph{d} (see Fig.~34 and Sect.~3.19.1 in Paper~I).       
\end{itemize}

\FUV\ data can be used to get an independent estimation of the \SFR\ of the galaxies. $UV$-emission probes star formation over time-scales of  
$\sim$100 Myr, the lifetime of the massive OB stars. We integrated the counts per second (CPS) within each galaxy or region and then applied  
$m_{FUV}=-2.5 \log ({\rm CPS}) + 18.82$ \citep{Morrissey05} to derive the magnitude in the \FUV-band. We compile the $m_{FUV}$ found for each galaxy  
in  Table~\ref{datos_fir}. We then corrected by extinction using $A_{FUV}=7.9 E(B-V)$ --the value of $E(B-V)$ adopted for every region was derived  
from our optical spectroscopy and it is compiled in Table~5 of Paper~I-- and then apply 
\begin{eqnarray}
\label{ffuv} f_{FUV} = 1.40\times10^{-15}\times10^{0.4(18.82-m^0_{FUV})},
\end{eqnarray}
where $m^0_{FUV}=m_{FUV}-A_{FUV}$ and $f_{FUV}$ is obtained in units of erg s$^{-1}$ cm$^{-2}$ \AA$^{-1}$. The value of $f_{FUV}$ computed for each  
galaxy is also shown in Table~\ref{datos_fir}. Once the \FUV\ luminosity is computed ($L_{FUV} = 4 \pi d^2 f_{FUV}$), the \FUV-based \SFR\ is derived  
applying the calibration provided by \citet{Salim07},
\begin{eqnarray}
\label{esfrfuv} SFR_{FUV} = 8.1\times10^{-41}L_{FUV}.
\end{eqnarray}

\subsection{X-ray data}

Finally, we also looked for the X-ray data available for our WR galaxy sample. Only four objects (HCG~31~AC, IRAS~08339+6517, Tol~9 and NGC~5253)  
have been observed at these high frequencies; their $X-ray$ luminosities are compiled in Table~\ref{rayosx}. Beside these data, we will also use (see  
Section~3.3) the WR galaxy sample that \citet{SS98a,SS98b}
%Stevens \& Strickland (1998a,b) 
observed in X-ray.

\begin{table}[t!]
\centering
  \caption{\footnotesize{X-ray data available for our WR galaxy sample.}}
  \label{rayosx}
  \tiny
  \begin{tabular}{l l  c}
    \tableline
	\noalign{\smallskip}
	Galaxy     &   $\log L_{\rm X, (0.2-2.0\, keV)}$ [\ergs]   &  Ref. \\
	\tableline
    \noalign{\smallskip}
HCG 31          &  40.88 $\pm$ 0.13      & SS98  \\
IRAS 08339+6517 &  41.45                 & SS98 \\
%Tol 9           &  $<$1.2$\times$10$^{-13}$& FFZ82 \\
Tol 9           &  $<$40.29$^a$               & FFZ82 \\   %40.43 sin corregir
NGC 5253        &  38.60 $\pm$ 0.18      & SS98\\  
  \tableline
  \end{tabular}
  \begin{flushleft}
  $^a$ This value was derived assuming the distance to Tol~9 and the flux quoted by FFZ82, $f_{\rm 0.5-3 keV}<1.2\times 10^{-13}$ \ergcms, and  
multiplying for 0.72 to correct for the X-ray range. \\ 
    {\sc References}:  FFZ82: Fabbiano, Feigelson \& Zamorani (1982); SS98: Stevens \& Strickland (1998b).
\end{flushleft}
\end{table}

\begin{table*}[t!]
\centering
  \caption{\footnotesize{\FUV, $U$, $B$, \Ha, $H$, \FIR, 15 $\mu$m, 60 $\mu$m and 1.4~GHz luminosities for all galaxies analyzed in this work.}}
  \label{lumi}
  \tiny
  \begin{tabular}{l@{\hspace{1pt}} c@{\hspace{3pt}}   c@{\hspace{3pt}}  c@{\hspace{3pt}}  c@{\hspace{3pt}}  c@{\hspace{3pt}}  c@{\hspace{3pt}}    
c@{\hspace{3pt}}c@{\hspace{3pt}}   c@{\hspace{3pt}}  }
  \noalign{\smallskip}
  \tableline
   \noalign{\smallskip}
Galaxy &  $L_{FUV}$ & $L_U$ & $L_B$ & $L_{\rm H\alpha}$ &  $L_H$  &$L_{\rm FIR}$ & $L_{\rm 15\, \mu m}$& $L_{\rm 60\, \mu m}$& $L_{\rm 1.4\, GHz}$ \\
   \noalign{\smallskip}
 & [10$^{39}$ \ergs\,\AA$^{-1}$] & [10$^8$ \Lo] & [10$^8$ \Lo] &   [10$^{40}$ \ergs] & [10$^8$ \Lo] &[10$^{42}$ \ergs] & [10$^{22}$ W Hz$^{-1}$] &  
[10$^{23}$ W Hz$^{-1}$] & [10$^{20}$ W Hz$^{-1}$]\\ 
   \noalign{\smallskip}
\tableline
\noalign{\smallskip}    
 %OBJETO  & LFUV  &  LU  (10$^8$) &  LB (10$^8$) & LHa (10$^{40}$) &  LFIR (10$^{42}$)  &  L15 (10$^{22}$) &  L60 (10$^{23}$)  &  L14 (10$^{20}$) \\

HCG 31       &  49$\pm$3    & 356   & 202$\pm$7  & 52$\pm$4     & 42.2  & 72$\pm$6$^a$ &  3.9$\pm$0.7$^a$  &   14.1$\pm$1.1$^a$ & 98$\pm$14\\
" AC         & 24.9$\pm$1.5 &  175   &  92$\pm$4 & 36.3$\pm$2.3  &  18.0 &   \nodata	    &     \nodata    &	 \nodata   &   79$\pm$11 \\
"      B     & 11.0$\pm$0.9 &  73.8 & 47.4$\pm$1.8	&  3.1$\pm$0.3 & 10.0 &    \nodata	    &     \nodata    &	 \nodata   	  &    7.5$\pm$1.1 \\
"      E     & 1.51$\pm$0.12 &  6.3  & 3.13$\pm$0.17   & 1.49$\pm$0.16& 0.586 & \nodata &    \nodata        &	\nodata    	     &    \nodata \\
"      F$^b$ & 1.52$\pm$0.15&  10.9 & 4.0$\pm$0.2   &    5.0$\pm$0.3 & 0.486 & \nodata&    \nodata        &	\nodata    	     &    \nodata \\
"      H     & 0.13$\pm$0.05 &  \nodata  &  \nodata   &    0.045$\pm$0.013 & \nodata & \nodata  &    \nodata        &	\nodata   &    \nodata \\
"      G     & 10.3$\pm$0.6 &  90.4 &  55$\pm$2	  &    6.4$\pm$0.4 & 13.1 &    \nodata  &   \nodata     &	 \nodata   	   &   11.9$\pm$1.8 \\
Mkn~1087     & \nodata & 1803   & 1127$\pm$83	  &   70$\pm$5   &   360 &   228$\pm$23   &   15$\pm$4    &   45$\pm$5	     &  178$\pm$9 \\
" N          & \nodata & \nodata   &   26.1$\pm$0.9  &   2.27$\pm$0.19    &   5.01 &  \nodata &    \nodata        &	\nodata   &    \nodata \\
Haro~15      & 84$\pm$5 & 614   &  347$\pm$13	  &   42$\pm$5	 &   124 &   62$\pm$6   &    11$\pm$3    &   12.2$\pm$1.1      &  160$\pm$9 \\
Mkn~1199     & \nodata& 479   &  291$\pm$11	  &   49$\pm$7	     &   344 &  116$\pm$6   & 9.8$\pm$1.1   &   23.8$\pm$1.2      &  126$\pm$4 \\
" NE         & \nodata & 21.1 & 16.6$\pm$0.6  & 0.90$\pm$0.23  &    21.7	 &     \nodata    &	\nodata   &   \nodata & \nodata \\
Mkn~5        & 0.274$\pm$0.018  & 4.21&   2.63$\pm$0.10  &  0.582$\pm$0.014  & 1.89 &  $<$0.30   &  $<$0.087  &0.036$\pm$0.007  &   3.9$\pm$0.8 \\
IRAS 08208+2816& 162$\pm$10 & 879  &  511$\pm$14 &  142$\pm$12	& 373 &  254$\pm$22  &  $<$5.5 	   &   49.7$\pm$4.0	   &  657$\pm$26 \\
IRAS 08339+6517& 79$\pm$4 &  1159  & 661$\pm$23	  &  120$\pm$6 &   592 &   198.6$\pm$1.8 &  18.3$\pm$1.8 &   42.6$\pm$0.3   &  249$\pm$37 \\
" Comp.      &  1.17$\pm$0.10  &  38      & 29.9$\pm$1.6 &  2.1$\pm$0.4 &   23.3 & \nodata  &    \nodata    &	\nodata    	     &   \nodata   \\
%POX 4       & &   48.3$\pm$0.5     &  38.4$\pm$1.2 MENDEZ ES ALTO   &    6.9$\pm$0.5    &  $<$24 	   &	1.56$\pm$0.14    &   10.4$\pm$1.2 \\
POX 4        & 7.0$\pm$0.4 & 104.7 &   51.1$\pm$0.5  &  23.3$\pm$1.9$^c$ &  17.9   &  6.9$\pm$0.5 &   $<$24  &	1.56$\pm$0.14    &   10.4$\pm$1.2 \\
" Comp.      & 0.667$\pm$0.075 &  2.42 & 2.17$\pm$0.07 & 0.188$\pm$0.018$^c$ & 1.1: & \nodata  &    \nodata    &	\nodata    	     &   \nodata   \\
UM 420       & 25.4$\pm$1.4 & 236   &  103$\pm$4	  &   47$\pm$3	& 57 &   142:	   &    \nodata        &	 27.7   	     &   74$\pm$34 \\
SBS 0926+606$^b$& 41.4$\pm$3.4 & 52.3  & 26.6$\pm$0.8  &  11.8$\pm$0.7 & 9.49  &    5.8$\pm$0.6    &    2.8  &	1.01$\pm$0.17    &   10$\pm$2 \\
" A & 6.4$\pm$0.4 &  28.1  &  12.8$\pm$0.4	  &    9.4$\pm$0.5   &  3.63 &    \nodata	       &     \nodata    &	\nodata    	     &   \nodata \\
" B &  35$\pm$3 &  24.2  & 13.8$\pm$0.4	  &    2.4$\pm$0.2     &   5.86 & \nodata	       &     \nodata    &	\nodata    	     &   \nodata \\

SBS 0948+532 &\nodata&  121.3  &  36.7$\pm$1.0	  &   78$\pm$3	&  15.8 &   \nodata	       &     \nodata       &	\nodata    	     & $<$38 \\
SBS 1054+365 &0.090$\pm$0.006 & 0.982 &  0.655$\pm$0.018 & 0.450$\pm$0.017 & 0.413 & 0.23$\pm$0.03 & $<$0.042  & 0.041$\pm$0.005 & 0.098$\pm$0.011 \\
SBS 1211+540 &0.076$\pm$0.005 &  0.608  &  0.316$\pm$0.008  & 0.141$\pm$0.005   & 0.127 &  \nodata	 &   \nodata    &	 \nodata     & $<$0.50  \\
SBS 1319+579 & 0.86$\pm$0.05 &  63.1   &  40.2$\pm$1.4	  &    2.38$\pm$0.15  & 28.6 &    5.3:  &     \nodata       &	 0.85   	     & $<$2.9  \\
SBS 1415+437 & 0.409$\pm$0.023 &  1.69  &  1.00$\pm$0.03   &    0.50$\pm$0.02    & 0.581 &   \nodata &     \nodata     &	 \nodata  & $<$0.052  \\
III Zw 107   & \nodata & 286  & 177$\pm$5  &   40.1$\pm$1.8   & 77.3 &   50$\pm$8    &  $<$7.3 	   &   10.4$\pm$1.5      &   62$\pm$4 \\
Tol 9        & 9.5$\pm$0.7  & 118    & 79$\pm$2 &   22.9$\pm$1.6  &  55.5 &  35.4$\pm$1.6  &   2.5$\pm$0.7  &	6.1$\pm$0.5     &   87.3$\pm$1.6 \\
Tol 1457-262$^b$& 38.6$\pm$2.6 & 319  &  182$\pm$5	  &   63$\pm$3&  109    & 82$\pm$6   &  $<$6.5 	   &   17.2$\pm$1.1      &  216$\pm$10 \\
" Obj~1      & 29.2$\pm$1.6 & 223 &   121$\pm$3 & 46.2$\pm$2.7  &  58.6 &   \nodata	       &     \nodata    &	\nodata    	     &   \nodata\\
" Obj~2      & 9.2$\pm$1.1 &  97.3 &   60.3$\pm$1.7 & 17.2$\pm$0.8 &  45.3 &  \nodata	       &     \nodata    &	\nodata    	     &   \nodata\\
" \#15       & 1.10$\pm$0.21 &  8.39 & 6.08$\pm$0.23 & 0.47$\pm$0.04 & 5.5: &    \nodata	   &     \nodata    &	\nodata    	     &   \nodata\\
Arp 252$^b$ & 47.4$\pm$3.3 & 711    & 435$\pm$13 &   90$\pm$6	& 423   &  361$\pm$19      &   38$\pm$5    &   79$\pm$4	     & 1968$\pm$60 \\
" ESO 566-8   & 38.5$\pm$2.7 & 597 & 350$\pm$10	 &   84$\pm$6	 &  313   & \nodata	       &     \nodata    &	\nodata    	     &   \nodata    \\
" ESO 566-7   & 8.9$\pm$0.6  &  114 & 85$\pm$3	 &   6.3$\pm$0.4 &  110   &  \nodata	       &     \nodata    &	\nodata     &   \nodata     \\
NGC 5253    & 2.35$\pm$0.15 & 36.6   &  22.9$\pm$0.2 &  4.4$\pm$0.2&  14.2   & 2.583$\pm$0.009  & 0.479$\pm$0.004 &  0.571$\pm$0.001  &   
1.64$\pm$0.07 \\
\noalign{\smallskip}
\tableline
  \end{tabular}
  \begin{flushleft}
   $^a$ As IRAS data do not allow to distinguish regions within the HCG~31 group, this value considers the flux of all galaxy members.\\
  $^b$ We are considering the flux of two galaxies: members F1 and F2 for HCG~31~F; galaxies~A and B for SBS~0926+606; \emph{obj~1} and \emph{obj~2}  
for Tol~1457-262; and ESO~566-8 (A) and ESO~566-7 (B) for Arp 252.\\
  $^c$ As we commented in Paper~I, the \Ha\ flux for POX~4 provided by \citet{ME99} seems to be overestimated, hence we consider here the value  
provided by \citet{GildePaz03}, that is 0.61 times smaller. We also scale our \Ha\ flux of POX~4~Comp using this factor. 
  \end{flushleft}
\end{table*}

\section{Analysis of the star formation rates}

\begin{table*}[t!]
\centering
  \caption{\footnotesize{\SFR\ values (in units of \Moy) derived for each galaxy using different luminosities and calibrations. The references are:  
S07 = \citet{Salim07}; G84 = \citet{G84}; K98 = \citet{K98};  R01 = \citet{RSVB01} ; C92 = \citet{Condon92} ; C02 = \citet*{CCB02} using the  
expression for $M>$5 \Mo. The last column indicates the assumed value of the \SFR\ we estimate to each galaxy considering all available data.}}
%Last column compiles the $q$ parameter defined in \S\ref{firradio} (ver Ecuación~\ref{agnq}******).}}
  \label{sfrg}
  \tiny
  \begin{tabular}{l c cc  cc@{\hspace{6pt}} c@{\hspace{6pt}} c@{\hspace{6pt}}c@{\hspace{6pt}}  c@{\hspace{8pt}}c}
  \tableline
   \noalign{\smallskip}
Galaxy    & $FUV$  & $U$ & \multicolumn{2}{c}{$B$} &     \Ha   &  \FIR &  15 $\mu$m &  60 $\mu$m    &  1.4 GHz &   Assumed  \\
\cline{4-5}
\noalign{\smallskip}
        &S07   & Eq.~\ref{sfru}  &   G84   &  Eq.~\ref{sfrb}  &   K98$^a$   &  K98  &  R01       &  C92          &  C02       & \SFR$^b$ \\
\tableline
\noalign{\smallskip}    

%OBJETO  & SFUV & SFRU &  SFRB & SFRB mia &	  SFRHa     &	 SFRFIR 	   &	SFR15	       &    SFR60  	      &   SFR14a	     \\	 
	
HCG 31   & 4.0$\pm$0.3 & 2.8 & 0.59 & 2.5 &  4.1$\pm$0.4 &   3.3$\pm$0.3 & 1.2$\pm$0.2 & 2.8$\pm$0.2 & 2.5$\pm$0.2 & 3.1 \\% 3.3  \\
   " AC (NGC~1741)  & 2.02$\pm$0.12 & 1.5 & 0.27  & 1.2  &  2.88$\pm$0.18 &  \nodata & \nodata  &  \nodata  & 1.98$\pm$0.03  & 2.0 \\ % 2.3   \\
  " B  & 0.89$\pm$0.07 & 0.68 &  0.14  & 0.67  &  0.24$\pm$0.02 & \nodata & \nodata  &  \nodata    &	 0.19$\pm$0.03    & 0.62 \\% 0.44  \\
  " E  & 0.122$\pm$0.009 & 0.073 &  0.009 & 0.057   &  0.118$\pm$0.013  & \nodata & \nodata  &  \nodata  & \nodata & 0.10 \\ %0.12 \\
  " F  & 0.123$\pm$0.013 & 0.12 & 0.012 & 0.071  &  0.40$\pm$0.03 & \nodata & 	\nodata    &    \nodata	    &\nodata    &  0.20 \\ % 0.26  \\
  " G (Mkn~1090) & 0.83$\pm$0.05 & 0.81 & 0.16  & 0.77  &  0.51$\pm$0.03 & \nodata & 	\nodata  &  \nodata	 &  0.30$\pm$0.04  & 0.49 \\ % 0.55  \\
  " H  & 0.011$\pm$0.004 & \nodata & \nodata  & \nodata  & 0.004$\pm$0.001 & \nodata & 	\nodata    &      \nodata	    &\nodata  & 0.008 \\
Mkn 1087 &\nodata & 12.1  & 3.27  & 11.8  &   5.6$\pm$0.4  & 10.3$\pm$1.1 & 4.6$\pm$1.3 & 8.8$\pm$1.0 & 4.5$\pm$0.2     & 6.3   \\
" N   & \nodata & \nodata & 0.076 & 0.39 & 0.180$\pm$0.015  & \nodata & \nodata &\nodata & \nodata & 0.12 \\
Haro 15  &6.8$\pm$0.4 & 4.6 &  1.00  &  4.1  &   3.3$\pm$0.4  &  2.8$\pm$0.3 & 3.2$\pm$0.9 & 2.4$\pm$0.2 & 4.0$\pm$0.2   & 3.6 \\ % 3.8   \\
Mkn 1199 &\nodata & 3.7 &  0.84  &  3.5  &   3.9$\pm$0.6  &	 5.2$\pm$0.3 & 3.0$\pm$0.3 & 4.7$\pm$0.2 & 3.16$\pm$0.10    & 3.7 \\ % 4.0   \\
" NE     & \nodata& 0.22  & 0.048  &  0.26  &   0.07$\pm$0.02 & \nodata & 	\nodata    &      \nodata	    & \nodata  & 0.05 \\
Mkn 5    & 0.0222$\pm$0.0014 & 0.050 & 0.008 & 0.049 & 0.046$\pm$0.011&	$<$0.014  &	$<$0.03   &  0.067$\pm$0.008  &  $<$0.10  & 0.040 \\ %0.045 \\
IRAS~08208+2816& 13.1$\pm$0.8 & 6.4 & 1.48& 5.8 &  11.3$\pm$0.9  &	11.4$\pm$1.0 &$<$1.68 &   9.7$\pm$0.8   &  16.4$\pm$0.7  & 11.6 \\ % 12.4  \\
IRAS~08339+6517& 6.4$\pm$0.4  & 8.2 & 1.92& 7.3 &  9.5$\pm$0.5  & 8.93$\pm$0.08& 5.6 $\pm$ 0.6 &  8.36 $\pm$ 0.06 &  6.2$\pm$0.9  & 7.0 \\ %7.9   \\
" Comp          & 0.095$\pm$0.008 & 0.37  & 0.087 &  0.44  & 0.17$\pm$0.02  & \nodata & 	\nodata    &      \nodata	    &\nodata  & 0.10 \\
POX 4   & 0.57$\pm$0.03 & 0.93 & 0.15 & 0.72  &  1.85$\pm$0.06 & 0.31$\pm$0.02 &$<$7.48 &  0.31$\pm$0.03 &  0.26$\pm$0.03 &  0.54 \\ %0.66  \\  
" Comp  &0.054$\pm$0.006 & 0.030 & 0.0063 & 0.041 & 0.012$\pm$0.004 & \nodata & 	\nodata    &   \nodata	 &\nodata & 0.031 \\ % 0.033 \\
UM 420  & 2.01$\pm$0.11 & 1.9 & 0.30  &  1.4  &  3.7$\pm$0.2  &	6.4$^c$	   &	\nodata   &  5.4$^c$	 &    1.9$\pm$0.8   &  2.1 \\ % 3.9  \\
SBS~0926+606 &  3.4$\pm$0.3 & 0.50 & 0.08  & 0.40 &  0.94$\pm$0.06 & 0.26$\pm$0.03 &$<$0.86  &    0.20$\pm$0.03 &  0.25$\pm$0.06 & 0.95 \\ %1.0 \\
      " A & 0.53$\pm$0.03 & 0.28  &  0.04 & 0.20 &  0.75$\pm$0.04 & \nodata  &	\nodata    &     \nodata	&    \nodata  & 0.52 \\ % 0.64  \\
      " B$^d$ & 2.82$\pm$0.25 & 0.25 &  0.04 & 0.22 &  0.19$\pm$0.02 & \nodata  &	\nodata  &    \nodata	&    \nodata     &      1.4?   \\
	  
SBS~0948+532 & \nodata & 1.1 & 0.11  & 0.53 &  6.2$\pm$0.2  & \nodata  &	\nodata     &    \nodata  &    $<$0.95       & 4.2 \\ % 6.2   \\
SBS~1054+365 & 0.0073$\pm$0.0005 & 0.013 & 0.0014& 0.019& 0.036$\pm$0.001&  0.016 &	$<$0.01  &   0.015  &  0.025$\pm$0.004  & 0.018 \\ % 0.020 \\
SBS~1211+540 & 0.0062$\pm$0.0004 & 0.009 & 0.0007& 0.009& 0.011$\pm$0.001& \nodata &  \nodata  &   \nodata  &    $<$0.01  & 0.007 \\ % 0.009 \\
SBS~1319+579 & 0.069$\pm$0.004   & 0.59 & 0.12  & 0.58 & 0.189$\pm$0.012 &  0.24	 &	  \nodata   &    0.17  &  $<$0.07  & 0.15 \\ %0.17  \\
SBS~1415+437 & 0.0331$\pm$0.0018 & 0.022 & 0.0029& 0.020& 0.039$\pm$0.002 & \nodata& \nodata  &  \nodata &    $<$0.01  & 0.030 \\ %0.036 \\
III~Zw~107   & \nodata & 2.3 & 0.51  & 2.2  &  3.19$\pm$0.15 &  2.3$\pm$0.4  & $<$2.24 &  2.0$\pm$0.3  &  1.52$\pm$0.09  & 2.0 \\ % 2.5   \\  
Tol 9        & 0.77$\pm$0.06 & 1.0 & 0.23 & 1.1 &  1.82$\pm$0.13 & 1.59$\pm$0.07 &0.8$\pm$0.2& 1.19$\pm$0.08& 2.18$\pm$0.04 & 1.3 \\ %1.5  \\
Tol 1457-262 &  3.2$\pm$0.2 & 2.6 & 0.53  & 2.3  &  5.0$\pm$0.3  & 3.7$\pm$0.3   &	$<$1.99   &   3.4$\pm$0.2 & 5.4$\pm$0.3 & 3.8 \\ %4.1   \\
     " Obj~1 & 2.37$\pm$0.13 & 1.8 & 0.35 & 1.6 &   3.7$\pm$0.2 & \nodata&	   \nodata     &     \nodata &  \nodata & 2.4 \\ % 3.0 \\
    "  Obj~2 & 0.74$\pm$0.09 & 0.87  & 0.17 & 0.83 &   1.37$\pm$0.06 & \nodata&	   \nodata   &     \nodata	 &  \nodata & 0.83 \\ % 1.1 \\
	 "  \#15 & 0.089$\pm$0.017& 0.09 & 0.018 & 0.10 & 0.038$\pm$0.003 & \nodata & \nodata&	   \nodata     &     \nodata & 0.06 \\ % 0.064 \\
Arp 252      & 3.84$\pm$0.26 & 5.3 & 1.26  & 5.0  &  7.2$\pm$0.5  & 16.2$\pm$0.9 & 11.6$\pm$1.4 & 15.4$\pm$0.8 & 49.2$\pm$1.5  & 10 \\			 
    " ESO 566-8 & 3.12$\pm$0.22 & 4.5 & 1.01 & 4.1  &  6.7$\pm$0.5 & \nodata&	   \nodata     &   \nodata	  & \nodata & 3.8 \\ % 5.0  \\
    " ESO 566-7 & 0.72$\pm$0.04 & 1.0 & 0.25 & 1.1 & 0.50$\pm$0.04  &  \nodata&	   \nodata     &   \nodata	  & \nodata &  0.53 \\ %0.61    \\
NGC 5253    &  0.190$\pm$0.010  & 0.36 & 0.07 & 0.35 &  0.348$\pm$0.017 & 0.12  &	0.15  &    0.11    &    0.041$\pm$0.002  & 0.14 \\ % 0.16 \\ 
\noalign{\smallskip}
\tableline
  \end{tabular}
  \begin{flushleft}
  $^a$ Considering the new correlation between the \SFR\ and the \Ha-luminosity provided by \citet{Calzetti07}, the \Ha-based \SFR s are 0.67 times  
the values shown here. \\  %$^b$ For $M>$5 \Mo. \\
% $^c$ For $M>$0.1 \Mo. 
$^b$ We consider the \Ha-based \SFR\ values provided by the \citet{Calzetti07} calibration to estimate the average \SFR\ (see text). \\
$^c$ The \FIR\ and 60~$\mu$m luminosities in UM~420 are overestimated because of the contribution of the foreground galaxy UGC~01809 (see Fig.~13 and  
Sect.~3.9.1 in Paper~I). \\
$^d$ SBS~0926+606~B shows extended \FUV\ emission, as it is seen in the right panel of Figure~\ref{galex}. The values of the \SFR\ provided by its  
\FUV\ and \Ha\ emission are very different. 
  \end{flushleft}
\end{table*}

\begin{figure*}[t!]
\centering
\begin{tabular}{cc}
\includegraphics[angle=270,width=0.45\linewidth]{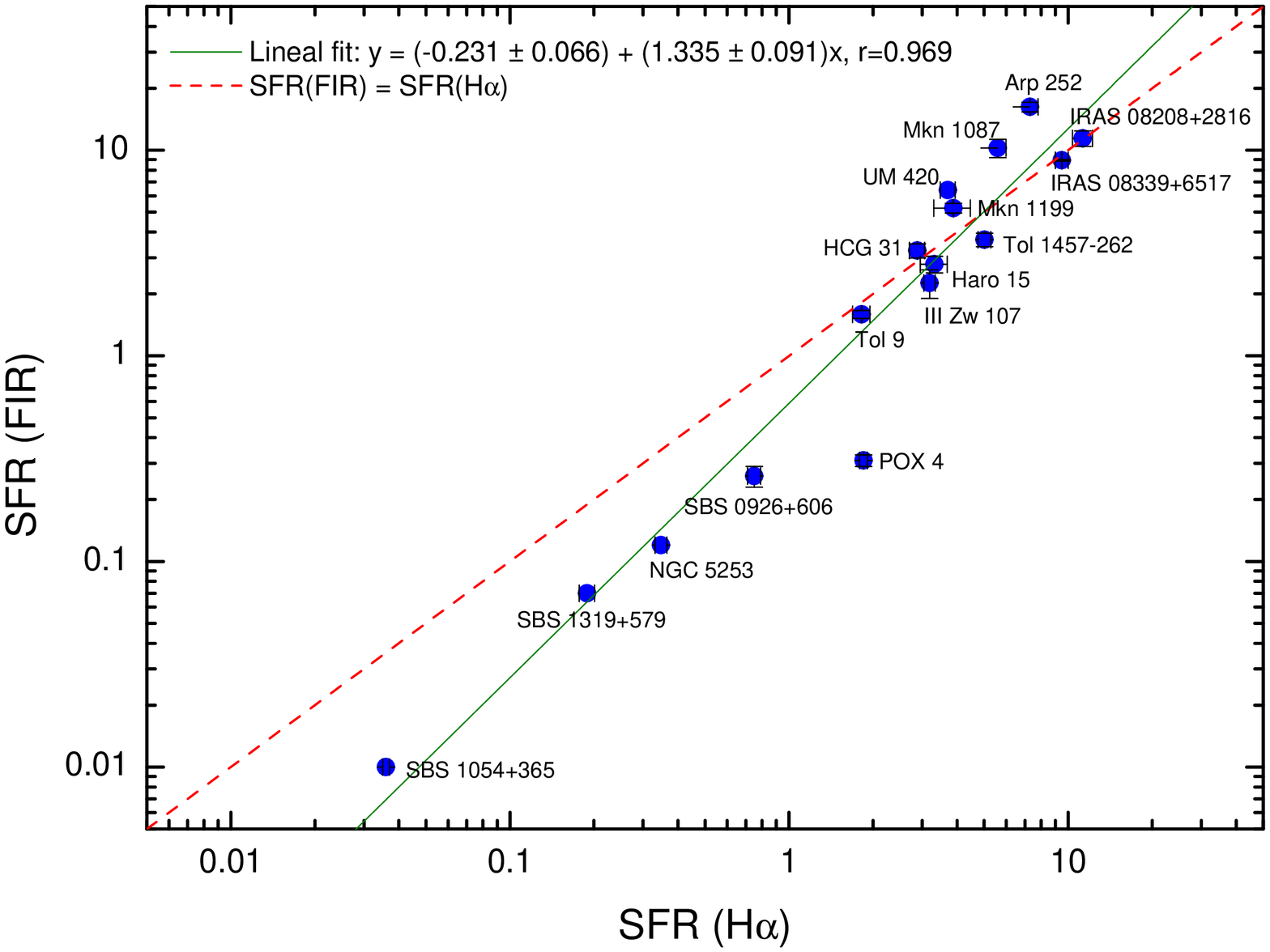} &
\includegraphics[angle=270,width=0.45\linewidth]{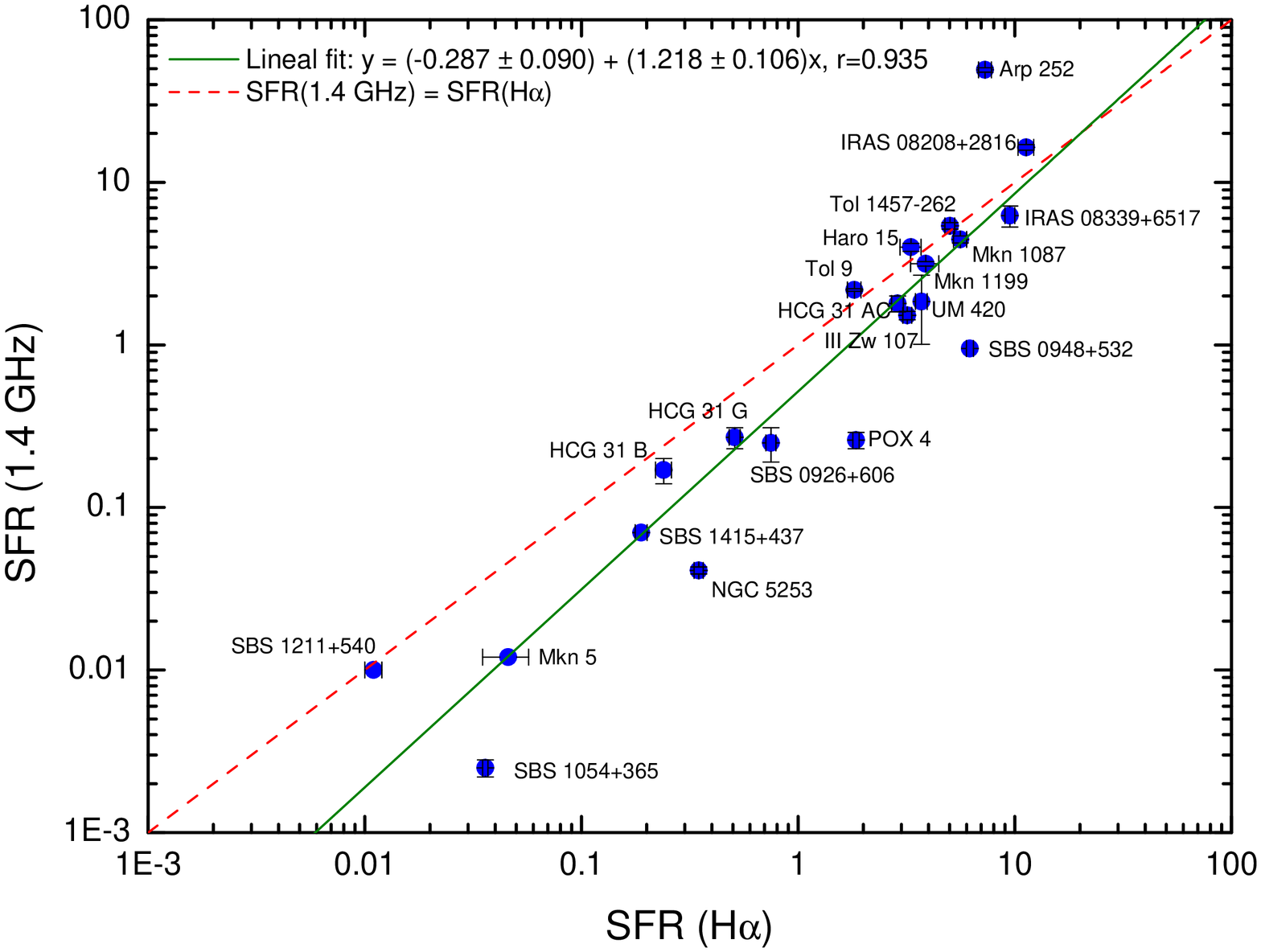} \\
\includegraphics[angle=270,width=0.45\linewidth]{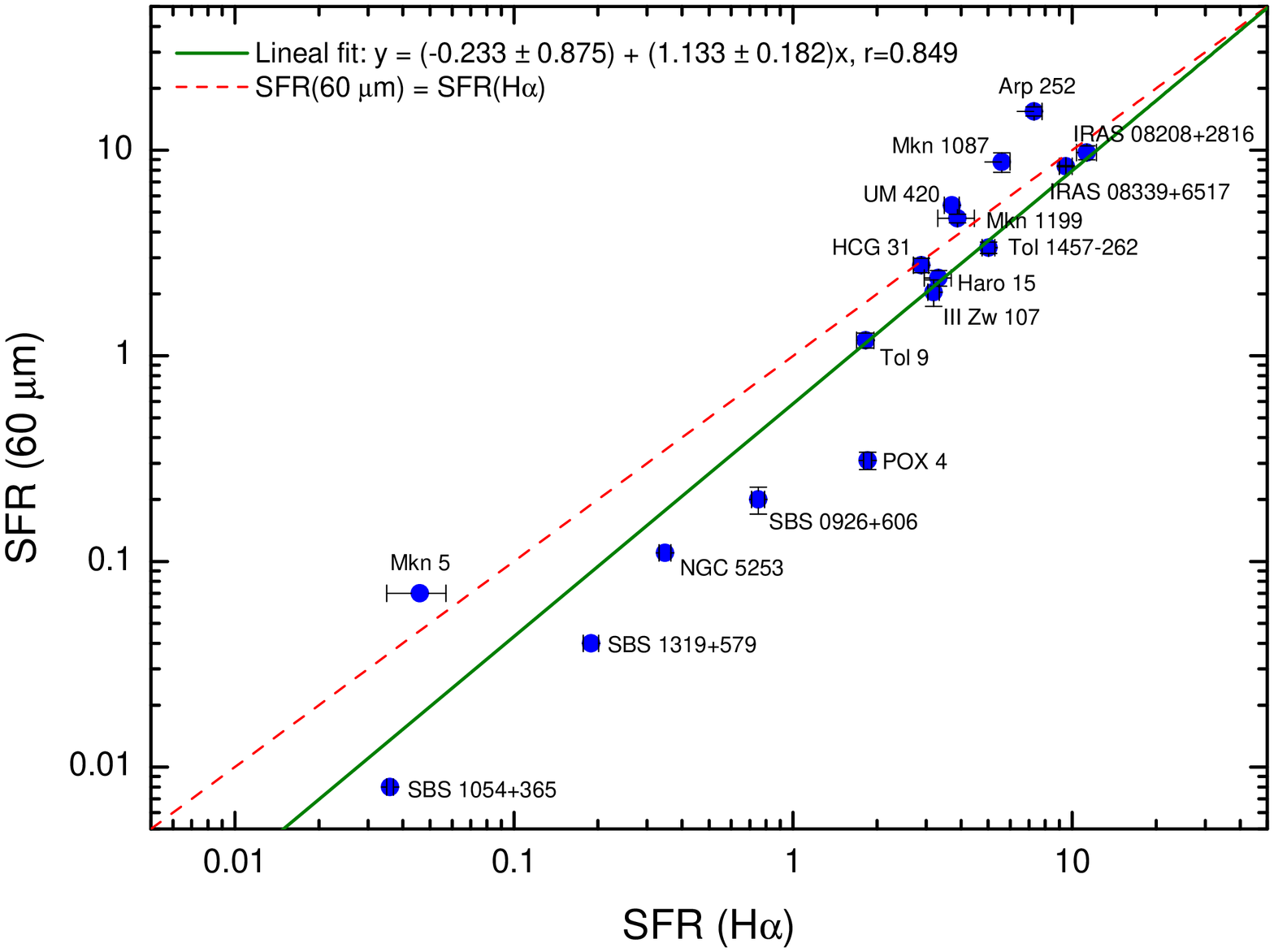} &
\includegraphics[angle=270,width=0.45\linewidth]{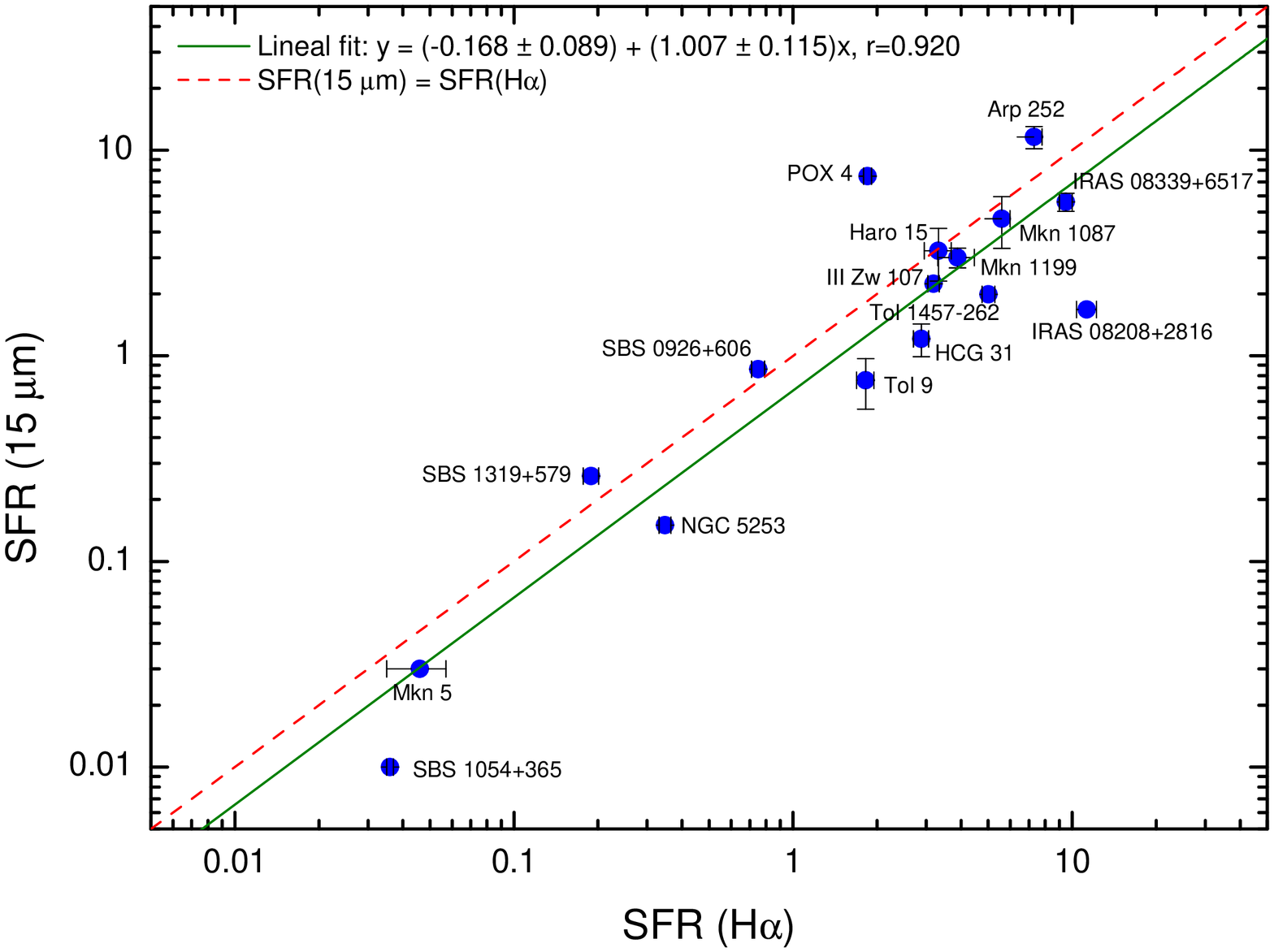} \\
\end{tabular}
\protect\caption[ ]{\footnotesize{Comparison between the \Ha-based \SFR\ --corrected for both extinction and [\ion{N}{ii}] contribution and assuming  
\citet{K98},  x-axis-- with the \SFR s derived using the \FIR, 15 $\mu$m, 60 $\mu$m and 1.4 GHz luminosities. The dotted red lines indicate the  
position with equal \SFR, the continuous green lines show a lineal fit to the data.}}
\label{sfr}
\end{figure*}

The star formation rate (\SFR), defined as the stellar mass formed per unit time, is the standard parameter used to quantify the star formation  
activity in galaxies. The determination of the \SFR\ is fundamental to get a proper understanding of the formation and evolution of the galaxies. As  
we said in the introduction, different techniques involving different data sets from $UV$ to radio 
%(i.e. Kennicutt 1998; Dopita et al. 2002; Calzetti et al. 2007, and references within them) 
often yield to different \SFR\ results. 
Part of the problem is related with the unknown amount of extinction within each particular galaxy \citep{Cal01}, such as the amount of dust  
obscuration depends on the galaxy mass, galaxy type, the chemical evolutionary state, gas content or even if the galaxy is interacting or merging  
with another independent object. As we explained in the previous section, \FIR\ and radio data provide an extinction-free estimation of the \SFR,  
while \FUV\ emission nicely traces the very young stellar component. Here, we analyzed all the available multiwavelength data for our sample of WR  
galaxies, including our reddening-corrected \Ha\ estimations (see Paper~I), to determine in a comprehensive way the \SFR\ within these objects.

Table~\ref{lumi} compiles all \FUV, $U$, $B$, \Ha, $H$, \FIR, 15~$\mu$m, 60~$\mu$m, and 1.4~GHz luminosities for the galaxies analyzed in this work. 
The $U$, $B$, and $H$ luminosities were computed from the reddening-corrected absolute magnitudes in the $U$, $B$ and $H$ bands (see Paper~I) using  
the standard equation $\log L_x =0.4\times(M_{x,\odot}-M_{x})$, and considering $M_{U,\odot}$=5.58 and $M_{B,\odot}$=5.48 \citep{Bessell98} and  
$M_{H,\odot}$=3.35 \citep{Colina96}.

We used the values listed in Table~\ref{lumi} to estimate the \SFR\ that each object is experiencing, following the different multi-wavelength  
techniques explained in the previous section. The values of the \Ha-based \SFR\ are extracted from Paper~I and consider the \citet{K98} calibration.  
Recently, \citet{Calzetti07} re-calibrated the relationship between the \Ha-luminosity and the \SFR; the \Ha-based values of the \SFR\ provided by  
\citet{Calzetti07} are 0.67 times the values derived using the \citet{K98} calibration. 
%in Sect.~2 (\FUV, \FIR\ and radio frequencies) and in Paper~I (\Ha).
 
Table~\ref{sfrg} compiles all \SFR\ values derived for each galaxy. From this table, it is evident that the agreement between values obtained using  
different methods is usually good, although sometimes we find clear discrepancies (i.e. POX~4, NGC~5253). 
%Taking into account all values, we compile in the last column 
%of Table~\ref{sfrg} the \SFR\ we assume for each galaxy. 
We note that for systems that involve two or more galaxies (HCG~31, SBS~0926+606, Tol~1457-262 and Arp~252) we list both the global and individual  
\SFR s, because the \FIR\ and the radio data do not have enough spatial resolution to distinguish the emission coming from different members, but  
\FUV\ and \Ha\ data do. We also considered HCG~31~F1 and F2 as a single entity (HCG~31~F) because the available \HI\ data include both \TDG\  
candidates.  

\begin{figure}[t!]
\centering
\includegraphics[angle=270,width=\linewidth]{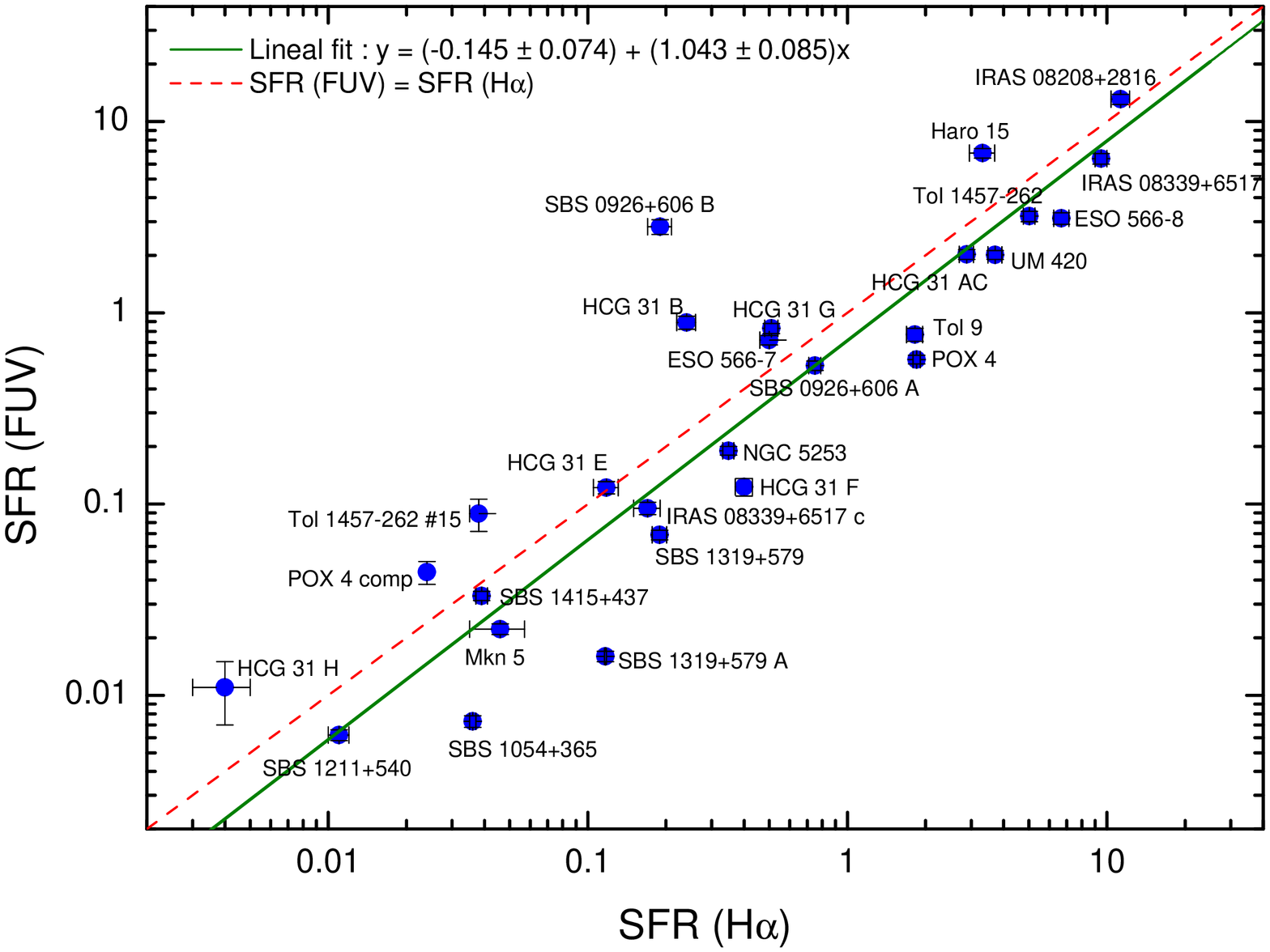}
\protect\caption[ ]{\footnotesize{Comparison between the \Ha-based \SFR\ --corrected for both extinction and [\ion{N}{ii}] contribution and assuming  
\citet{K98}, x-axis-- with the \SFR s derived using the \FUV\ luminosities. The dotted red line indicates the position with equal \SFR, while the  
continuous green line shows a fit to the data.}}
\label{sfrfuv}
\end{figure}

\begin{figure*}[t]
\begin{tabular}{cc}
\includegraphics[angle=270,width=0.45\linewidth]{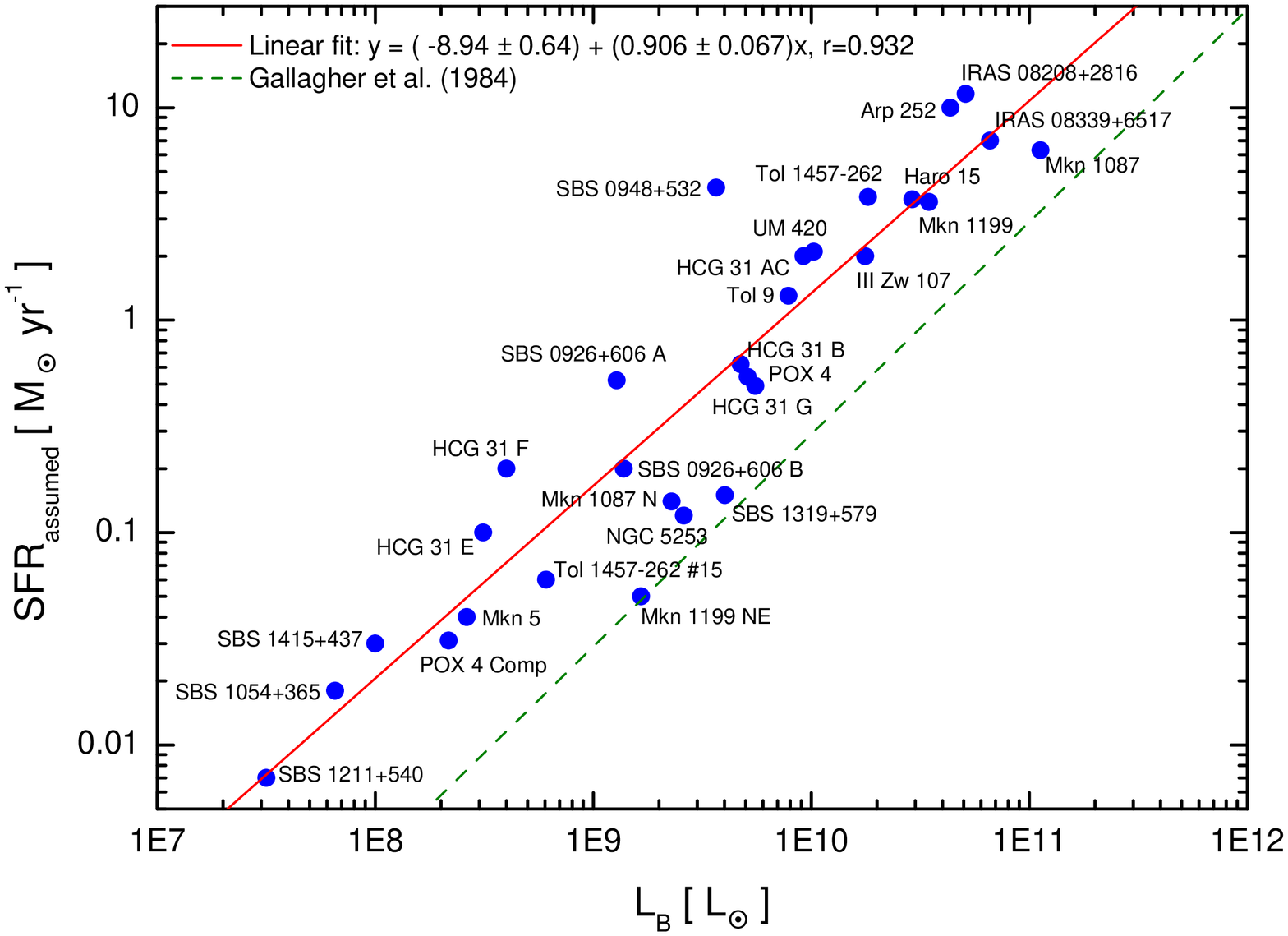} &
\includegraphics[angle=270,width=0.45\linewidth]{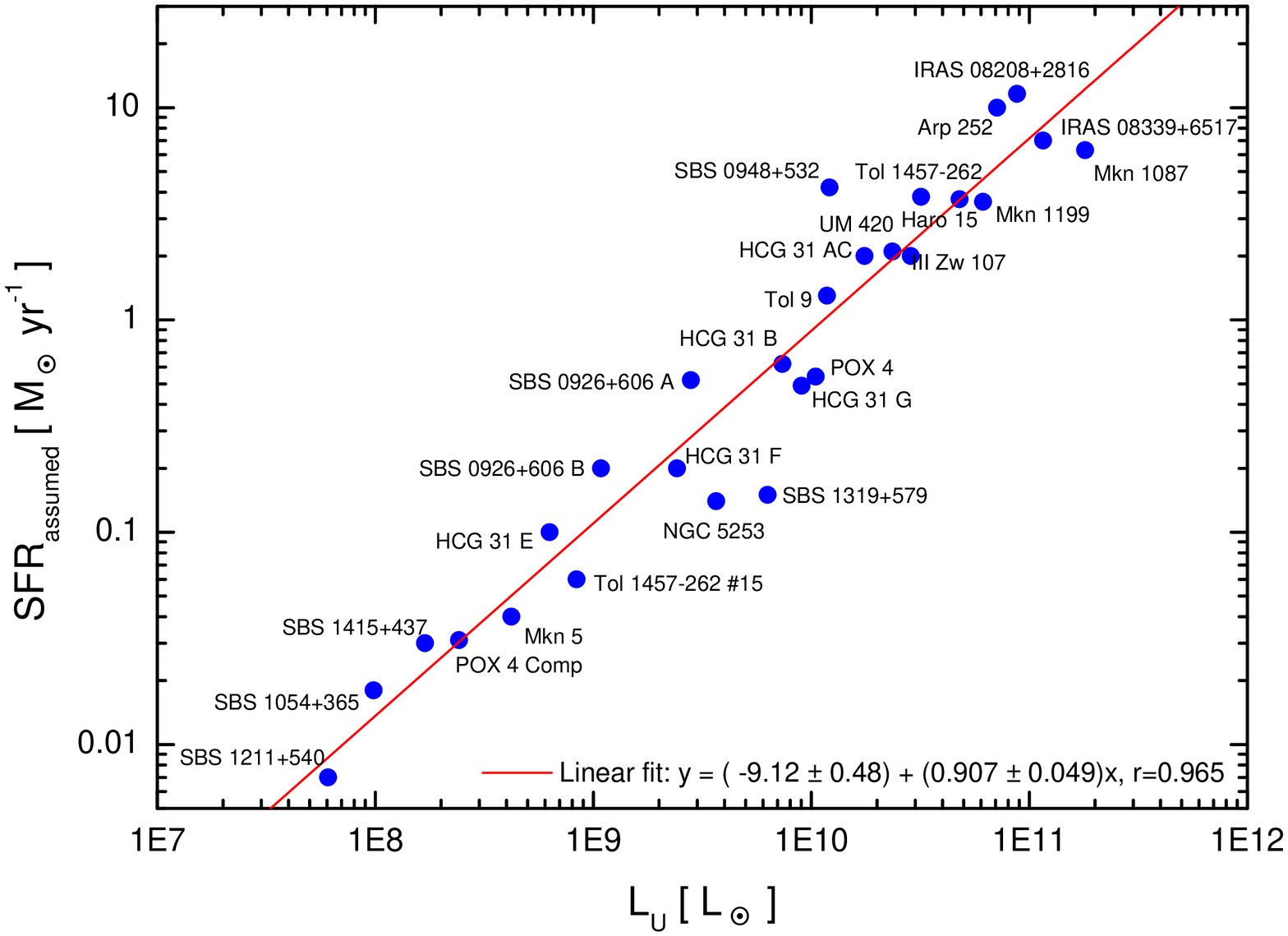} \\
\end{tabular}
\protect\caption[ ]{\footnotesize{Assumed \SFR\ vs. $B$-luminosity (left panel) and $U$-luminosity (right panel) for the analyzed galaxies.  
Luminosities are plotted in solar units. The best fit (in logarithm scale) to our data are plotted with a continuous red line. The previous  
calibration given by Gallagher et al. (1984) between the \SFR\ and the $B$-luminosity is shown by a discontinuous green line.}}
\label{sfrlb}
\end{figure*}

Figure~\ref{sfr} compares our \Ha-based \SFR\ (corrected for both extinction and [\ion{N}{ii}] contribution as we explained in Appendix~C of Paper~I)  
with the \SFR\ estimations derived from  \FIR, 15~$\mu$m, 60~$\mu$m and 1.4~GHz luminosities. The diagram involving  $L_{\rm 15\, \mu m}$ seems to  
show a higher scatter at higher \SFR, but this calibration is more uncertain. As a particular case, Arp~252 always shows a disagreement between the  
\SFR\ derived from \Ha\ and other parameters, remarking with the 1.4 GHz luminosity. The main object within Arp~252 is the bright galaxy ESO~566-8.  
This behavior, together the fact that the \FIR-radio-continuum relation is not satisfied in this system (see below) strongly suggest that ESO~566-8  
has some activity different to its starbursting nature (an \AGN\ or a radio-galaxy), something we already commented when we analyzed this system (see  
Sect.~3.19.2 of Paper~I). The rest of the objects agree rather well when comparing values obtained from different calibrations. As previous authors  
pointed out (i.e. Dopita et al. 2002; James et al. 2005), the correction of the \Ha\ fluxes for both extinction and [\ion{N}{ii}] emission is vital  
to get a reliable estimation of the SFR using \Ha-images.

Although the agreement between the \Ha-based \SFR\ and the \SFR s derived using \FIR\ and radio luminosities is good, we observe that the values  
provided using the \Ha\ luminosity are slightly higher than those estimated using the other calibrations. The difference seems to be higher at lower  
\Ha-luminosities.  A linear fit to the data (green continuous lines in Figure~\ref{sfr}) confirms this trend. The zero-points of the fits (0.59,  
0.52, 0.58, 0.68 for the \Ha-\FIR, \Ha-1.4~GHz, \Ha-60~$\mu$m, and \Ha-15~$\mu$m relations, respectively) indicate that, for \SFR=1 \Mo\,yr$^{-1}$,  
the value of the \SFR\ provided by \Ha-luminosity is $\sim$0.6 times the \SFR\ values estimated using the other relations. Bell (2003) concluded that  
both radio and \FIR\ luminosities underestimate the \SFR\ for low-luminosity galaxies because the non-thermal emission seems to be suppressed by a  
factor of 2--3 in dwarf objects. However, the difference is not significative if we use the \citet{Calzetti07} calibration instead of the \citet{K98}  
calibration to derive the \Ha-based \SFR. 

The comparison of the \FUV-based with the \Ha-based \SFR\ (Figure~\ref{sfrfuv}) also shows a good agreement: except for some few objects (remarkably  
SBS~0926+606~B\footnote{As we commented before, the \FUV\ emission observed in SBS~0926+606~B is much more extended that the \Ha\ emission, and hence  
the derived \SFR\ is more than one order of magnitude higher using the \FUV\ than the \Ha\ emission.}) both relations provide similar values. We also  
observe that the \FUV-based \SFR s seem to be slightly lower than the \Ha-based \SFR s. A linear fit to the data (shown in Figure~\ref{sfrfuv} with a  
continuous green line and with a correlation coefficient of $r$=0.927) indicates that the \FUV-based \SFR\ is, on average, $\sim$0.71 times the  
\Ha-based \SFR. This value is similar to the factors found before when comparing the \Ha-based \SFR\ with the \FIR- and radio-based \SFR s.  
Interestingly, all these numbers are coincident with the ratio between the \citet{K98} and the \citet{Calzetti07} calibrations to the \SFR\ using the  
\Ha\ flux, $SFR_{\rm C07}$(\Ha)/$SFR_{\rm K98}$(\Ha)=0.67. We therefore conclude that the new \Ha-based calibration provided by \citet{Calzetti07}  
should be preferred over the widely-used \citet{K98} calibration when computing the \SFR\ from \Ha\ luminosities. The \SFR\ estimated for each object  
considering all available multiwavelength data, and listed in last column of Table~\ref{sfrg}, has been computed considering the \citet{Calzetti07}  
value.
%0.71
Finally, we must say that there are increasing evidences that the \Ha\ luminosity underestimates the \SFR\ relative to the \FUV\ luminosity in dwarf  
galaxies with \SFR$\leq$0.01 \Moy\ (i.e., Lee et al. 2009) and hence the \FUV-based \SFR\ should be preferred over the \Ha-based \SFR\ in those  
systems.

\begin{figure*}[t]
\begin{tabular}{cc}
\includegraphics[angle=270,width=0.45\linewidth]{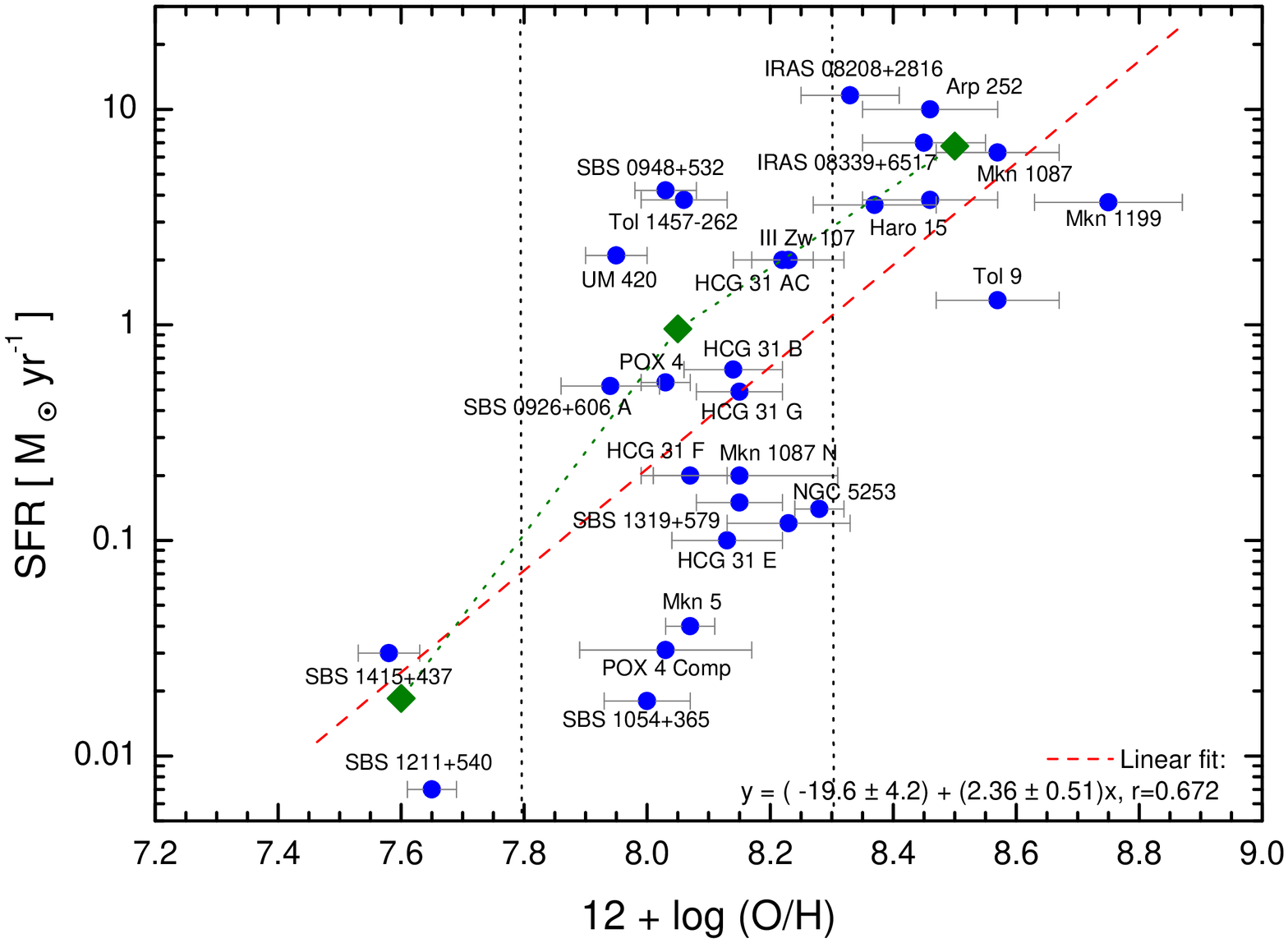} &
\includegraphics[angle=270,width=0.45\linewidth]{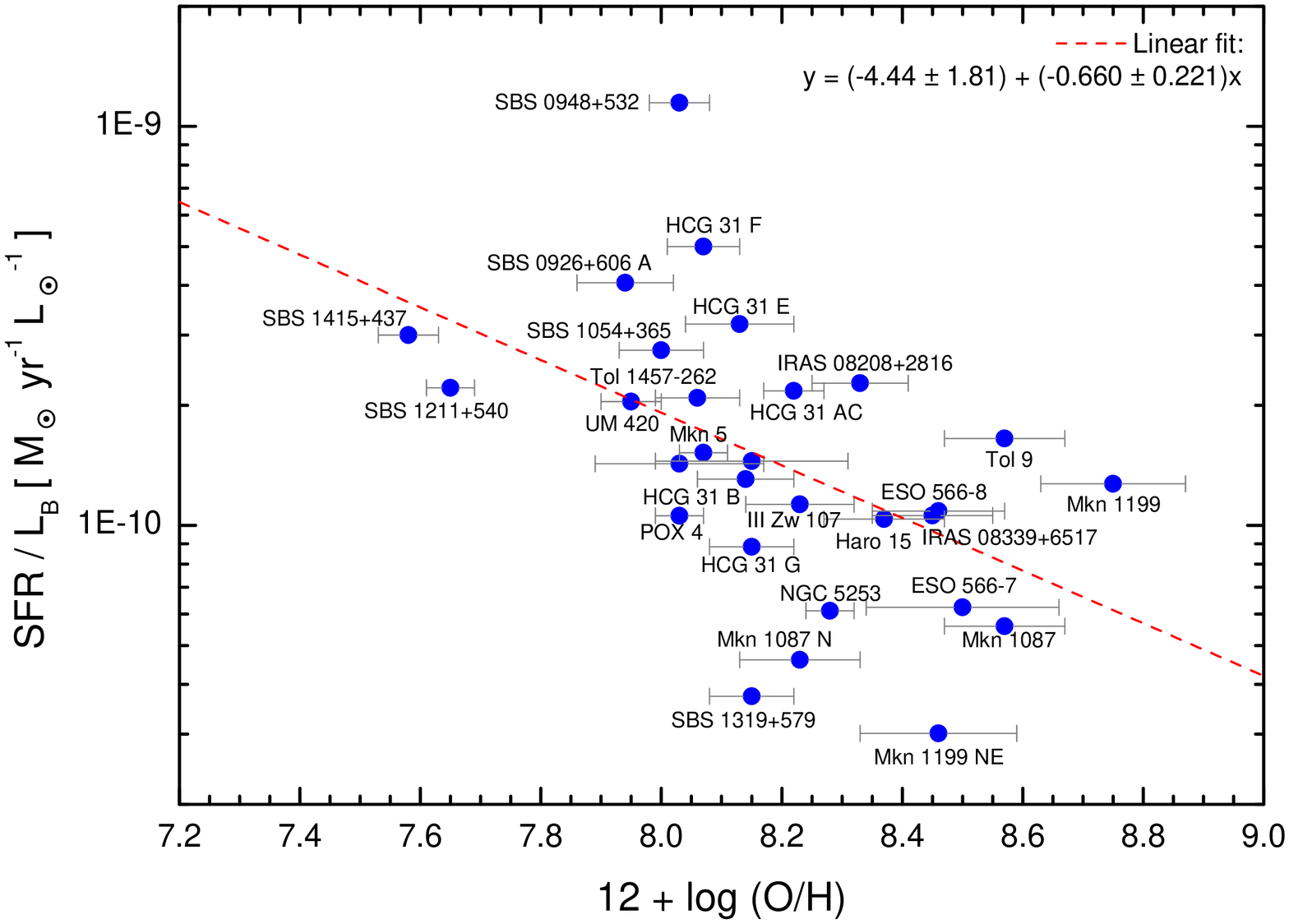} \\
\end{tabular}
\protect\caption[ ]{\footnotesize{\SFR\ vs \abox\ (\emph{left}) and \SFR/$L_B$ vs \abox\  (\emph{right}) for our sample of WR galaxies. The  
red-dotted line indicates a fit to our data. Green diamonds in left panel plot the average value obtained in the low, intermediate and high-metallicity regimes.}}
\label{sfrabox}
\end{figure*}

\subsection{$L_B$-\SFR\ and $L_U$-\SFR\ relations for starburst galaxies}

Just for comparison, we also estimated the \SFR\ from the $B$-luminosity using the calibration provided by Gallagher et al. (1984). 
$SFR_B$ represents the star formation activity occurred in the last few hundreds of \Myr, while the rest of the calibrations are tracing the massive  
stars and the nebular emission of the gas that only last for some few tens of \Myr. For our galaxy sample, $SFR_B$ is always lower than the \SFR\  
derived from the other calibrations, as we should expect because of the starbursting nature of the analyzed galaxies. 
%Only one of them, UM~420, seems to have a $SFR_B$ similar to that derived 
%from the \Ha, \FIR\ and 1.4 GHz luminosities. That may indicate that the galaxy 
%UM~420 has been forming stars at a relatively high rate for at least 100 \Myr.  
The value of the \SFR$_B$ in Mkn~1087 using Gallagher et al. (1984) equation is only half of that estimated from other calibrators, remarking its  
Luminous Blue Compact Galaxy (\LCBG) nature \citep{LSER04b}. 

We used our data to establish a new relation between the \SFR\ and the $B$-luminosity, that should be applied only in starburst galaxies 
%(similar to those analyzed here) 
and just as a first estimation of the actual \SFR. The left panel of Figure~\ref{sfrlb} shows the relation between $L_B$ (in solar units) and the  
assumed \SFR\ for all our galaxies. Despite some clear discrepancies between some galaxies that show very different \SFR\ for a similar  
$B$-luminosity (for example, just compare members G and F of HCG 31), we observe a good agreement, having galaxies with higher $B$-luminosities  
higher star-formation activity. The discrepancies are consequence of the different star-formation histories of the galaxies (relative contribution  
and age of the underlying stellar population, metallicity, age of the most recent star-formation event). A linear fit to our data provides the  
relation
\begin{equation}
\label{sfrb}
%SFR_{B,{\rm starbursts}} = 9.057\times10^{-9} L_B^{0.825}, r=0.906
SFR_{B,{\rm starbursts}} = 1.148\times10^{-9} L_B^{0.906}
\end{equation}
being $L_B$ expressed in units of \Lo. The correlation coefficient of this fit is $r$=0.932. The third column in Table~\ref{sfrg} compiles the  
$SFR_B$ computed for each galaxy using this equation. The relation obtained by Gallagher et al. (1984) gives values one order of magnitude lower than  
those we obtain with our new calibration. 

We also computed a relation  between the \SFR\ and the $U$-luminosity for our sample galaxies. The right panel of Figure~\ref{sfrlb} shows such  
relation. A linear fit to the data yields
\begin{equation}
\label{sfru}
%SFR_{U,{\rm starbursts}} = 3.467\times10^{-9} L_U^{0.847}, r=0.934
 SFR_{U,{\rm starbursts}} = 7.59\times10^{-10} L_U^{0.907},
\end{equation}
being $L_U$ expressed in units of \Lo. This fit has somewhat smaller scatter than our derived \SFR-$L_B$ relation, resulting in a correlation  
coefficient $r$=0.965. However, the slope for both calibrations (0.894$\pm$0.070 and 0.900$\pm$0.054 
%(0.825$\pm$0.080 and 0.847$\pm$0.069 
for \SFR-$L_B$ and \SFR-$L_U$, respectively) are similar.  
We remark that Equations~\ref{sfrb} and \ref{sfru} can not be applied to galaxies with no strong star-formation activity, because the \SFR\ value  
derived from them will be overestimated (as it is happening in Mkn~1199~NE or IRAS~08339+6517~comp).

\subsection{Comparison of \SFR\ and metallicity}

Left panel of Figure~\ref{sfrabox} compares the assumed \SFR\ with the oxygen abundance computed for each galaxy. 
We estimated the average \SFR\ values in the low (\abox$<$7.8), intermediate (7.8$<$\abox$<$8.3) and high (\abox$>$8.3) metallicity regimes, the we plot 
in this figure using green diamonds. As we see, the dispersion in the intermediate-metallicity range is quite high, but that is just a consequence of the star-formation
history of each particular galaxy (see Paper~IV), as in this metallicity regime lies both very dwarf objects (i.e., Mkn~5, SBS~1054+365) and large and bright star-forming galaxies 
(i.e., Tol~1456-262, III~Zw~107) which share a relatively similar chemical history. Besides the large dispersion in the intermediate-metallicity regime, it is clear that galaxies with higher  
metallicity have higher global star formation rates. That is a consequence of the building of the galaxies, because more massive objects are more  
metal-rich than less massive galaxies (see below) and, hence, when the starburst is initiated, galaxies with higher mass (and with higher  
metallicities) will create stars at a higher rate than those found in smaller objects. The comparison of the \SFR\ per $B$-luminosity, \SFR/$L_B$  
with the metallicity (Figure~\ref{sfrabox}, right) also
shows a tremendous dispersion for 12+log(O/H) between 8.0 and 8.2.
However, we observe that \SFR/$L_B$ decreases with increasing oxygen abundance (the red-dotted line shows a fit to the data), indicating that  
galaxies with lower metallicity (and, therefore, less massive objects) have stronger star-forming bursts than those found in higher metallicity (more  
massive) objects.
%because \SFR/$L_B$ increases with decreasing oxygen abundance. 
%The most discordant point in Figure~\ref{sfrabox} (right) is UM~420, that 
%has a low \SFR/$L_B$ for its metallicity. As we said above, this may indicate that 
%UM~420 have had a high and constant \SFR\ in the last 100 Myr, perhaps the 
%star-formation activity is not so strong now and it started to decrease, 
%as their \Ha, FIR and 1.4 GHz luminosities suggest. 
%On the other hand, 
SBS~0948+532 has the highest \SFR/$L_B$ in our sample, indicating the strength of the starburst, as we saw when analyzed its photometric properties  
(see Sect.~3.11 in Paper~I). On the other hand, SBS~1319+579 has a very low \SFR/$L_B$ in comparison with \BCDG s of similar characteristics,  
indicating the peculiarity of this galaxy. We will see below that other properties of SBS~1319+579 show additional discrepancies with the average  
behavior in \BCDG s, suggesting that the star-formation activity has been somewhat suppressed in this object. For example, the gas depletion  
timescale is extremely long for a starburst galaxy, ($\tau\sim$12.7 Gyr, see Table~\ref{radiofirt}).   

\subsection{A $L_X$-\SFR\ relation for starburst galaxies}

Although several relations between the X-ray luminosity and the star formation rate have been proposed (i.e., Ranalli et al. 2003; Lou \& Bian 2005)  
they seem not to be appropriate for young starbursting systems. For example, as we explained in the analysis of the \LCBG\ IRAS~08339+6517  
\citep*{LSEGR06}, the relation provided by \citet{RCS03} 
%(Equation~\ref{eRCS03}) 
gives a very high \SFR\ value (61.8 \Moy) in comparison with the estimations obtained using other frequencies (6--8 \Moy). \citet{SS98a} showed that  
the X-ray luminosities in WR galaxies are substantially higher that those found in non-WR galaxies with similar $B$-luminosity. That is a consequence  
of the higher rate of superbubbles and supernova explosions in WR galaxies. 

\begin{figure}[t!]
\includegraphics[angle=270,width=\linewidth]{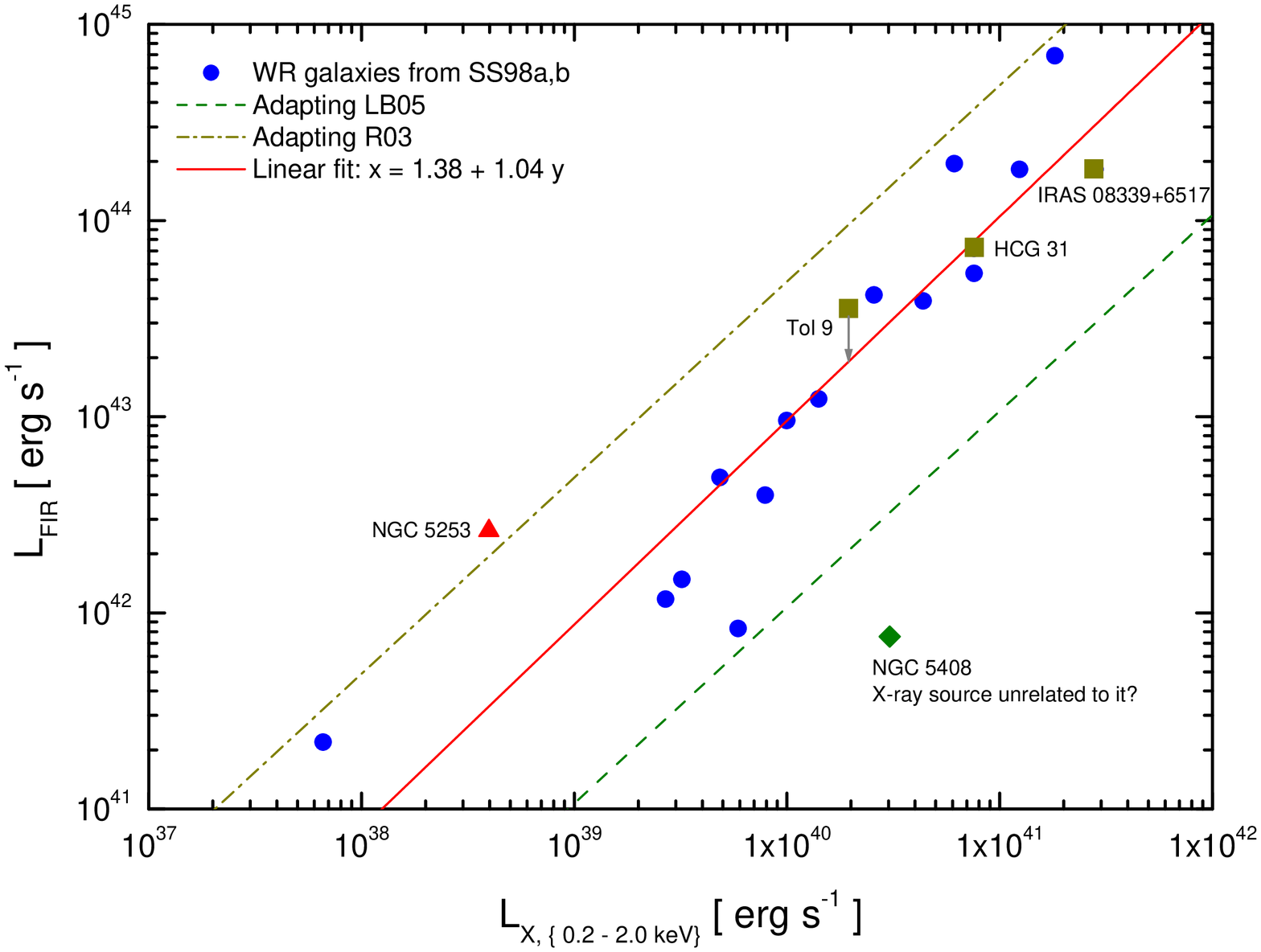}
\protect\caption[ ]{\footnotesize{X-ray luminosity in the 0.2--2.0 keV range vs. \FIR\ luminosity for the sample of WR galaxies analyzed by Stevens  
\& Strickland (1998a,b). The red continuous line is the best fit to the data, excluding the values for NGC~5253 and NGC~5408. The green discontinuous  
line is the relation obtained using the \SFR-$L_X$ calibration provided by \citet{Lou05}, while the yellow dotted-dashed line is the relation  
obtained from the \citet{RCS03} calibration. The three additional WR galaxies of our sample for which X-ray data are available (HCG~31,  
IRAS~08339+6517, and Tol~9) are indicated with dark yellow squares.}}
\label{gsfrx}
\end{figure}

We have used the sample of WR galaxies analyzed by Stevens \& Strickland (1998a,b) to get a tentative calibration between \SFR\ and $L_X$ for this  
kind of objects. These authors obtained X-ray data in the 0.2--2.0 keV range using the satellite ROSAT. We have checked which of these galaxies also  
possess \FIR\ data from the \IRAS\ satellite, and established a relation between $L_{FIR}$ and $L_X$, as it is shown in Figure~\ref{gsfrx}. Only 18  
galaxies have available data for both luminosities. NGC~5253 was included in the Stevens \& Strickland (1998a,b) analysis, but they indicated that  
the X-ray emission in this object is very peculiar. The X-ray emission measured in NGC~5408 may be unrelated with the galaxy. Hence, neglecting the  
contribution of these two galaxies, the linear fit to the data gives 
\begin{eqnarray}
L_{FIR} = 24 \times L_X^{1.04}, 
%\log L_{FIR} = (6.85 \pm 6.19) + (0.902 \pm 0.154)\log L_X,
\end{eqnarray}
being the correlation coefficient $r$=0.929. Considering the calibration given by \citet{K98} between $L_{FIR}$ and \SFR\ (Equation~\ref{esfrfir}),  
we find the following calibration between soft X-rays (0.2--2.0 keV) and \SFR: 
\begin{eqnarray}
\label{misfrx} SFR_X = 1.08 \times 10^{-42} L_X^{1.04}. 
%SFR_X = 3.186 \times 10^{-37} L_X ^{0.9}.
\end{eqnarray}
As it can be seen in Figure~\ref{gsfrx}, our new \SFR-$L_X$ calibration agrees better with the observational data that the relations explained  
before. Indeed, \citet{Lou05} relation gives values around one order of magnitude lower than those expected from the \FIR\ luminosity, but   
\citet{RCS03}  calibration provides values almost one order of magnitude higher than the actual ones. Using our new relation and the available X-ray  
data for our WR galaxies (see Table~\ref{rayosx}), we derive a $SFR_X$ of 3.5, 13.8, and 0.86 \Moy\ for HCG~31, IRAS 08339+6517, and Tol~9,  
respectively. These values agree well with the actual \SFR\ estimated for each object (see Table~\ref{sfrg}).

\begin{table}[t!]
\centering
  \caption{\footnotesize{Additional \FIR\ and radio properties.}}
  \label{radiofirt}
  \tiny
  \begin{tabular}{l  cr  c }
  \tableline
   \noalign{\smallskip}
Galaxy        &   $q^a$   &     $log R^b$         &  $\tau_g$ \\ % SNR$^{-1}$ &       \\ % & $\alpha_{IR}^b$  & $T_{dust}$  \\
              &           &                     &  [Gyr] \\ %  [yr]     &         \\ % &                  &  [K]   \\
\tableline
\noalign{\smallskip}    

HCG 31 AC       & 2.39  &  1.19    & 1.92   \\  %&   50 &  2.76  &  42 \\
HCG 31 B        &\nodata&  1.25    &  5.3   \\  %&  613 &\nodata & \nodata\\
HCG 31 F        &\nodata&  \nodata & 2.9   \\  %&  337 &\nodata & \nodata \\
HCG 31 G        &\nodata&  1.11    & 2.3    \\  %&  329 &\nodata & \nodata \\
Mkn 1087        & 2.53  &  1.30    & 3.3    \\  %&   21 &  2.87  & 42 \\
Haro 15         & 2.01  &  1.48    & 1.9    \\  %&   44 &  2.20  & 42\\
Mkn 1199        & 2.39  &  1.31    & 0.40   \\  %&   34 &  2.41  & \\
Mkn 5           &\nodata&  \nodata & 2.1   \\  %& 2500 &\nodata & \\
IRAS~08208+2816 & 2.01  &  1.57    & \nodata\\  %&   12 &  2.04  & \\
IRAS~08339+6517 & 2.33  &  1.21    & 0.89   \\  %&   16 &  2.36  & \\
POX 4           & 2.25  &  0.93    & 4.2    \\  %&  259 &  2.04  & \\  
UM 420          & 2.71  &  1.08    & \nodata\\  %&   31 & $>$0.46& \\
SBS~0926+606    & 2.18  &  0.79    & 2.0 (A), 5.6 (B)    \\ % &  449 & $>$1.61& \\
SBS~0948+532    &\nodata& \nodata  & \nodata\\  %&   22 &\nodata & \\
SBS~1054+365    & 2.79  & $-$0.10  & 4.0    \\ % & 4350 & $>$1.92& \\
SBS~1211+540    &\nodata& \nodata  & 3.5  \\ % & 2.9  &\nodata & \\
SBS~1319+579    & 2.15  & \nodata  & 12.7   \\ % &  674 &\nodata & \\
SBS~1415+437    &\nodata& \nodata  &  3.5   \\ % & 3460 &\nodata & \\
III~Zw~107      & 2.34  &  1.06    &  3.5   \\ % &   54 &  2.03  & \\  
Tol 9           & 2.03  &  1.16    &  1.95  \\ % &   79 &  2.54  & \\
Tol 1457-262    & 2.00  &  1.43    &  1.51  \\ % &   30 &  2.34  & \\
Arp 252         & 1.69  &  2.28    & \nodata\\ % &   18 &  1.56  & \\
NGC 5253        & 2.62  &  0.33    &  0.97  \\ % &  709 &  1.03  & \\ 
\noalign{\smallskip}
\tableline
  \end{tabular}
  \begin{flushleft}
  $^a$ The logarithmic ratio of \FIR\ to radio flux density parameter, $q$, is defined in Equation~\ref{agnq}.\\
  $^b$ $R$ is the non-thermal to thermal ratio, derived from the 1.4~GHz and \Ha\ fluxes.\\
  $^c$ The gas depletion timescale is defined as $\tau_g=1.32$\MHi/\SFR\ (Skillman et al. 2003)
  \end{flushleft}
\end{table}

\begin{figure*}[t!]
\centering
\includegraphics[angle=270,width=\linewidth]{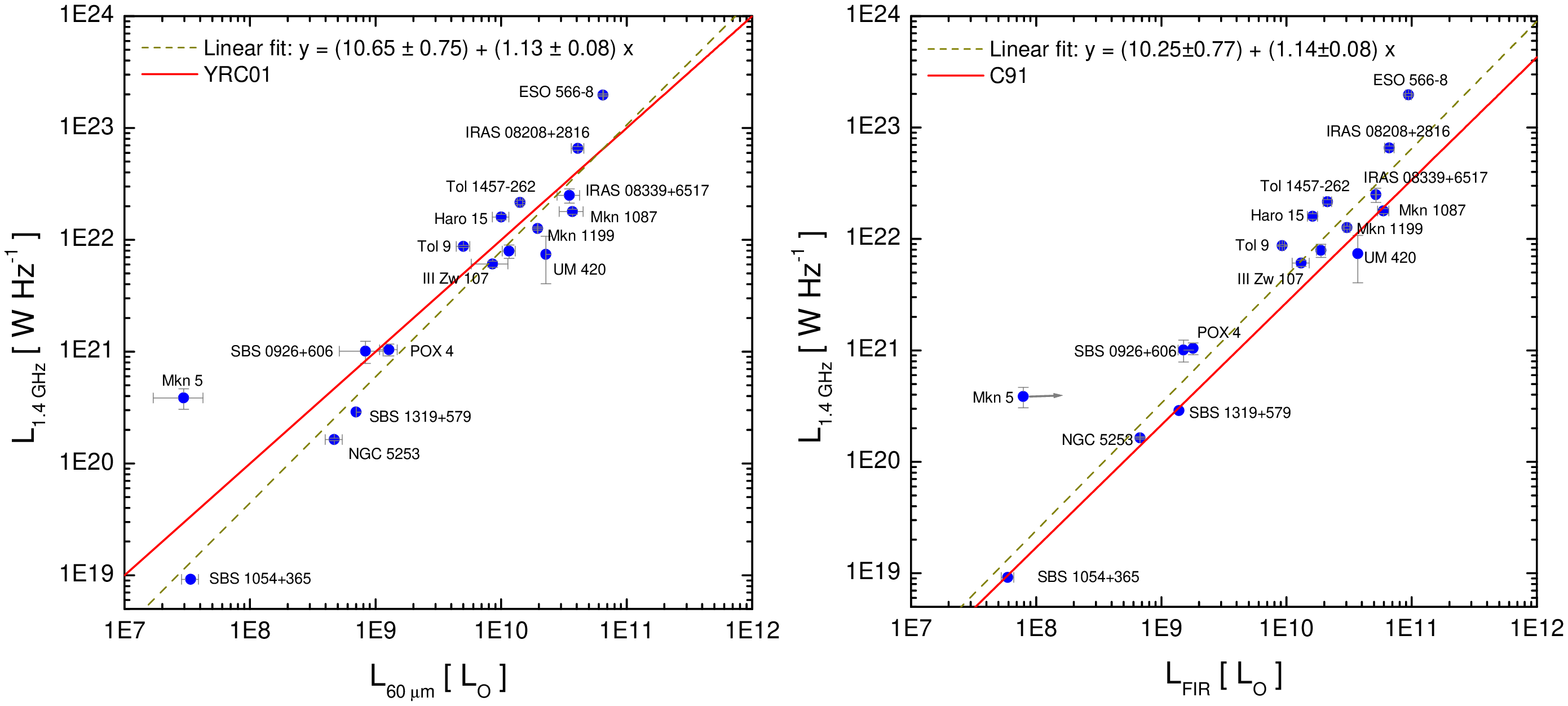}
\protect\caption[ ]{\footnotesize{1.4 GHz radio-continuum luminosity vs. the 60$\mu$m luminosity (left) and the \FIR\ luminosity (right). The  
relations derived by \citet{YRC01} (Equation~\ref{eYRC01}, left diagram) and \citet{Condon91} (Equation~\ref{eC91}, right diagram) are plotted with a  
red continuous line. The best linear fits to our data (in logarithmic units) are shown with a yellow dashed line.}}
\label{radiofir}
\end{figure*}

\section{\FIR/radio correlation}

%At the beginning of the 70s, van der Kruit (1971, 1973) found a correlation 
%between mid-infrared and radio luminosities. Harwit \& Pacini (1975) proposed 
%that the infrared is thermal reradiation from dusty \ion{H}{ii} regions, while 
%the 1.4 GHz luminosity is dominated by synchrotron radiation from relativist 
%electrons accelerated in supernova remanents (SNRs) from the same population 
%of massive stars that heat and ionize the \ion{H}{ii} regions. 
We have used the luminosity data shown in Table~\ref{lumi} to check if our WR galaxies follow the \FIR/radio correlation. As it was shown by Condon et  
al. (1992), the \FIR/radio correlation is much tighter for starbursts than for active galaxies. Figure~\ref{radiofir} (left) plots the 1.4~GHz  
luminosity vs. the 60~$\mu$m luminosity for our sample galaxies and the relation between both quantities found by \citet{YRC01},
\begin{eqnarray}
\label{eYRC01} L_{1.40\,\rm{GHz}}\ [{\rm W\ Hz^{-1}}] = 10^{12} L_{60\ \mu m}\ [L_{\odot}],
\end{eqnarray}  
%indicating the ongoing massive star formation (Popescu et al. 2000 A\&A 362, 138) 
%and that it is not powered by an AGN. $L_{60\, \mu m}$ can be derived in solar 
%units from the $f_{60}$ flux density using the expression given by 
%Yun, Reddy \& Condon (2001):
%\begin{eqnarray}
%\log  L_{60\, \mu m} = 6.014+2\log D+\log f_{60}. 
%\end{eqnarray}
while Figure~\ref{radiofir} (right) shows $L_{\rm 1.4\, GHz}$ vs. the total \FIR\ luminosity and the relation given by \citet{Condon91},
\begin{eqnarray}
\label{eC91} \log \, L_{1.49\,\rm{GHz}} [{\rm W\ Hz^{-1}}] = 1.1 \log \, L_{FIR}\ [L_{\odot}] +  10.45.
\end{eqnarray}
Bell (2003) remarked that the radio-\FIR\ correlation is linear not because both radio and \FIR\ emission track \SFR, but rather because they fail to  
track \SFR\ in independent, but coincidentally quite similar, ways. Further analysis (i.e., Hunt et al. 2005) also found that this relation is not  
hold for some low-metallicity or young starbursts galaxies. However, as it is seen in Figure~\ref{radiofir}, all analyzed objects except Mkn~5 (which  
has a very uncertain value for \FIR) and Arp~252 (ESO 566-8 hosts some kind of nuclear activity) follow well both relations. This indicates that the  
galaxies are starbursting systems and are not active galaxies (Seyfert or AGNs). We already reached this conclusion when we analyzed the diagnostic  
diagrams involving several emission-line ratios (see Paper~III). Figure~\ref{radiofir} includes a linear fit (in logarithmic scale) to our data  
(neglecting Mkn~5, for which the \FIR\ values have high uncertainties). The relation given by \citet{Condon91} seems to be slightly displaced with  
respect our observational data, although we also see some small discrepancies in the \citet{YRC01} relation for the faintest objects.

\begin{figure}[t!]
\centering
\includegraphics[angle=270,width=\linewidth]{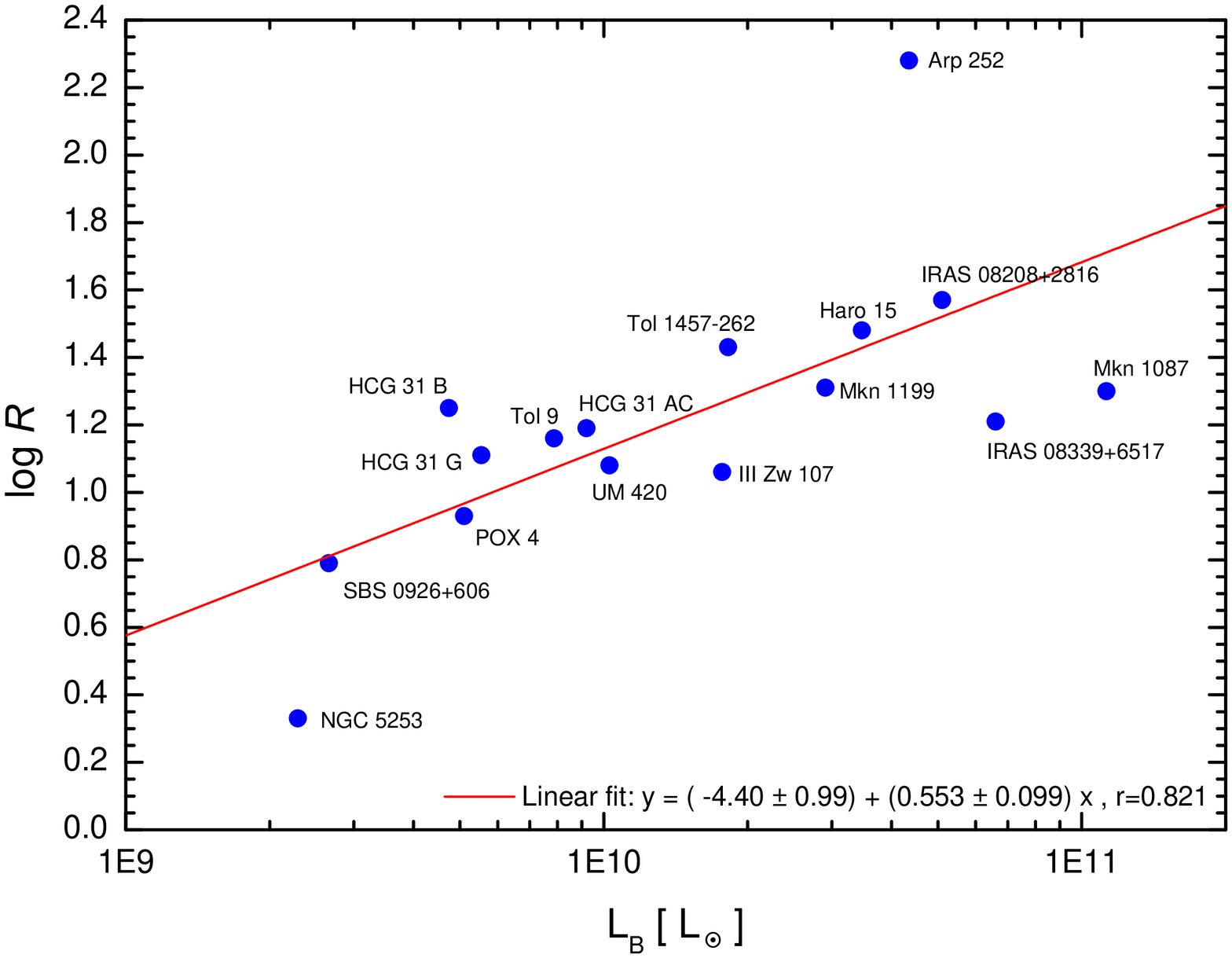}
\protect\caption[ ]{\footnotesize{Comparison of the $B$-luminosity (in solar units) and the logarithmic non-thermal to thermal ratio, $\log R$, for  
our sample galaxies. A linear fit is shown with a continuous red line.}}
\label{logr}
\end{figure}

The non-\AGN\ nature of our sample of WR galaxies is also supported by the analysis of the $q$ parameter an the \FIR\ spectral index. The $q$  
parameter is defined as the logarithmic ratio of \FIR\ to radio flux density,
\begin{eqnarray}
\label{agnq} q \equiv \log \frac{F_{FIR} ({\rm W\ m^{-2}})\ / \ 3.75 \times 10^{12}\ {\rm Hz}}{S_{\rm 1.4\ GHz}\ ({\rm W\ m^{-2}\ Hz^{-1}})},
\end{eqnarray}
and it is very robust for most galaxy populations: $<q>$ = 2.34 $\pm$ 0.19 (Condon et al. 1991; Yun et al. 2001). Galaxies with $q<1.8$ are more than  
3 times more radio loud than the mean for the star-forming galaxies, so they can be classified as AGN-powered. As it seen in Table~\ref{radiofirt},  
all galaxies except Arp~252 have a $q$ value similar to that expected for starburst galaxies.

\begin{table*}[t!]
\centering
  \caption{\footnotesize{Keplerian mass (\Mkep), dynamical mass (\Mdyn), neutral gas mass (\MHi), ionized gas mass (\MHii), warm dust mass (\Mdust),  
mass of the ionizing star cluster (\Mstar), total stellar mass ($M_{stars}$), and baryonic mass (\Mbar) of the galaxies analyzed in this work.}}
%The last two columns compile the 
%total stellar mass derived from the $H$-band luminosity assuming 
%$M_{stars}/L_{H}$=0.8 and the total baryonic mass 
%computed following \Mbar=$M_{stars}+1.32M_{\rm H\,I}$. }}
  \label{masas}
  \tiny
  \begin{tabular}{l  cc  cc cc  cc}
  \tableline
   \noalign{\smallskip}
Galaxy      &   \Mkep     &  \Mdyn &      \MHi   &        \MHii    &    \Mdust$^a$ &   \Mstar     & $M_{stars}$  & \Mbar  \\
            & [10$^8$ \Mo]& [10$^8$ \Mo]& [10$^8$ \Mo]& [10$^6$ \Mo]& [10$^6$ \Mo] & [10$^6$ \Mo] & [10$^8$ \Mo] & [10$^8$ \Mo] \\		    
\tableline
\noalign{\smallskip}   
   HCG 31    & \nodata   &  \nodata&  74.6         &  7.8$\pm$0.5  &  2.56      &  37.8$\pm$2.8  &   33.8   & 141  \\
   "      AC &  340      &  850    &  36.4 	       &  5.4$\pm$0.3  &   \nodata  &  18.2$\pm$1.2	 &   14.4   &  62.5   \\
   "      B  &   26      &   54	   &  19.4	       &  0.45$\pm$0.04&   \nodata  &   8.2$\pm$0.7  &    8.0   &  33.6    \\
   "      E  & \nodata   & 13$^b$  & \nodata       &  0.22$\pm$0.02&   \nodata  &   2.6$\pm$0.3 &    0.47   & \nodata\\
   "      F  &    3.0    &   15	   &   6.13	       &  0.75$\pm$0.04&   \nodata  &  0.40$\pm$0.03 &   0.39  &   8.5 \\
   "      G  &   21      &   68	   &  19.4	       &  0.96$\pm$0.05&   \nodata  & 	7.0$\pm$0.4  &   10.5  &   36.1   \\
   Mkn 1087  &  560      & 1800    & 156           &  10.4$\pm$9.7 &    7.8     &   129$\pm$9    &   288   &  494  \\
   "       N &   2.2     & 45$^b$  & \nodata       & 0.337$\pm$0.028& \nodata   &  3.34$\pm$0.28 &     4.0 & \nodata\\
   Haro 15   &  121      &  365    &  55$\pm$18    &  6.2$\pm$0.7  &    2.1     & 	77$\pm$9     &   99.2  &  172   \\
   Mkn 1199  &   82      & 1650    &  12.2	       &  7.3$\pm$1.1  &    3.1     &   140$\pm$21   &   275   &  291   \\
   " NE      &  2.9      &  70$^b$ & \nodata       &  0.13$\pm$0.03& \nodata    &   1.7$\pm$0.4  &   17.4  & \nodata\\
   Mkn 5     &   21      &   36	   & 0.72$\pm$0.09 & 0.081$\pm$0.010&   0.099   &0.300$\pm$0.012 &    1.5  &   2.5 \\
IRAS 08208+2816&  39     & 600$^b$ & \nodata       &  21.1$\pm$1.7 &    8.84    &  166$\pm$14    &   298   & \nodata\\
IRAS 08339+6517&  100    &  370	   &  53$\pm$6     &  14.5$\pm$0.7 &    3.91    &  226$\pm$10    &   476   &  546   \\
" Comp       &     80    &  100    &  7.0$\pm$0.9  &  0.26$\pm$0.03& \nodata    &  4.1$\pm$0.5   &   18.6  &  27.9    \\ 
POX 4        &    5.0    &   76	   &  21	       &  5.70$\pm$0.18&    0.093   &  9.8$\pm$0.3   &   14.3  &  29.1   \\
" Comp       & \nodata   & \nodata & \nodata       &   0.05         & \nodata   &  0.79$\pm$0.07 &   0.88  & \nodata \\
	UM 420   &   21      &  200$^b$ & \nodata       &  6.9$\pm$0.4  &    5.0$^c$ & 13.8$\pm$0.9   &   45.6  & \nodata\\
SBS 0926+606 &  \nodata  & \nodata & 17.7$\pm$7.2  &  1.75$\pm$0.10&    0.37    & 8.4$\pm$0.6    &  7.59   &  30.9 \\  
" A          &  \nodata  &     23  &  9.6$\pm$3.6  &  1.40$\pm$0.07&   \nodata  & 3.59$\pm$0.17  &    2.9  &  15.5   \\
" B          &  \nodata  &     45  &  8.1$\pm$3.6  &  0.35$\pm$0.03&   \nodata  &  4.8$\pm$0.4   &    4.7   & 15.4   \\
SBS 0948+532 &   21      &  90$^b$  & \nodata       &  11.6$\pm$0.4 &   \nodata  & 18.8$\pm$0.7   &   12.6$^d$&\nodata\\
SBS 1054+365 &    0.78   &    15   &  0.61$\pm$0.06&0.067$\pm$0.003&    0.012   & 0.23$\pm$0.01  &    0.33 &    1.13 \\
SBS 1211+540 &    1.13   &     1.14&  0.24$\pm$0.04&0.021$\pm$0.001&   \nodata  & 	 0.05	     &  0.102$^d$& 0.420  \\
SBS 1319+579 &   86      &  140    & 16.4          &  0.35$\pm$0.02&    0.30    & 0.78$\pm$0.05  &   22.9   &  44.5   \\
SBS 1415+437 &    2.5    &    4.9  &  0.96$\pm$0.07& 0.074$\pm$0.003&  \nodata  & 0.14$\pm$0.01  &    0.46 &    1.74 \\
  III Zw 107 &    8.2    &  180	   &  67$\pm$12    &  6.0$\pm$0.3  &    1.25    & 	74$\pm$3     &   61.8  &  150   \\
      Tol 9  &   12      & 580$^e$ &  22$\pm$2     &  3.4$\pm$0.2  &    2.41    &  27.4$\pm$1.9  &   44.4   &  106   \\
Tol 1457-262 &   \nodata & 950     &  47	       &  9.4$\pm$0.5  &    1.83    &  41$\pm$2      &   87.2   &  149   \\
" Obj 1      & 62        & 290$^b$ &  \nodata	   &  6.9$\pm$0.4  &    \nodata &  23.4$\pm$1.5  &   46.9   &  \nodata   \\
" Obj 2      &  \nodata  & 150$^b$ &  \nodata      & 2.56$\pm$0.12 &  \nodata   &  14.4$\pm$0.7  &   36.2  &  \nodata \\
Arp 252      & \nodata   &\nodata  &  \nodata      &  13.4$\pm$0.9 &   6.32     & 82.3$\pm$4.9   & 338    & \nodata \\
" ESO 566-8    &   73      & 440$^b$ & \nodata     & 12.5$\pm$0.8  & \nodata    &  45$\pm$3      &  250   & \nodata\\
" ESO 566-7    &    4      & 150$^b$ & \nodata     & 0.93$\pm$0.07 & \nodata    &  16.8$\pm$1.2  &   88   & \nodata\\
NGC 5253     & \nodata   &  83$^e$ & 1.63$\pm$0.10 & 0.65$\pm$0.03 &  0.042     &  2.22$\pm$0.11 &   11.4   &   13.6   \\
\noalign{\smallskip}
\tableline
  \end{tabular}
  \begin{flushleft}
   $^a$ This value is for the entire system: all galaxies in the HCG~31 group, members A and B in SBS~0926+606, all galaxies in Tol~1457-262, and  
ESO~566-8 and ESO~566-7 in Arp~252. We neglect the contribution of the \FIR\ emission in dwarf objects associated to larger galaxies (companion  
objects surrounding Mkn~1087, Mkn~1199, IRAS~08339+6519, and POX~4).\\
  $^b$ Tentative value of \Mdyn\ computed using Equations~\ref{emdynmb}--\ref{emdynmj} and \ref{ecmdyn}.\\
  $^c$ The warm dust mass is very probably overestimated because of the \FIR\ contribution of the foreground galaxy~UGC 01809.\\
  $^d$ $M_{stars}$ computed assuming $V-J\sim0.8$ and $J-H\sim0.3$.\\
  $^e$ The detailed analysis of the \HI\ gas and its kinematics will be presented elsewhere (LS+10).
  \end{flushleft}
\end{table*}

Table~\ref{radiofirt} also compiles the non-thermal to thermal ratio, $R$, of the galaxies with available 1.4~GHz radio-continuum data. The thermal  
flux at 1.4~GHz was computed applying Equation~\ref{ftermico}. The majority of the galaxies show the typical value for star-forming galaxies, $\log R=  
1.3 \pm 0.4$ \citet{DPKC02}. The low value in $R$ found in POX~4 and NGC~5253 may be because the \Ha\ flux has been overestimated, although the  
situation of NGC~5253 is far from clear (L\'opez-S\'anchez et al. 2010a). The value obtained for SBS~1054+365 is not reliable, we consider that or the  
1.4~GHz flux was underestimated (very probably) or the \Ha\ flux was overestimated. However, the high value found in Arp~252 (the emission comes  
mainly from ESO~566-8), $\log R=2.28$, is real and indicates that the thermal flux at 1.4~GHz is less than 0.5\%. As reported by several authors  
(i.e., Klein, Wielebinski \& Thuan 1984; Klein, Weiland \& Brinks 1991; Bell 2003), dwarf galaxies seem to have a lower non-thermal-to-thermal  
emission ratio than normal spiral galaxies. The values obtained for the $R$ parameter in our galaxy sample tend to be lower at lower $B$-luminosities,  
as it is shown in Figure~\ref{logr}. The difference between dwarf and larger galaxies is often interpreted as higher efficiency of cosmic-ray  
confinement in more massive galaxies (e.g., Klein et al. 1984; Price \& Duric 1992; Niklas, Klein \& Wielebinski 1997; Bell 2003).

%Finally, the \FIR\ spectral index, $\alpha_{IR}$,
%\begin{eqnarray}
%\alpha_{IR}\equiv \frac{\log (f_{60\, \mu m}/f_{25\, \mu m})}{\log (60/25)},
%\end{eqnarray}
%(Condon et al 2002) distinguishes between very warm ($T>100$ K) dust near a luminous 
%AGN and cooler ($T<50$ K) dust heated by an extended starburst of similar luminosity 
%(Rodr\'{\i}guez Espinosa et al. 1996).
%%, so it can also be used to separate AGNs and starbursts. Some radio sources powered 
%%by AGNs can be recognized by their large-scale structures 
%%(jets and lobes extending beyond the optical galaxies) 
%%or very high brightness temperatures $T\gg 10^5$ K at 1.4 GHz (Condon 1992). 
%Sources with $\alpha_{IR}<$ 1.5 are being powered primarily by AGNs, but as it is shown 
%in Table~\ref{radiofirt}, all the galaxies analyzed here (except****) have  $\alpha_{IR}<$ 2???.

\section{Analysis of the masses}

\begin{figure*}[t!]
\begin{tabular}{cc}
\includegraphics[angle=270,width=0.45\linewidth]{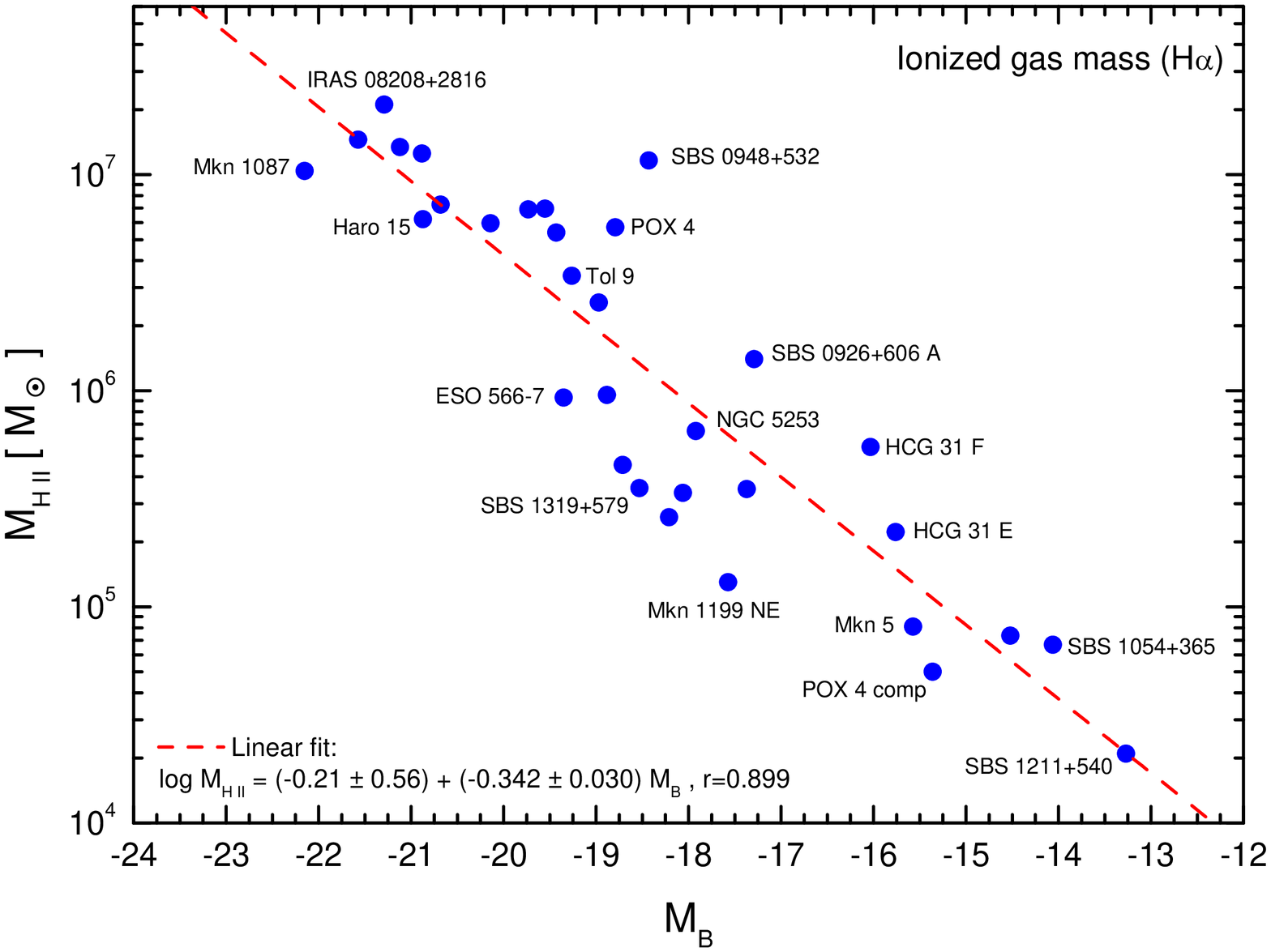} &
\includegraphics[angle=270,width=0.45\linewidth]{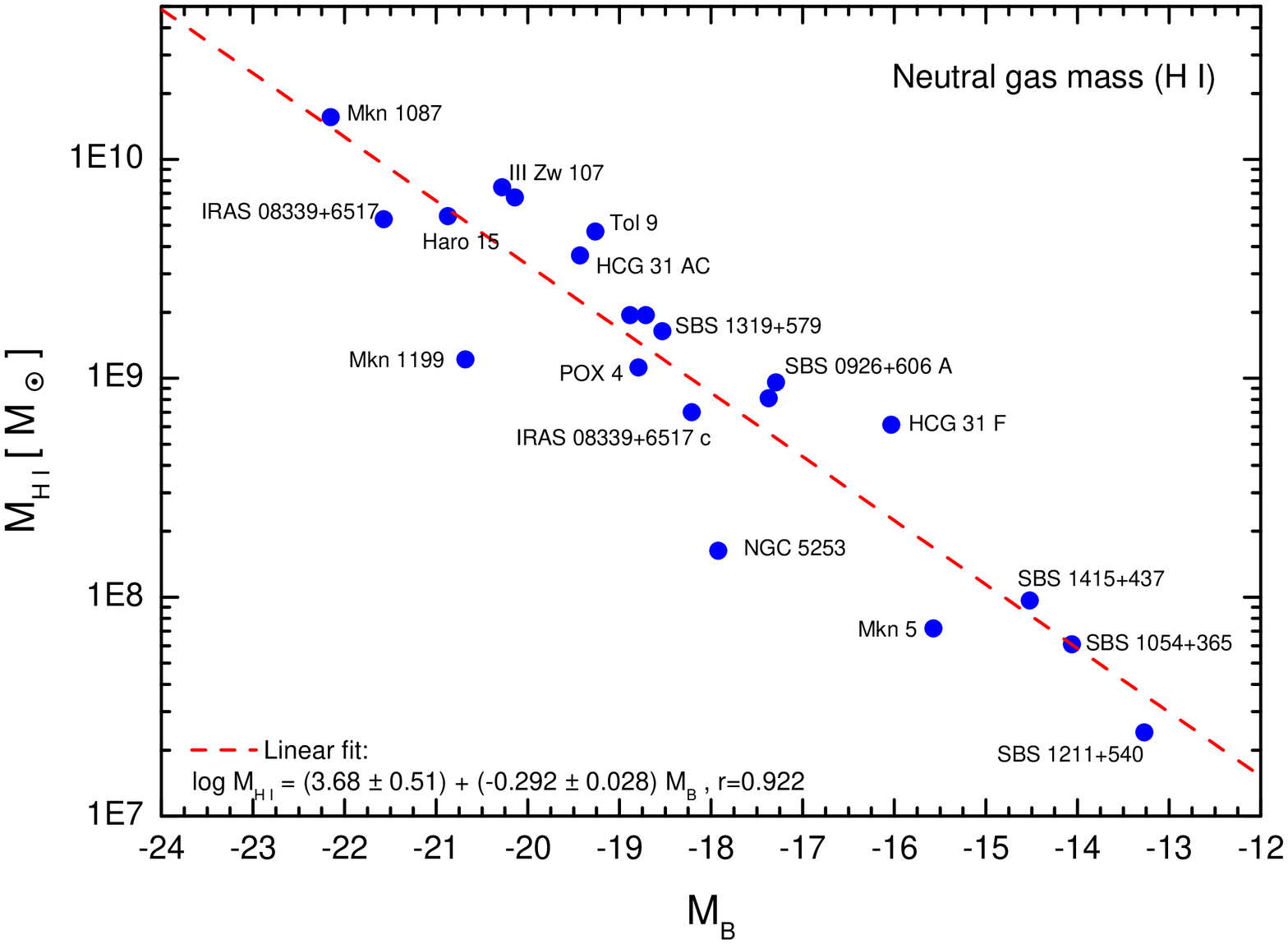} \\
\includegraphics[angle=270,width=0.45\linewidth]{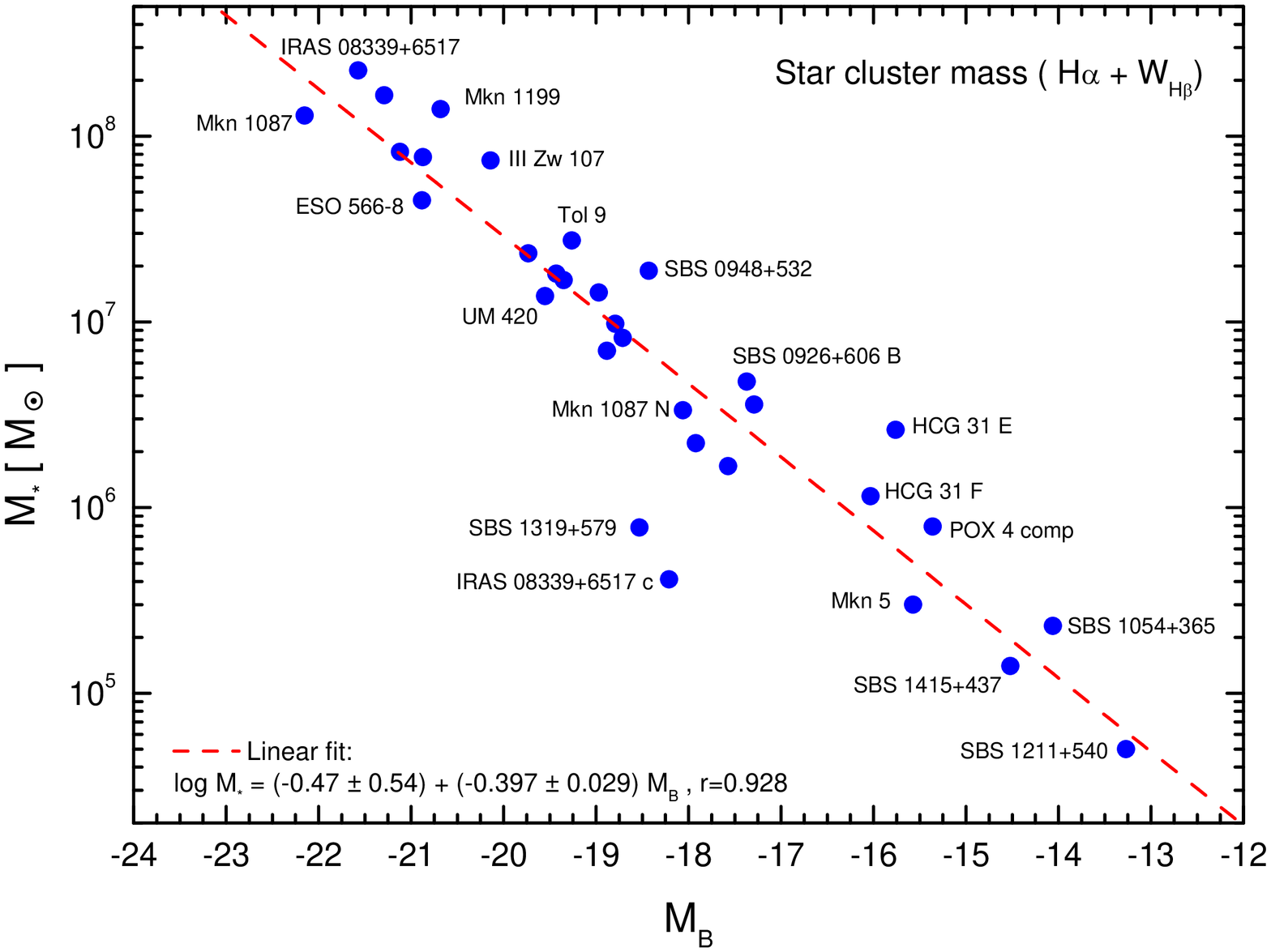} &
\includegraphics[angle=270,width=0.45\linewidth]{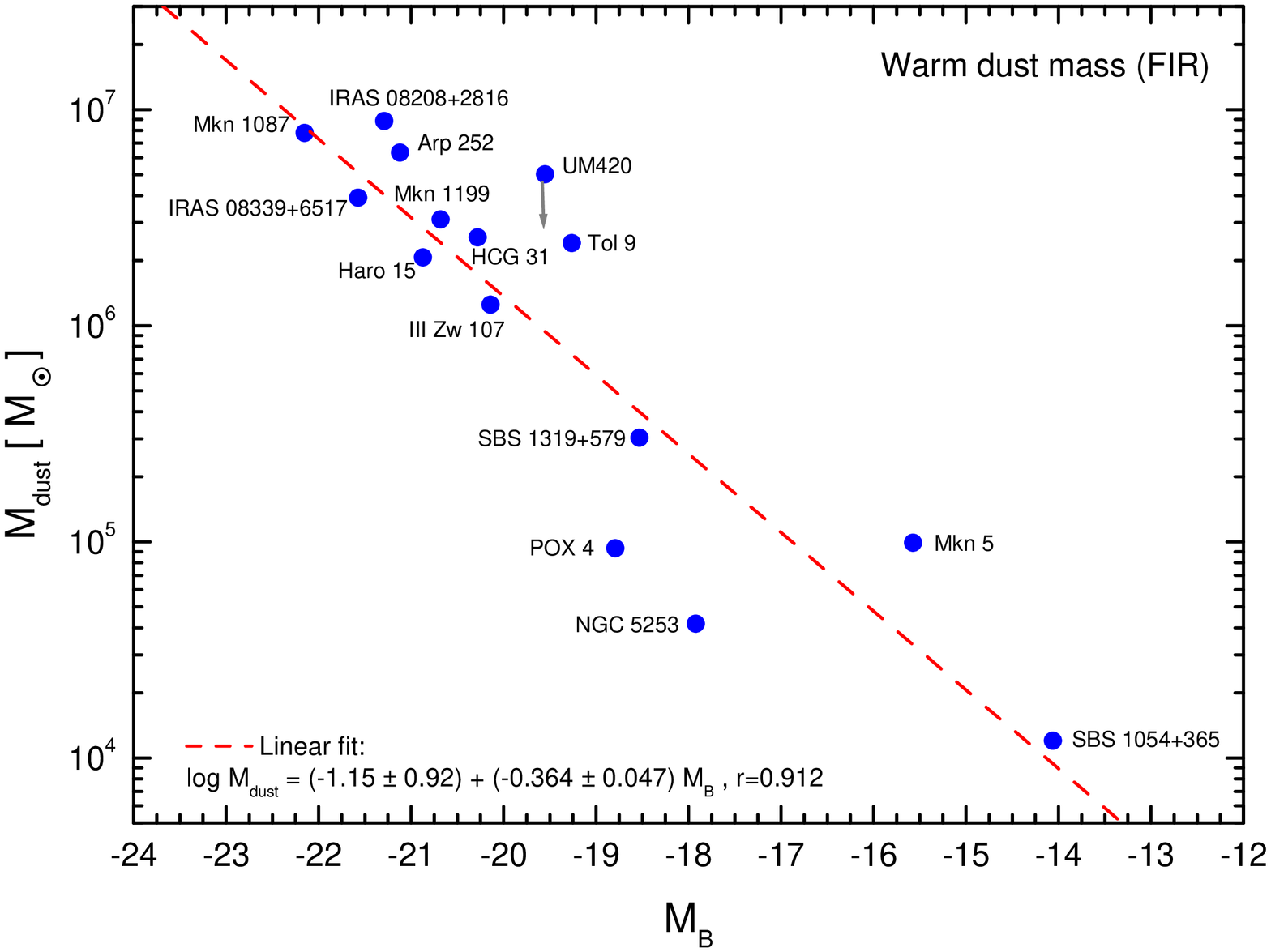} \\
\end{tabular}
\protect\caption[ ]{\footnotesize{Ionized gas mass (\MHii), neutral gas mass (\MHi), mass of the ionizing star cluster (\Mstar), and warm dust mass  
(\Mdust) vs. the absolute $B$ magnitude for the analyzed galaxies. Linear fits to the data are shown with a dashed red line.}} 
%Some objects have been labeled.}}
\label{masasmb}
\end{figure*}

\begin{figure*}[t!]
\includegraphics[angle=270,width=\linewidth]{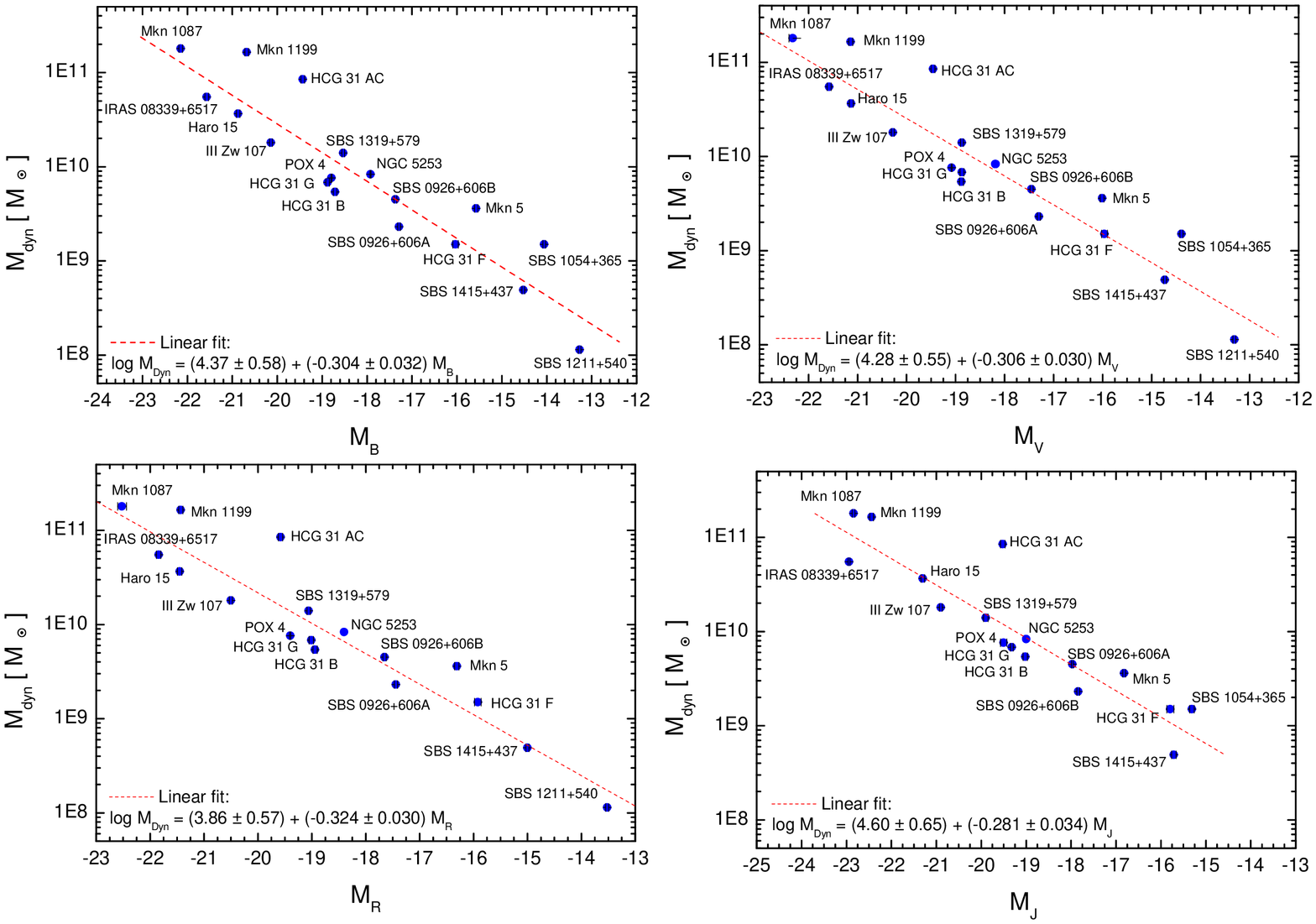}
\protect\caption[ ]{\footnotesize{Dynamical mass (\Mdyn) vs. absolute $B$, $V$, $R$ and $J$ magnitudes for the galaxies analyzed in this work. A  
linear fit is shown with a dashed red line.}}
\label{masascolor}
\end{figure*}

For this work, we have estimated the ionized gas mass --\MHii, using the \Ha\ images presented in Paper~I--; neutral gas mass --\MHi, using \ion{H}{i}  
data at 21~cm compiled from the \mbox{literature--;} mass of the ionizing star cluster --\Mstar, using \Ha\ and \WHb, see Paper~I--; warm dust mass  
--\Mdust, using the \FIR\ fluxes--; Keplerian mass --\Mkep, via the kinematics of the ionized gas--; and dynamical mass --\Mdyn, using the \HI\  
\mbox{kinematics--.} All these data are compiled in Table~\ref{masas}. 
%while Table~\ref{mlratios} shows all the mass-to-$B$ luminosity ratios. 
The estimation of \Mkep\ and \Mdyn\ for each galaxy was explained in Paper~II. We just remember that, as the extension of the neutral gas is usually  
larger than the stellar component, our estimations of \Mdyn\ are very probably underestimated. Furthermore, no-rotational movements would yield in a  
overestimation of the total mass. Only interferometric \HI\ analysis can definitely provide a more precise determination of the dynamical mass of each  
system. However, we may use our \Mdyn\ values as a rough estimation of the total mass of the systems. Their comparison with \Mkep, \MHi, \MHii,  
\Mdust\ and their associated mass-to-light ratios will give clues about the galaxy type, dynamics and the fate of the neutral gas.

We first compare all mass determinations with the optical luminosity of the galaxies. Figure~\ref{masasmb} shows the relations between \MHii, \MHi,  
\Mdust\ and \Mstar\ with the absolute $B$-magnitude. As we should expect, and besides some scatter, all mass determinations clearly increase with  
increasing optical luminosity. We have performed a linear fit to the data; the results are:
\begin{eqnarray}  
\label{emasas} \log M_{\rm H\,II} = (-0.21 \pm 0.56) - (0.342 \pm 0.030)M_B, \\
\log M_{\rm H\,I} = (3.69 \pm 0.52) - (0.292 \pm 0.028)M_B,\\
\log M_{\rm dust} = (-1.15 \pm 0.92) - (0.364 \pm 0.047)M_B,\\ 
\log M_{\star} = (-0.47 \pm 0.54) - (0.397 \pm 0.029)M_B,
\end{eqnarray} 
with correlation coefficients of 0.899, 0.922, 0.912 and 0.928, respectively. 
Some deviations to the fits are found in Mkn~1199 (that possesses a relatively low \MHi), SBS~0948+532 (with a very high \MHii), POX~4 (it seems to be  
\Mdust\ deficient, while its ionized gas mass may be overestimated), UM~420 (its high \Mdust\ is very probably consequence of the contamination of the  
\FIR\ emission by the foreground galaxy UGC 01809, see Sect.~3.9 in Paper~I and Sect.~3.9 in Paper~II), SBS~1319+579 and IRAS~08339+6517~Comp (that  
have a very low \Mstar\ for their absolute $B$-magnitude) and NGC~5253 (that is both \MHi\ and \Mdust\ deficient). 

Figure~\ref{masascolor} plots the dynamical mass (that represents the total mass of the galaxy) versus the absolute magnitude in several broad-band  
filters ($B$, $V$, $R$ and $J$). We find a clear correlation between these quantities, a linear fit to the data yields:
\begin{eqnarray}  
\label{emdynmb} \log M_{dyn} = (4.37 \pm 0.58) - (0.304 \pm 0.032)M_B, \\
\label{emdynmv} \log M_{dyn} = (4.28 \pm 0.55) - (0.306 \pm 0.030)M_V,\\
\label{emdynmr} \log M_{dyn} = (3.86 \pm 0.57) - (0.324 \pm 0.030)M_R,\\
\label{emdynmj} \log M_{dyn} = (4.60 \pm 0.65) - (0.281 \pm 0.034)M_J,
\end{eqnarray}
with correlation coefficients $r$ of 0.922, 0.931, 0.940 and 0.907, respectively. Notice that slopes in all fits are quite similar. The most important  
deviations to these fits are found in clearly interacting systems (Mkn~1199 and HCG~31~AC) but also in Mkn~5 and SBS~1054+365. 

%We would like to remark the importance of the diagrams shown in Figure~\ref{masascolor}, 
%because it is not common to find in the literature a comprehensive and detailed analysis 
%of a sample of galaxies for which both the total (dynamical) mass and the 
%reddening-corrected luminosity in optical and \NIR\ filters has been performed. 

\begin{figure}[t!]
\includegraphics[angle=270,width=\linewidth]{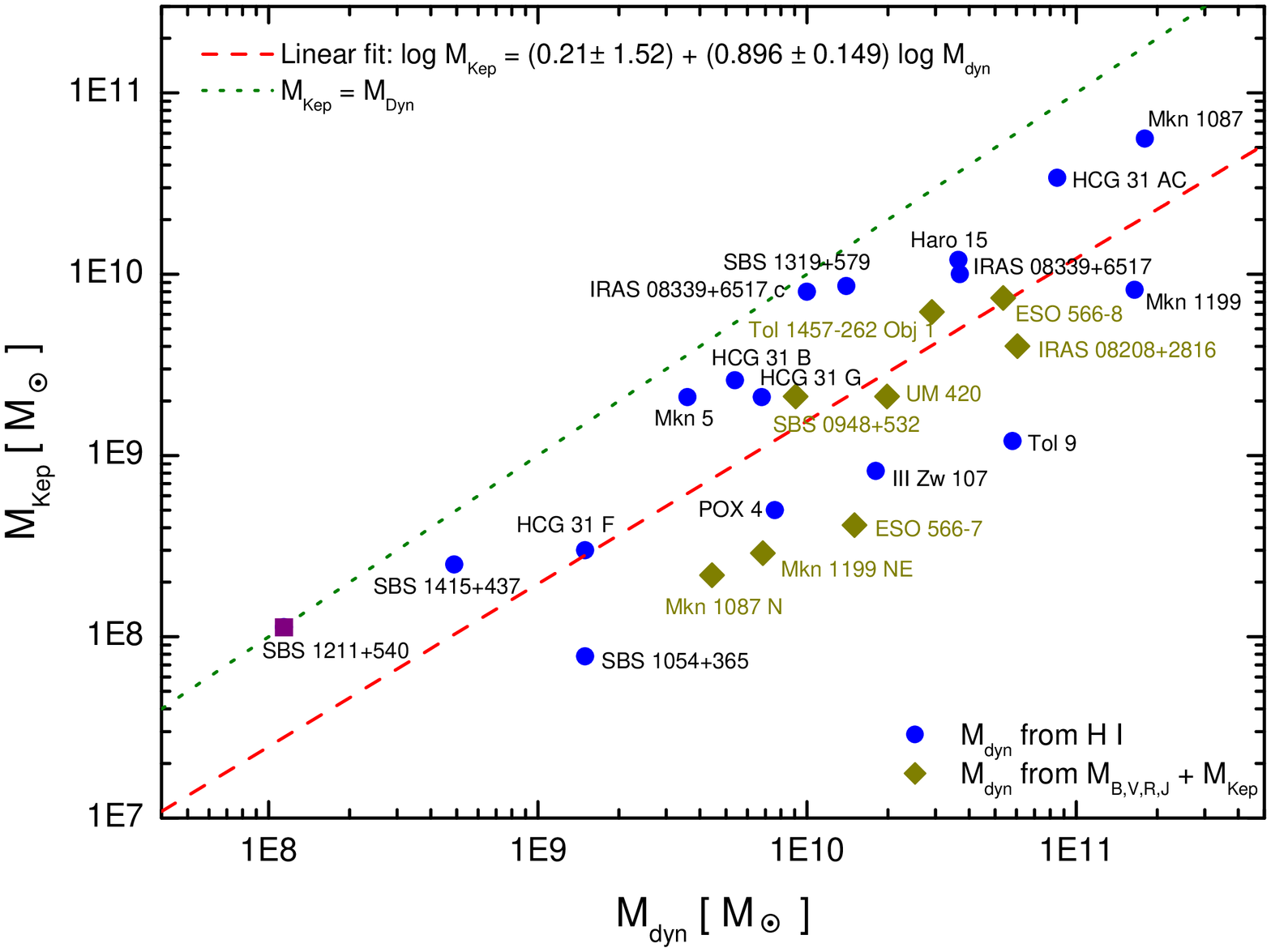}
\protect\caption[ ]{\footnotesize{Keplerian mass (\Mkep) vs. dynamical mass (\Mdyn) for the galaxies studied in this work. 
Blue circles plot galaxies with a direct estimation of \Mdyn\ using the \HI\ data, 
while the dark yellow diamonds indicate when \Mdyn\ was derived 
using Equations~\ref{emdynmb}--\ref{emdynmj} and \ref{ecmdyn}. 
The dashed red line is a linear fit to the \Mdyn\ derived from the \HI\ data excluding SBS~1211+540. 
The dotted green line indicates \Mkep=\Mdyn.}}
\label{mkepdyn}
\end{figure}

We have compared the Keplerian mass (derived from the kinematics of the ionized gas) with the dynamical mass (estimated from the kinematics of the  
neutral gas).  Figure~\ref{mkepdyn} plots both sets of values. 
%indicating galaxies with a direct estimation of \Mdyn\ (blue circles) 
%and objects for which the dynamical mass was estimated using 
%the absolute $B$-magnitude and Equation~\ref{emdynmb} (dark yellow diamonds). 
As we expected, \Mkep\ is lower than \Mdyn\ for almost all cases (\Mkep=\Mdyn\ is shown by a dotted green line in Figure~\ref{mkepdyn}). Although the  
dispersion is high --and we remember that \Mkep\ and/or \Mdyn\ may be overestimate because of interaction features-- we have performed a tentative fit  
to the data, that yield
\begin{eqnarray}  
%\log M_{Kep} = 0.21 + 0.898 \log M_{dyn},
\label{ecmdyn}M_{dyn} = 0.584 \times M_{Kep}^{1.114}.
\end{eqnarray}
with a correlation coefficient $r=0.827$. This relation is included in Figure~\ref{mkepdyn} as a red dashed line. 
As we explained in Sect.~3.13 of Paper~II, \Mkep\ in SBS~1211+540 has been 
probably overestimated, so we did not consider this point in the analysis. 
The relation indicates that \Mkep\ is between 12\% (for \Mdyn=10$^{11}$ \Mo) and 24\% (for \Mdyn=10$^8$ \Mo)
%is between 25 and 10\% 
the total dynamical mass. Hence, as we should expect, more massive galaxies have a higher \Mdyn/\Mkep\ ratio, indicating that the kinematics of the  
ionized gas is not appropriate to derive the total dynamical mass in those objects.  
%being the difference higher at higher masses. 
%We may re-write the relation as
%\begin{eqnarray}  
%\label{ecmdyn}M_{dyn} = 0.584 \times M_{Kep}^{1.114}.
%\end{eqnarray}

Using Equations~\ref{emdynmb}--\ref{emdynmj} and \ref{ecmdyn} we have computed a tentative value for the dynamical mass in the galaxies with lack of  
\HI\ data. We have included the results in Table~\ref{masas}, and plotted these points in Figure~\ref{mkepdyn} (dark yellow diamonds). As we can see,  
they match well with the positions of the galaxies for with \Mdyn\ was derived from \HI\ data, but we will not consider these points in the subsequent  
analysis.

We prefer to use our \NIR\ data to derive a proper value of the stellar mass of all the galaxies
%Assuming a mass-to-light ratio, the stellar mass of the galaxies can be also estimated. 
Following the description provided by \citet{Kirby08}, we may assume a $H$-band mass-to-light ratio of $M_{stars}/L_H$=0.8 to compute the stellar  
mass, $M_{stars}$, from the $H$-luminosity (that is compiled for all objects in Table~\ref{lumi}). This assumption is well supported by both  
observations \citep{Bell03,Kirby08} and theory \citep{deJong96}, and considers a 12~Gyr old solar metallicity stellar population with a constant  
star-formation rate and Salpeter initial mass function. Hence, the $H$-band mass-to-light ratio may be somewhat overestimated for our young galaxies.  
Combining the $H$-band derived stellar mass and the \HI\ mass (we neglect the ionized gas, molecular gas, and dust contributions), the total baryonic  
mass, \Mbar, can be computed via
\begin{eqnarray}  
M_{bar} = M_{stars} + 1.32 M_{\rm H\,I},
\end{eqnarray}
where the factor 1.32 corrects the \HI\ mass for the presence of helium. The derived values for both \Mbar\ and $M_{stars}$ are compiled in last  
columns in Table~\ref{masas}. For SBS~0948+532 and SBS~1211+540, that lack of \NIR\ colors, we have assumed that $V-J\sim0.8$ and $J-H\sim0.3$ to  
derive the $H$-band luminosity.

\begin{figure}[h!]
\includegraphics[angle=270,width=\linewidth]{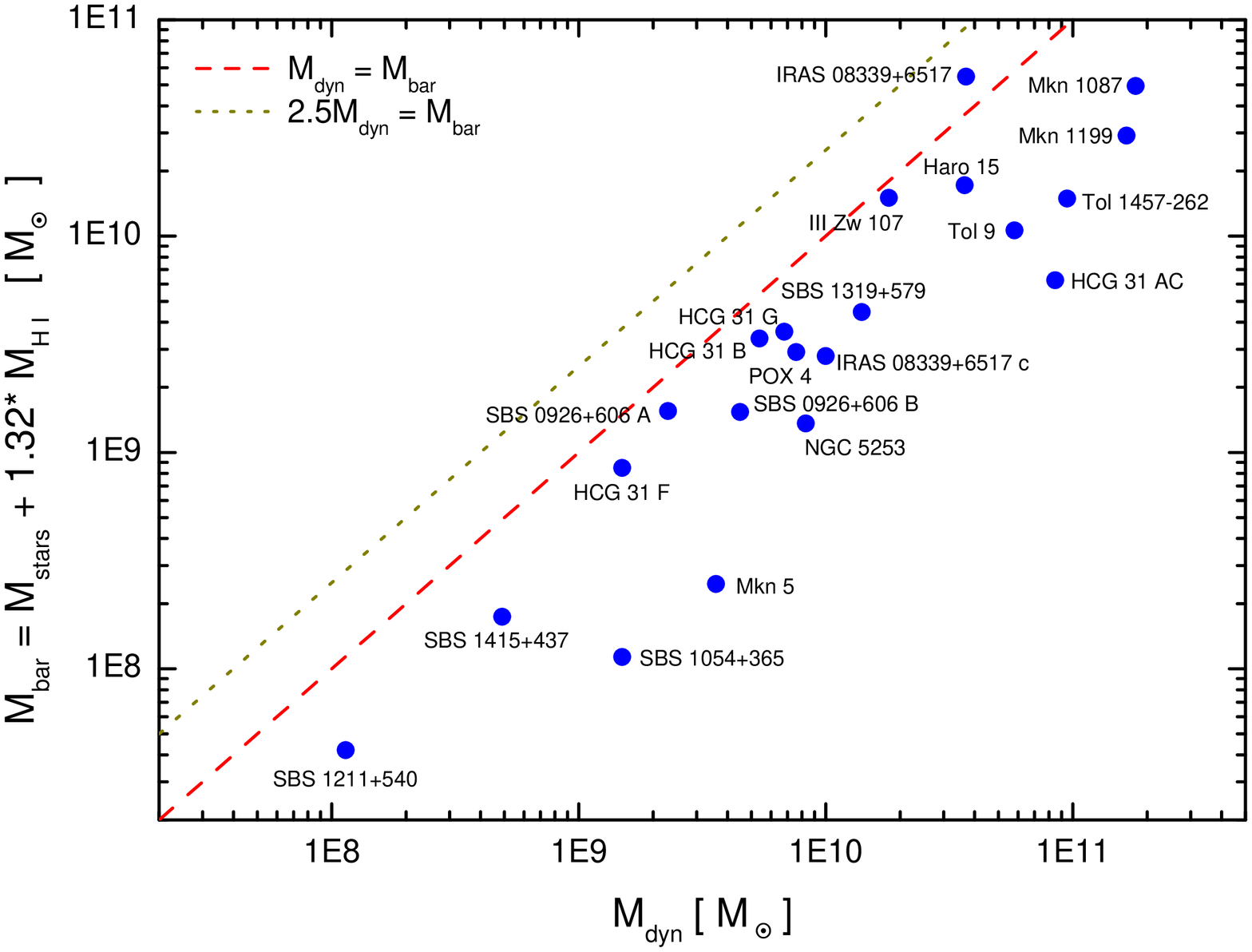}
\protect\caption[ ]{\footnotesize{Comparison between the baryonic mass (\Mbar) and the dynamical mass (\Mdyn) for our sample galaxies. The dashed red  
line indicates \Mbar=\Mdyn, the dotted yellow line indicates the position of \Mbar=\Mdyn\ if \Mdyn\ is computed assuming that the \HI\ size of the  
galaxies is 2.5 their optical extend. 
Note that strongly interacting systems (HCG~31~AC or  Tol~1457-262) lie away from the observed main trend.}}  
%hence something similar may be happening in SBS~1054+365, Mkn~5 and NGC~5253.}}
\label{mbar}
\end{figure}

As we should expect, the comparison between the dynamical and the baryonic masses (Figure~\ref{mbar})  
indicates that \Mdyn\ is always larger than \Mbar\ --except for IRAS~08339+6517, that has expelled a considerable fraction of its neutral gas to the  
intergalactic medium and shows a disturbed \HI\ kinematics \citep{CSK04} with a long tidal stream that makes impossible to get a good estimation of  
\Mdyn\ \citep{LSEGR06}--.
Besides the uncertainties in \Mdyn, this indicates the presence of dark matter in all systems.  The dark matter contribution would be even higher if,  
as we said, our values of \Mdyn\ are underestimated because of the uncertainty in the extension of the \HI\ disk (in all cases, except in those  
galaxies for which interferometric data were available, we used the maximum of the radius of the optical extent to compute \Mdyn). The dotted yellow  
line in Fig~\ref{mbar} indicates the position of \Mdyn=\Mbar\ if \Mdyn\ is computed assuming that the extension of the neutral gas is 2.5 times the  
size of the optical extent. Indeed, only interferometric \HI\ maps and a detailed analysis of the rotation of the neutral gas (i.e., de Blok et al.  
2008; Westmeier et al. 2010) can provide a better estimation of the dynamical masses of galaxies. This issue is even more important if interactions  
are disturbing the rotation pattern of the \HI\ gas. A clear example of this is Tol~9 within the Klemola~13 group. Our interferometric \HI\ map  
\citep{LS08b,LS+10b} shows that the neutral gas cloud in which this \BCG\ is embedded includes not only Tol~9 but also some nearby dwarf galaxies.  
Indeed, this \HI\ cloud seems to rotate as a single entity, and shows a long tidal tail in direction to other galaxies of the group. However, the  
maximum of the \HI\ emission is located exactly at Tol~9.

In any way, the important aspect to emphasize here is that, besides the unknown amount of dark matter, strongly interacting systems, as HCG~31~AC or  
Tol~1457-262, lie away from the observed main trend, having dynamical masses that are almost 14 and 7 times their baryonic masses. Mkn~1087, Mkn~1199  
and Tol~9, that are in clear interaction with nearby objects, also have larger \Mdyn\ than expected. Consequently, SBS~1054+365 and Mkn~5, that  
clearly have dynamical masses which are more than one order of magnitude higher than those expected for dwarf galaxies with \Mbar$\sim10^8$ \Mo, may also have  
highly perturbed \HI\ kinematics. The same situation may be happening in NGC~5253, which has a dynamical mass that is almost an order of magnitude higher than
that expected for a galaxy with \Mbar$\sim 10^9$ \Mo.

\section{Mass-metallicity relations} 
 
Our data set allows to investigate the mass-metallicity ($M-Z$) relation of star-forming galaxies.  The relationship between metallicity and stellar  
mass provides key clues about galaxy formation and evolution; however commonly the luminosity is used instead of the mass to analyze such correlations  
(i.e, Paper~IV and references within). Observationally, the $M-Z$ relation arises because low mass galaxies have larger gas fractions than higher mass  
galaxies (i.e., Boselli et al. 2001; Kewley \& Ellison 2008). Theoretically, the mean stellar metallicity of the galaxies increases with age as a  
consequence of the chemical enrichment of the ISM, while the stellar mass increases with time as galaxies undergo merging processes (i.e. Somerville  
\& Primack 1999; Calura, Matteucci \& Menci 2004). 
Once the \NIR\ luminosities or the optical-\NIR\ SED are known, $M_{stars}$ can be estimated relatively well using stellar evolutionary synthesis  
models, as we explained in the previous section. Hence, the main problem to study the $M-Z$ relation lies in all the uncertainties involving the  
determination of an accurate oxygen abundance, such as different methods yield to very different results (see Paper~IV and Kewley \& Ellison 2008).  
Here, the oxygen abundance of the majority of the galaxies was computed using the direct method, but Pilyugin (2001a) calibration has been applied to  
compute the metallicity of some few massive objects (Mkn~1087, Haro~15, IRAS~08339+6517, ESO~566-7, ESO~566-8), as we explained in Paper~IV.

\begin{figure}[t]
\includegraphics[angle=270,width=\linewidth]{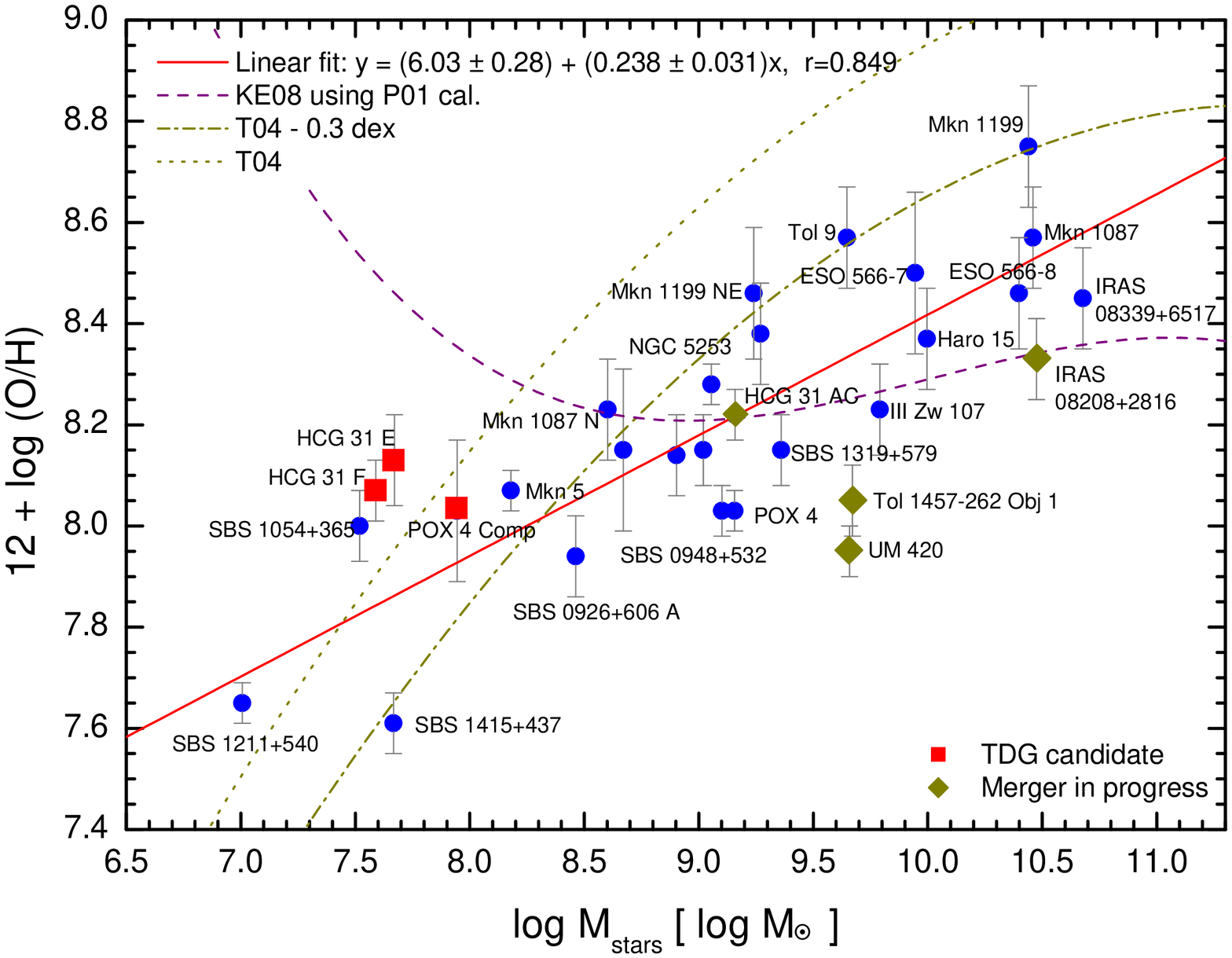}
\protect\caption[ ]{\footnotesize{Relation between $M_{stars}$ and the oxygen abundance for our galaxy sample. A linear fit to the data is shown with  
a dashed red line. The \TDG\ candidates are plotted with a red square, while galaxies in the process of merging are shown with a yellow diamond. A  
linear fit to the data without considering these two groups of objects is shown with a continuous red line. Some previous $M_{stars}-Z$ relations are  
also plotted: \citet{Tremonti04} with a yellow dotted line, \citet{Tremonti04} corrected by a factor of 0.3 dex in oxygen abundance --dotted-dashed  
yellow line--, and the \citet{KE08} relation considering the \citet{P01b} empirical calibration to derive the metallicity --dashed pink line--. These  
relations are only valid for $\log M_{stars}\geq 8.5$.}}
\label{aboxmass}
\end{figure}

\begin{figure*}[t]
\begin{tabular}{cc}
\includegraphics[angle=270,width=0.48\linewidth]{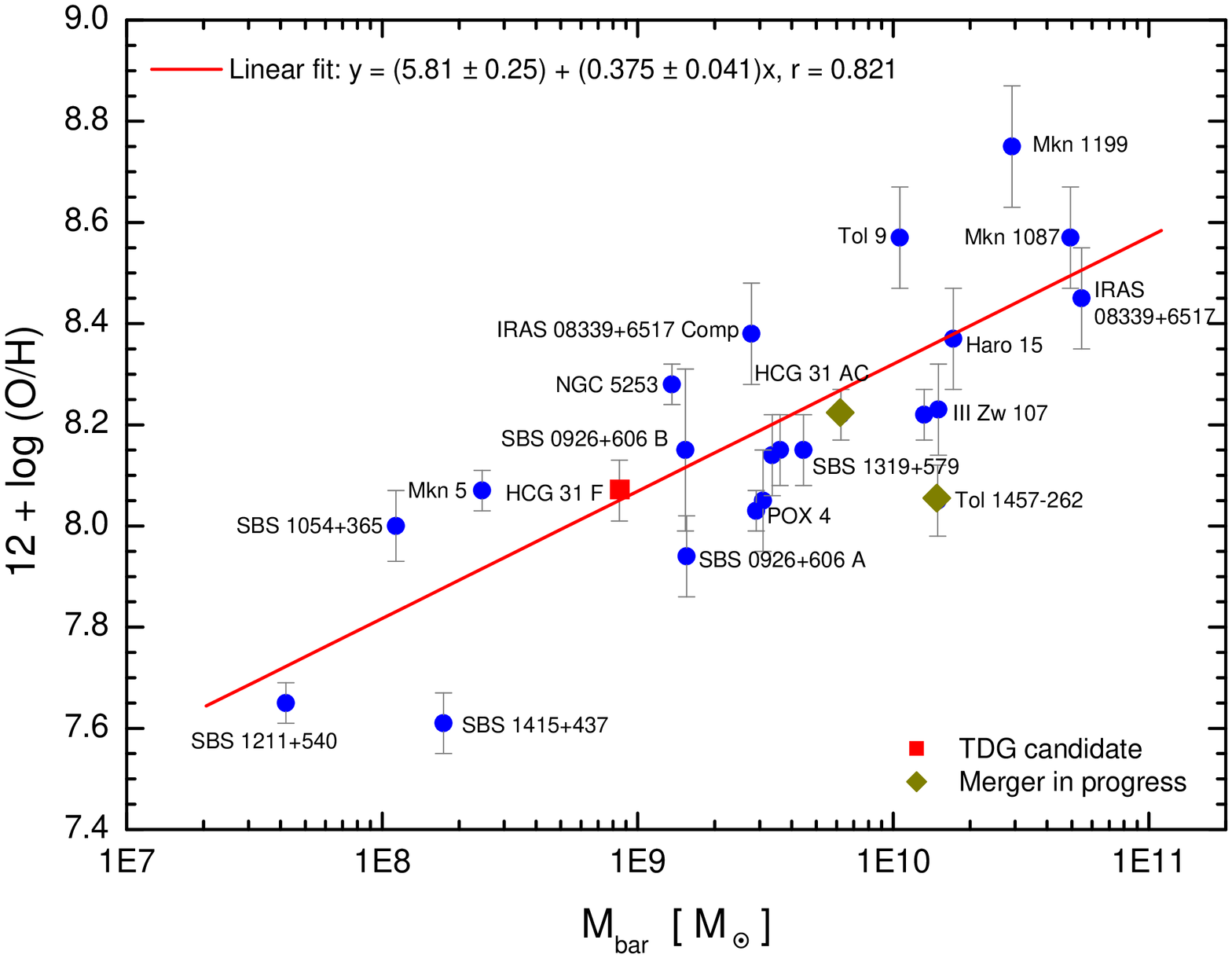} &
\includegraphics[angle=270,width=0.48\linewidth]{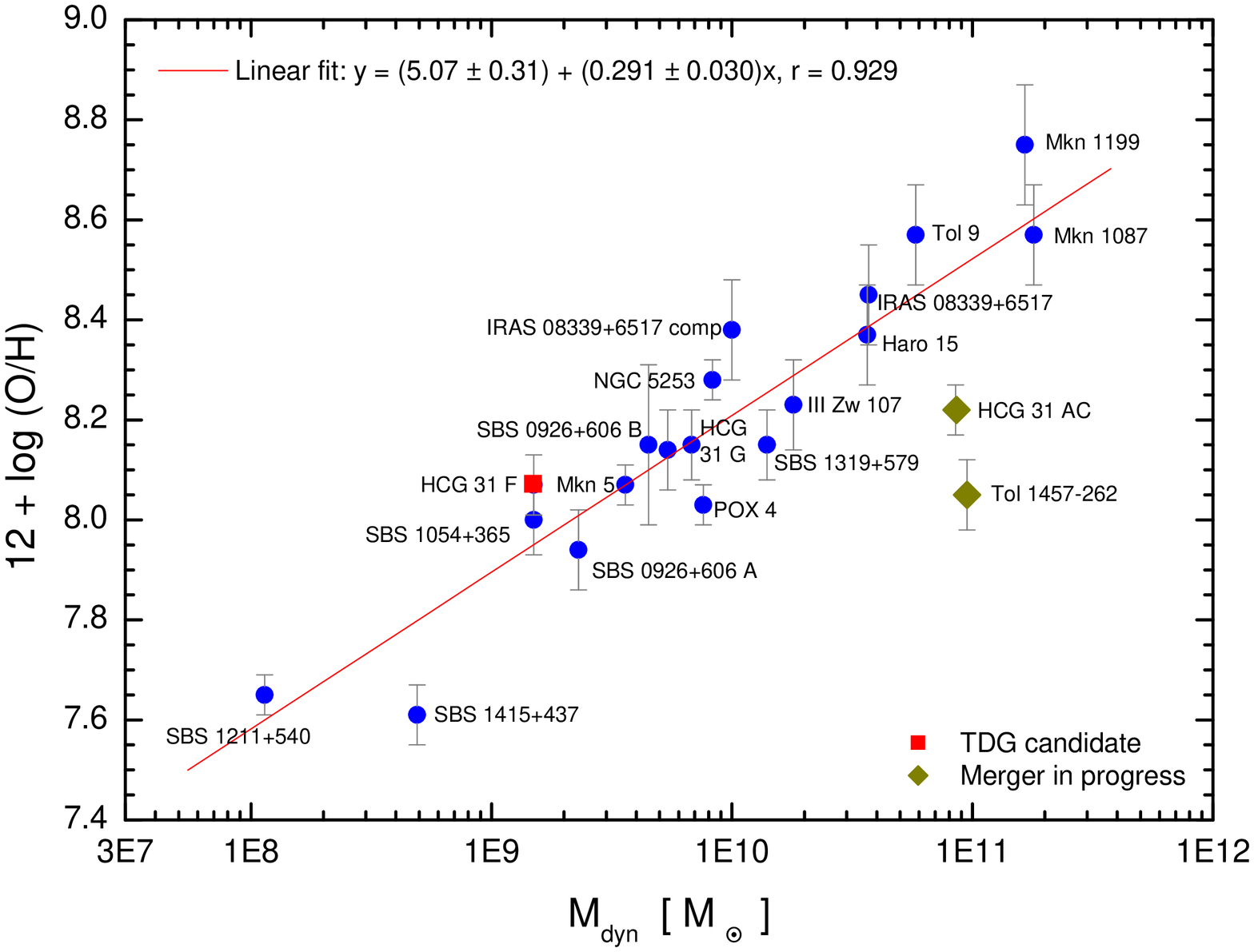} \\
\end{tabular}
\protect\caption[ ]{\footnotesize{Relation between $M_{bar}$ (left panel) and \Mdyn\ (right panel) with the oxygen abundance for our galaxy sample.   
The \TDG\ candidate HCG~31~F is plotted with a red square, while HCG~31~AC and Tol~1457-262, that are galaxies in the process of merging, are shown  
with a yellow diamond. A linear fit to the data without considering these problematic objects is shown with a continuous red line in both diagrams.}}
\label{aboxdyn}
\end{figure*}

Figure~\ref{aboxmass} shows the relations between the stellar mass and the oxygen abundance, and Figure~\ref{aboxdyn} shows the relations between the  
baryonic mass (left panel) and the dynamical mass (right panel) with the oxygen abundance. From Figures~\ref{aboxmass} and \ref{aboxdyn} is quite  
evident that a $M-Z$ relation is satisfied for our sample galaxies. Although there is still a considerable dispersion for some objects, the comparison  
with the luminosity-metallicity relation (see Fig.~17 and Sect.~5 of Paper~IV) suggests a closest correlation when using the stellar, baryonic, or the  
dynamical masses than the absolute optical/\NIR\ magnitudes.

The $M_{stars}-Z$ diagram (Fig.~\ref{aboxmass}) shows a large dispersion for galaxies with \abox$\sim$8, as there are dwarf galaxies with  
$M_{stars}\sim4\times10^7$ \Mo\ (HCG~31~F and E, SBS~1054+365) and large systems with \mbox{$M_{stars}\sim5\times10^9$ \Mo}\ (UM~420, Tol~1457-262  
Obj~1) within this metallicity range. The origin of this dispersion is that the low-mass systems are \TDG\ candidates, that have higher oxygen  
abundance than that expected for their mass (this is not the case of SBS~1054+365), while the high-mass objects are very probably a merger of two  
independent galaxies (and hence their oxygen abundance is much lower than the expected for a single, more massive galaxy). Neglecting the \TDG\  
candidates and the galaxies in the processing of merging, a linear fit to the data yields to
\begin{eqnarray} 
\label{mstarsabox} x = (6.03 \pm 0.28) + (0.238 \pm 0.031) \log M_{stars},
\end{eqnarray}
being x=\abox. This relation has a correlation coefficient of $r$=0.849 and it is plotted in Fig.~\ref{aboxmass} with a red continuous line. Our  
$M_{stars}-Z$ relation is quite different from previous relations given in the literature. For comparison, the $M_{stars}-Z$ diagram includes the  
relations provided by \citet{Tremonti04} --dotted yellow line-- and \citet{KE08} using the \citet{P01b} calibration to compute the oxygen  
abundance\footnote{As we concluded in Paper~IV, this calibration provides the best results to the oxygen abundances derived for our galaxies, that  
were mainly computed following the direct \Te\ method.} --dashed pink line--. \citet{Tremonti04} derived the oxygen abundances using theoretical  
photoionization models, that overestimate between 0.2 and 0.4 dex the metallicity derived from the direct \Te\ method (see Paper~IV). Hence, we also  
plot in Fig.~\ref{aboxmass} the \citet{Tremonti04} relation corrected by a factor of 0.3 dex in oxygen abundance (dotted-dashed yellow line). As we  
see, \citet{Tremonti04} relation is steeper than that derived here, and does not agree well with our data. On the other hand, the \citet{KE08}  
relation is quite flat in comparison with our observational data. These authors commented that for masses between $3\times10^8$ and \mbox{$10^{11}$  
\Mo,} the metallicity of their sample galaxies rises only $\sim$0.2 dex on average, but we find a variation of $\sim$0.7 dex in the same mass  
interval. 

Interestingly, \citet{KE08} did not derive any $M_{stars}-Z$ relation using oxygen abundances determined with the \Te\ method, such as the SDSS  
catalog contains very few metal-poor and starbursting galaxies, and the scatter of the available data is huge. These authors finally concluded that  
the choice of the metallicity calibration has the strongest effect on the $M-Z$ relation, because a considerable variation in shapes and  
$y$-intercepts are found. Many of their fits suggested a flatter $M-Z$ relation at higher masses, something that was previously noticed by  
\citet{Tremonti04}. These authors explained this issue as a consequence of effective galactic winds that remove metals from the low-mass galaxies  
($M\leq10^{10.5}$ \Mo). Although our data do not allow to explore this issue at high masses, we do not see any trend of this effect.

It is very interesting to note that Tol~9, in which we detect a clear example of galactic wind, has a relatively low stellar mass for its expected  
metallicity. Probably that is indicating the strength and youth of the star-formation phenomena in this \BCG. We should expect that the position of  
this object in the $M_{stars}$-O/H diagram will move to higher masses and lower metallicities if the star-formation processes continue and the fresh  
new material is expelled far from the galaxy via the effect of galactic winds.     

The linear fits to the $M_{bar}-Z$ and \Mdyn$-Z$ relations, that are plotted with red continuous lines in Figure~\ref{aboxdyn} and do not consider  
\TDG\ candidates and mergers in progress, are
\begin{eqnarray}  
x = (5.81 \pm 0.25) + (0.375 \pm 0.041) \log M_{bar}, \\
x = (5.07 \pm 0.31) + (0.291 \pm 0.030) \log M_{dyn},
\end{eqnarray}
being x=\abox. The correlation coefficients are \mbox{$r=0.821$} and $r=0.929$ for the  $M_{bar}-Z$ and $M_{dyn}-Z$ relations, respectively. The  
slopes of all the $M-Z$ relations agree relatively well. We note the tightness of the  \Mdyn$-Z$ relation: except the mergers in progress (HCG~31~AC  
and Tol~1457-262), all galaxies are found relatively close to the relation. This indicates that the dark matter content of the galaxies also increases  
with the metallicity, in agreement with the predictions of the evolutionary galaxy models.    

In summary, the scatter observed in the luminosity-metallicity and in the mass-metallicity relationships of star-forming galaxies are consequence of  
both the nature and the star-formation histories experienced by these objects. Only a detailed analysis of each system can give the clues needed to  
understand the evolution of the global properties in star-forming galaxies and their comparison between dwarf, normal and massive galaxies. 

\begin{table*}[t!]
\centering
  \caption{\footnotesize{Mass-to-light ratios of all mass estimations compiled in Table~\ref{masas}. The \Mbar/\Mdyn, $M_{stars}/M_{bar}$,  
$M_{gas}/M_{bar}$, $M_{gas}/M_{stars}$, and $M_{dust}$/$M_{gas}$ ratios are also listed in the last columns.}}
  \label{mlratios}
  \tiny
  \begin{tabular}{l@{\hspace{8pt}}  c@{\hspace{9pt}}   c@{\hspace{5pt}}  c@{\hspace{5pt}}c@{\hspace{9pt}} c@{\hspace{9pt}}c@{\hspace{9pt}}   
c@{\hspace{9pt}}c@{\hspace{9pt}}  c@{\hspace{9pt}}c@{\hspace{9pt}} c@{\hspace{9pt}} c@{\hspace{6pt}}c}
  \tableline
   \noalign{\smallskip}
Galaxy      &   $\frac{M_{\rm H\,I}}{L_B}$     &   $\frac{M_{\rm H\,II}}{L_B}$      &    $\frac{M_{\rm dust}}{L_B}$        &   $\frac{M_{\star}}{L_B}$      
& $\frac{M_{Kep}}{L_B}$       &    $\frac{M_{dyn}}{L_B}$     
          &  $\frac{M_{stars}}{L_B}$             &    $\frac{M_{bar}}{L_B}$      &   $\frac{M_{bar}}{M_{dyn}}$  & $\frac{M_{stars}}{M_{bar}}$ 
		  & $\frac{M_{gas}}{M_{bar}}$     & $\frac{M_{gas}}{M_{stars}}$    & $\frac{M_{dust}}{M_{gas}}$ \\

	  \noalign{\smallskip}
	  
	    & $\big[\frac{M_{\odot}}{L_{\odot}}\big]$ &  $\big[10^4\frac{M_{\odot}}{L_{\odot}}\big]$   & $\big[10^4\frac{M_{\odot}}{L_{\odot}}\big]$ &  
$\big[10^4\frac{M_{\odot}}{L_{\odot}}\big]$  & $\big[\frac{M_{\odot}}{L_{\odot}}\big]$ & $\big[\frac{M_{\odot}}{L_{\odot}}\big]$ &  
$\big[\frac{M_{\odot}}{L_{\odot}}\big]$ & $\big[\frac{M_{\odot}}{L_{\odot}}\big]$ & & & &  &[$10^4$]    \\    
		
		\noalign{\smallskip}
\tableline
\noalign{\smallskip}   
   HCG 31    &   0.369   &   3.86 &  1.27   & 18.7 & \nodata & \nodata & 0.167 &  0.655 & \nodata&  0.255 &  0.744 &  2.91  &  2.60   \\
   "      AC &   0.395   &   5.86 & \nodata & 19.8 &   3.7   &   9.23  & 0.156 &  0.678 & 0.073  &  0.231 &  0.769 &  3.34  & \nodata \\
   "      B  &   0.409   &   0.96 & \nodata & 17.3 &   0.55  &   1.14  & 0.169 &  0.709 & 0.622  &  0.238 &  0.762 &  3.20  & \nodata \\
   "      E  & \nodata   &   7.09 & \nodata & 83.7 & \nodata & 4.15$^a$& 0.150 & \nodata& \nodata& \nodata& \nodata& \nodata& \nodata \\
   "      F  &   1.53    &  13.7  & \nodata & 28.6 &   0.75  &   3.73  & 0.097 &  2.11  & 0.565  &  0.046 &  0.954 & 20.8   & \nodata \\
   "      G  &   0.350   &   1.72 & \nodata & 12.6 &   0.38  &   1.23  & 0.189 &  0.651 & 0.531  &  0.290 &  0.710 &  2.44  & \nodata \\
   Mkn 1087  &   0.138   &   0.92 &  0.69   & 11.4 &   0.50  &   1.60  & 0.256 &  0.438 & 0.274  &  0.583 &  0.417 & 0.715  & 3.78    \\
   "        N& \nodata   &   1.29 & \nodata & 12.8 &   0.084 &   2.78  & 0.153 & \nodata& \nodata& \nodata& \nodata& \nodata& \nodata \\
   Haro 15   &   0.159   &   1.79 &  0.60   & 22.2 &   0.35  &   1.05  & 0.286 &  0.495 & 0.471  &  0.577 &  0.423 & 0.732  & 2.85    \\
   Mkn 1199  &   0.0419  &   2.49 &  1.07   & 48.0 &   0.28  &   5.67  & 0.945 &  1.00  & 0.177  &  0.945 &  0.055 & 0.059  & 19.2    \\
   "     NE  & \nodata   &   0.78 & \nodata & 10.1 &   0.17  & 4.22$^a$& 1.05  & \nodata& \nodata& \nodata& \nodata& \nodata& \nodata \\
   Mkn 5     &   0.273   &   3.08 &  3.76   & 11.4 &   8.0   &  13.7   & 0.575 &  0.936 & 0.068  &  0.614 &  0.386 & 0.628  & 10.4    \\
IRAS 08208+2816 &\nodata &   4.14 &  1.73   & 32.5 &   0.076 & 1.18$^a$& 0.585 & \nodata& \nodata& \nodata& \nodata& \nodata& \nodata \\
IRAS 08339+6517& 0.0805  &   2.19 &  0.59   & 34.2 &   0.151 &   0.56  & 0.720 &  0.827 & 1.48   &  0.871 &  0.129 & 0.148  &  5.57   \\
  " Comp     &   0.234   &   0.87 & \nodata & 1.37 & \nodata & 3.34    & 0.623 &  0.932 & 0.279  &  0.669 &  0.331 & 0.496  & \nodata \\ 
POX 4        &   0.219   &  11.2  &  0.183  & 19.1 &   0.098 &   1.49  & 0.281 &  0.570 & 0.383  &  0.492 &  0.508 & 1.03   &  0.63   \\
" Comp       & \nodata   &   2.30 & \nodata & 36.4 & \nodata & \nodata & 0.406 & \nodata& \nodata& \nodata& \nodata& \nodata& \nodata \\
UM 420       & \nodata   &   6.76 & 4.9$^b$ & 13.4 &   0.204 & 1.94$^a$& 0.444 & \nodata& \nodata& \nodata& \nodata& \nodata& \nodata \\
SBS 0926+606 &   0.664   &   6.58 &  1.39   & 31.6 & \nodata & \nodata & 0.285 &  1.16  & \nodata&  0.246 &  0.754 & 3.97   & 1.59   \\	
"           A&   0.746   &  10.9  & \nodata & 28.0 & \nodata &   1.79  & 0.226 &  1.21  & 0.676  &  0.187 &  0.813 & 4.35   & \nodata \\
"           B&   0.587   &   2.54 & \nodata & 34.5 & \nodata &   3.26  & 0.340 &  1.11  & 0.342  &  0.305 &  0.695 & 2.28   & \nodata \\
SBS 0948+532 & \nodata   &  31.7  & \nodata & 51.4 &   0.57  & 2.46$^a$& 0.345 & \nodata& \nodata& \nodata& \nodata& \nodata& \nodata \\
SBS 1054+365 &   0.929   &  10.2  &  1.83   & 35.1 &   1.19  &  22.9   & 0.505 &  1.73  & 0.076  &  0.292 &  0.798 & 2.43   & 1.50   \\
SBS 1211+540 &   0.762   &   6.61 & \nodata & 15.8 &   3.57  &   3.61  & 0.321 &  1.33  & 0.368  &  0.242 &  0.758 & 3.13   & \nodata \\
SBS 1319+579 &   0.408   &   0.88 &  0.75   & 1.94 &   2.14  &   3.48  & 0.669 &  1.11  & 0.318  &  0.514 &  0.486 & 0.946  & 1.40   \\
SBS 1415+437 &   0.964   &   7.35 & \nodata & 14.0 &   2.50  &   4.90  & 0.465 &  1.74  & 0.355  &  0.268 &  0.732 & 2.74   & \nodata \\
  III Zw 107 &   0.378   &   3.37 &   0.71  & 41.8 &   0.046 &   1.02  & 0.349 &  0.848 & 0.834  &  0.412 &  0.588 & 1.43   & 1.42   \\
   Tol 9$^e$ &   0.595   &   4.32 &   3.06  & 34.8 &   0.152 &   7.38  & 0.564 &  1.35  & 0.183  &  0.418 &  0.582 & 1.39   & 3.90   \\
Tol 1457-262 &   0.258   &   5.17 &   1.01  & 22.5 & \nodata &   5.22  & 0.479 &  0.820 & 0.157  &  0.584 &  0.416 & 0.711  & 2.95   \\
" Obj~1      &  \nodata  &   5.70 & \nodata & 19.3 &  0.051  & 2.40$^a$& 0.387 & \nodata& \nodata& \nodata& \nodata& \nodata& \nodata \\
" Obj~2      &  \nodata  &   4.25 & \nodata & 23.9 & \nodata & 2.49$^a$& 0.601 & \nodata& \nodata& \nodata& \nodata& \nodata& \nodata\\ 
Arp 252      & \nodata   &   3.01 &   1.45  & 18.9 & \nodata & \nodata & 0.775 & \nodata& \nodata& \nodata& \nodata& \nodata& \nodata \\
" ESO 566-8  & \nodata   &   3.57 & \nodata & 12.9 &   0.21  & 1.54$^a$& 0.716 & \nodata& \nodata& \nodata& \nodata& \nodata& \nodata \\
" ESO 566-7  & \nodata   &   1.09 & \nodata & 19.6 &   0.047 & 1.75$^a$& 1.03  & \nodata& \nodata& \nodata& \nodata& \nodata& \nodata \\
NGC 5253     &   0.070  &   2.84 &   0.182 & 9.67 & \nodata &   3.60  & 0.496 &  0.594 & 0.164  &  0.841 &  0.159 & 0.189  & 1.94   \\
\noalign{\smallskip}
\tableline
  \end{tabular}
  \begin{flushleft}
  $^a$ Using the tentative value of \Mdyn\ computed using Eq~\ref{emdynmb}.
  \end{flushleft}
\end{table*}

\section{Schmidt-Kennicutt relation}

\begin{figure}[t]
\includegraphics[angle=270,width=\linewidth]{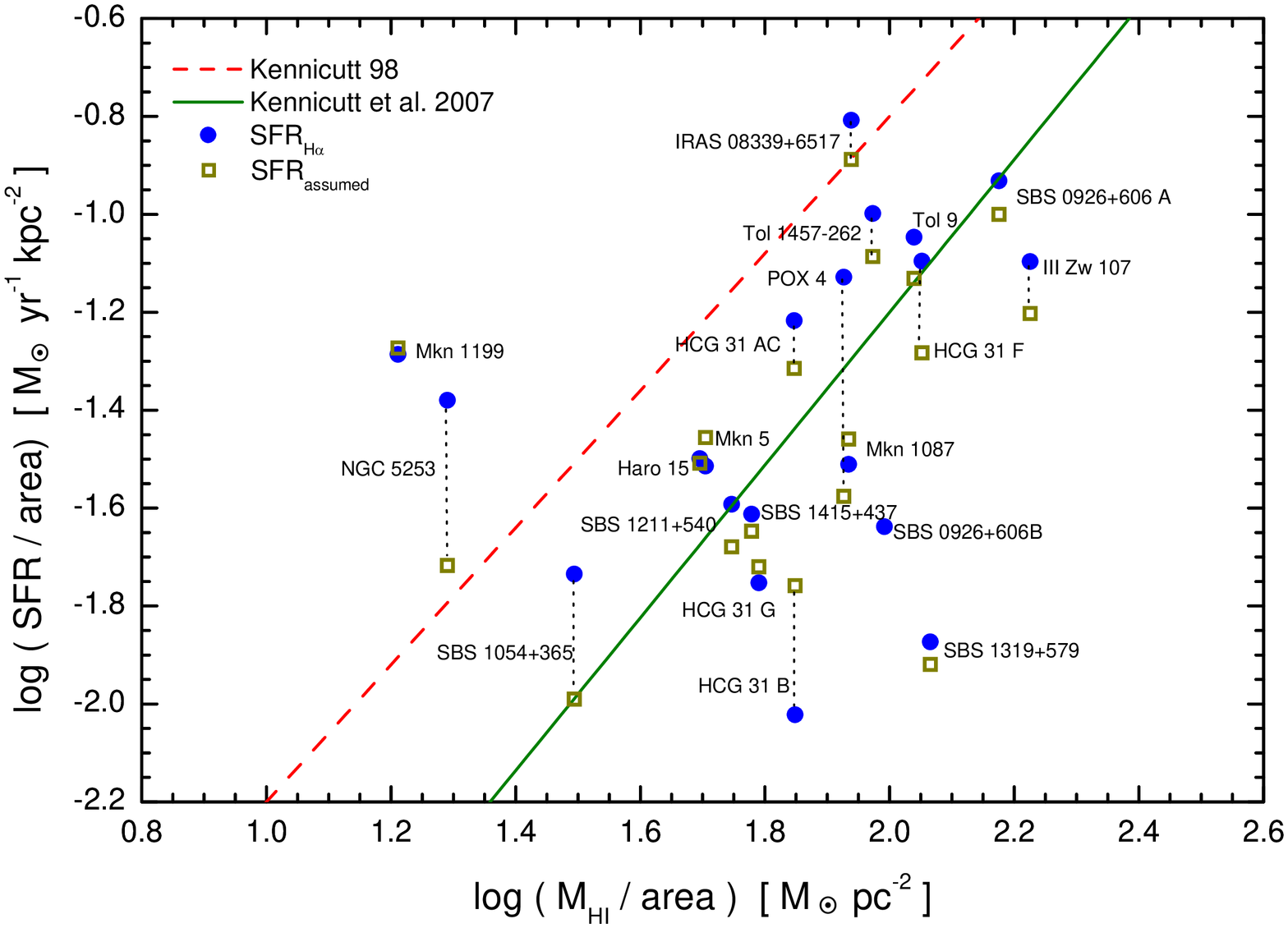}
\protect\caption[ ]{\footnotesize{Relation between the \SFR/area and the \HI\ gas density for our galaxy sample. We show two values per galaxy, one  
assuming the \Ha-based \SFR\ following the \citet{K98} calibration (blue circles) and other considering the \SFR\ assumed combining all  
multi-wavelength data. The solid green line is the best fit to the M~51 data \citet{K07}; the dashed red line is the relation for integrated values of  
star-forming galaxies derived by \citet{K98}.}}
\label{sfra}
\end{figure}

We now investigate if the studied galaxies do obey the Schmidt-Kennicutt scaling laws of star formation. %(Kennicutt 1998). 
It is well known  that there exist a tight correlation between the average \SFR\ per unit area and the mean surface density of the cold gas on global  
galactic scales. Such correlation is usually parameterized via a power-law relation \citep{Schmidt59,Schmidt63,K98,K07}
%(Schmidt 1959, 1963; Kennicutt 1998; Kennicutt et al. 2007), 
that has proven to be very useful as an input scaling law for analytical and numerical models of galaxy evolution (e.g., Kay et al. 2002).

Figure~\ref{sfra} shows, on logarithm scale, the \SFR\ per unit area versus the surface density of the \HI\ gas (\MHi/area) for all the galaxies for  
which we have \HI\ measurements. We plot two values for each galaxy, one assuming the \Ha-based \SFR\ using \citet{K98} calibration (blue circles) and  
other considering the \SFR\ assumed combining all multi-wavelength data (last column in Table~\ref{sfrg}). 
Almost both values are quite similar in all galaxies except in some objects, remarkably POX~4 and NGC~5253. The majority of the galaxies are located  
close to the relation given by \citet{K07}, which is the best fit to the data of star-forming regions within the nearby Sbc galaxy M~51. These authors  
also included the molecular gas to get this relation, but we have not considered it in our galaxy sample. 
The assumption of neglecting the molecular gas is valid in low-mass, low-metallicity galaxies, because of both the difficulty of detecting CO and the  
uncertainties of the correspondence between CO and H$_2$ in low-metallicity objects (i.e., Wilson 1995; Taylor, Kobulnicky \& Skillman 1998; Braine et  
al. 2004). However, we should expect some molecular gas contribution in more massive galaxies, as IRAS~08339+6517, Mkn~1087 and Mkn~1199.

From Fig.~\ref{sfra}, it is evident that our data agree much better with the relation given by  \citet{K07} than with the relation obtained by  
\citet{K98} for star-forming galaxies (and not regions within galaxies). Interestingly, a recent study of the star-formation activity  within UV-rich  
regions found in the outskirts of the galaxy pair NGC~1512/1510 \citep{KoribalskiLS09} yield to this same result.
We note that we plot the \Ha-based \SFR\ using the \citet{K98} calibration because both \citet{K98} and \citet{K07} relations use this calibration. In  
case of using the \Ha-based \SFR\ derived from the \citet{Calzetti07} calibration, the agreement of our data with the \citet{K07} relation will be  
even better. 
Some clear disagreements with the scaling laws of star formation are Mkn~1199 and NGC~5253 (both seem to be \HI\ deficient; the molecular gas  
component in Mkn~1199 would not explain its position in the diagram, as it would require that $\sim$40\% of the total neutral hydrogen mass is H$_2$)  
and %HCG~31~B and
SBS~1319+579 (that show lower \SFR\ than that predicted considering their \HI\ gas amount).  We then conclude that some external factors are indeed  
affecting the \emph{normal} star-formation activity in these three galaxies.

\section{Analysis of the mass-to-light ratios}

\begin{figure*}[t]
\begin{tabular}{cc}
\includegraphics[angle=270,width=0.45\linewidth]{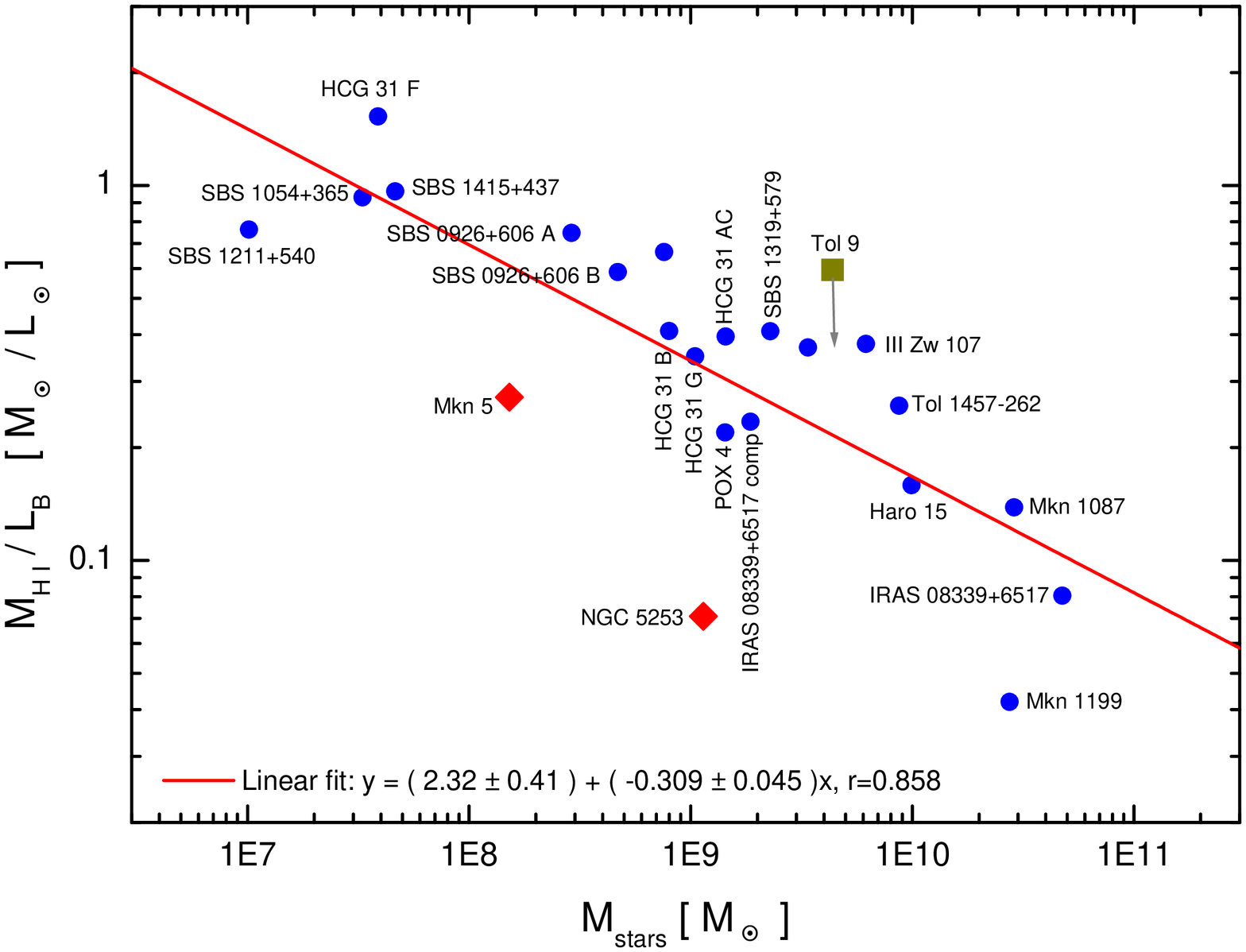} &
\includegraphics[angle=270,width=0.45\linewidth]{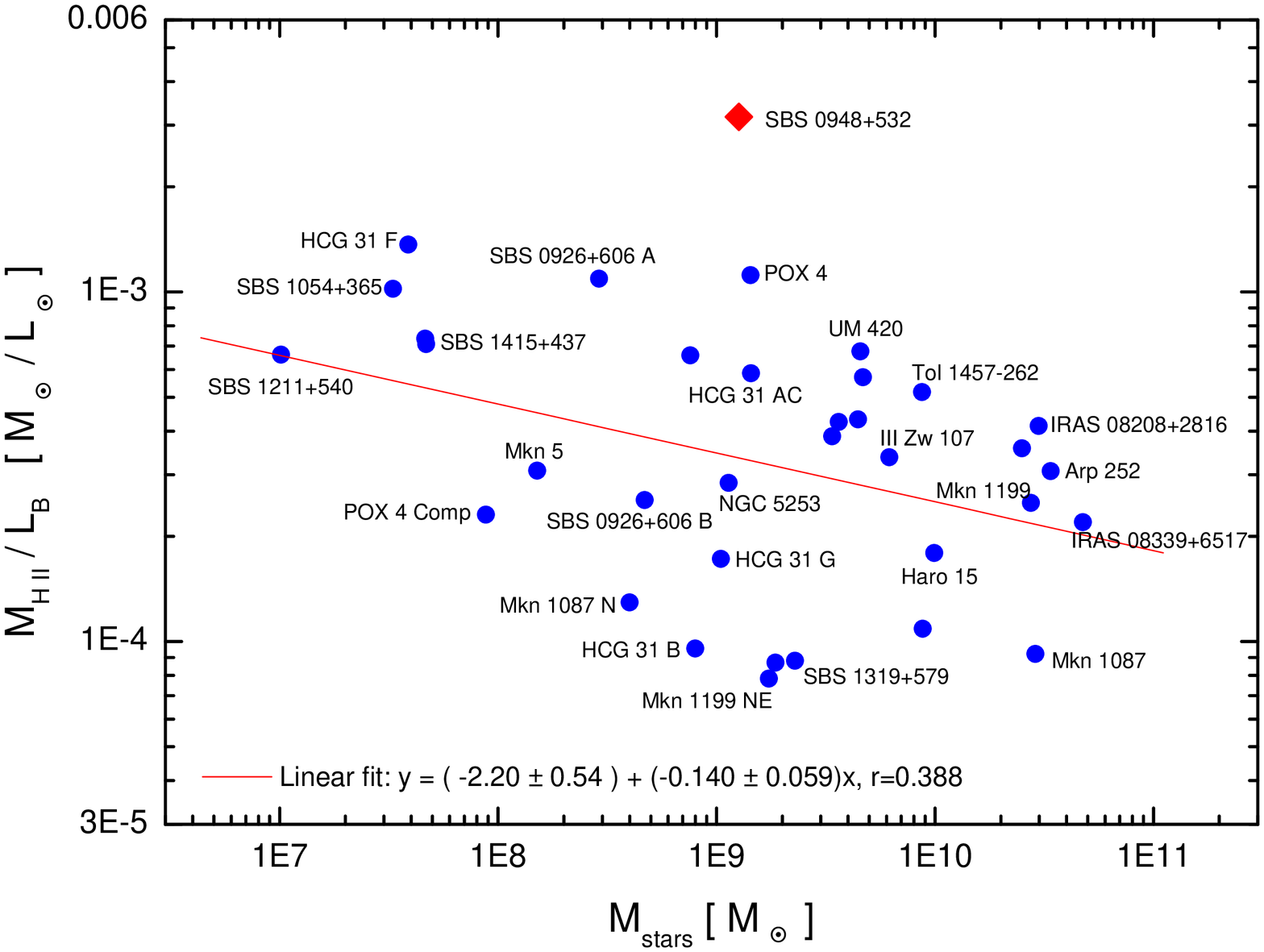} \\
\includegraphics[angle=270,width=0.45\linewidth]{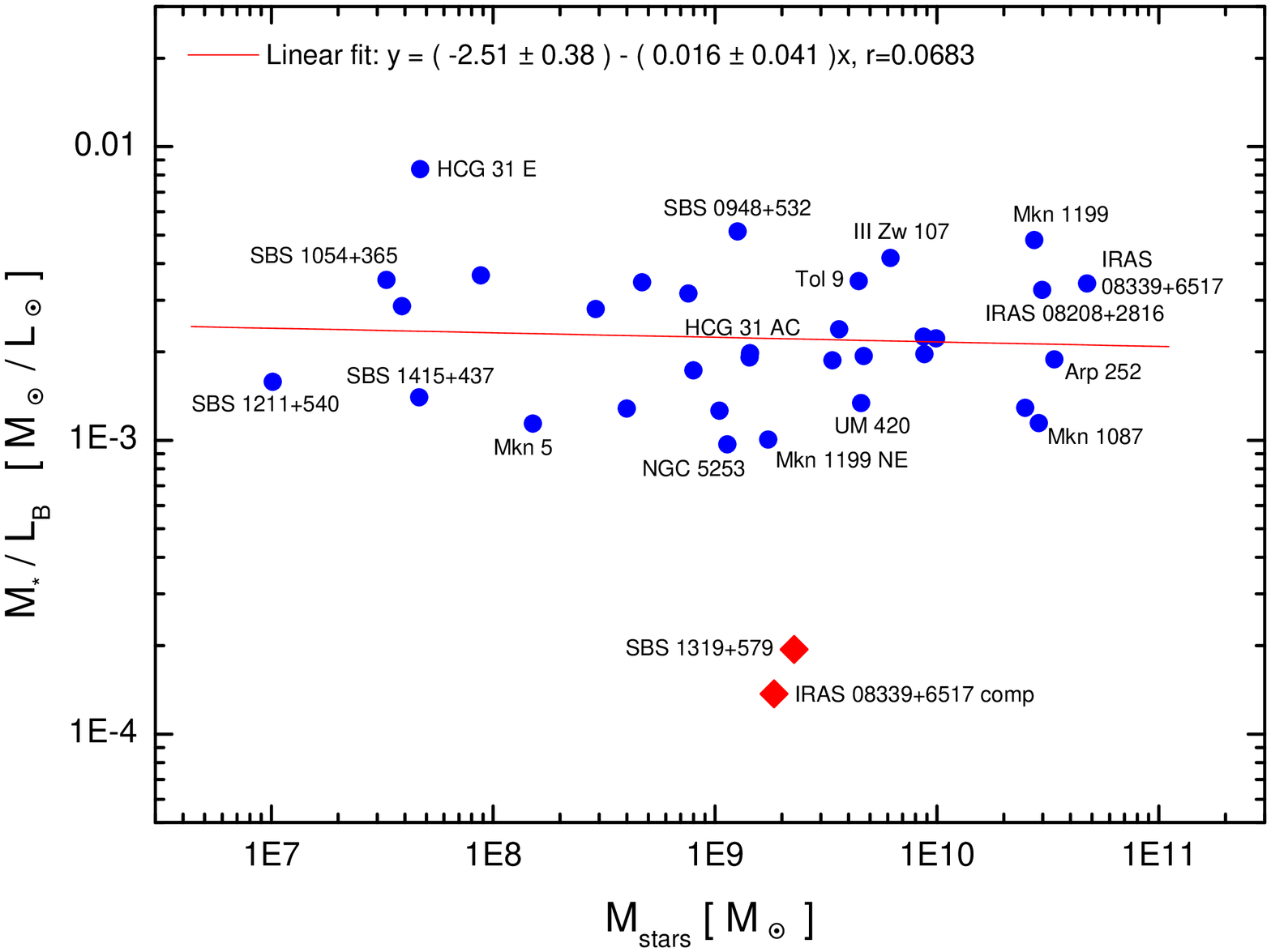} &
\includegraphics[angle=270,width=0.45\linewidth]{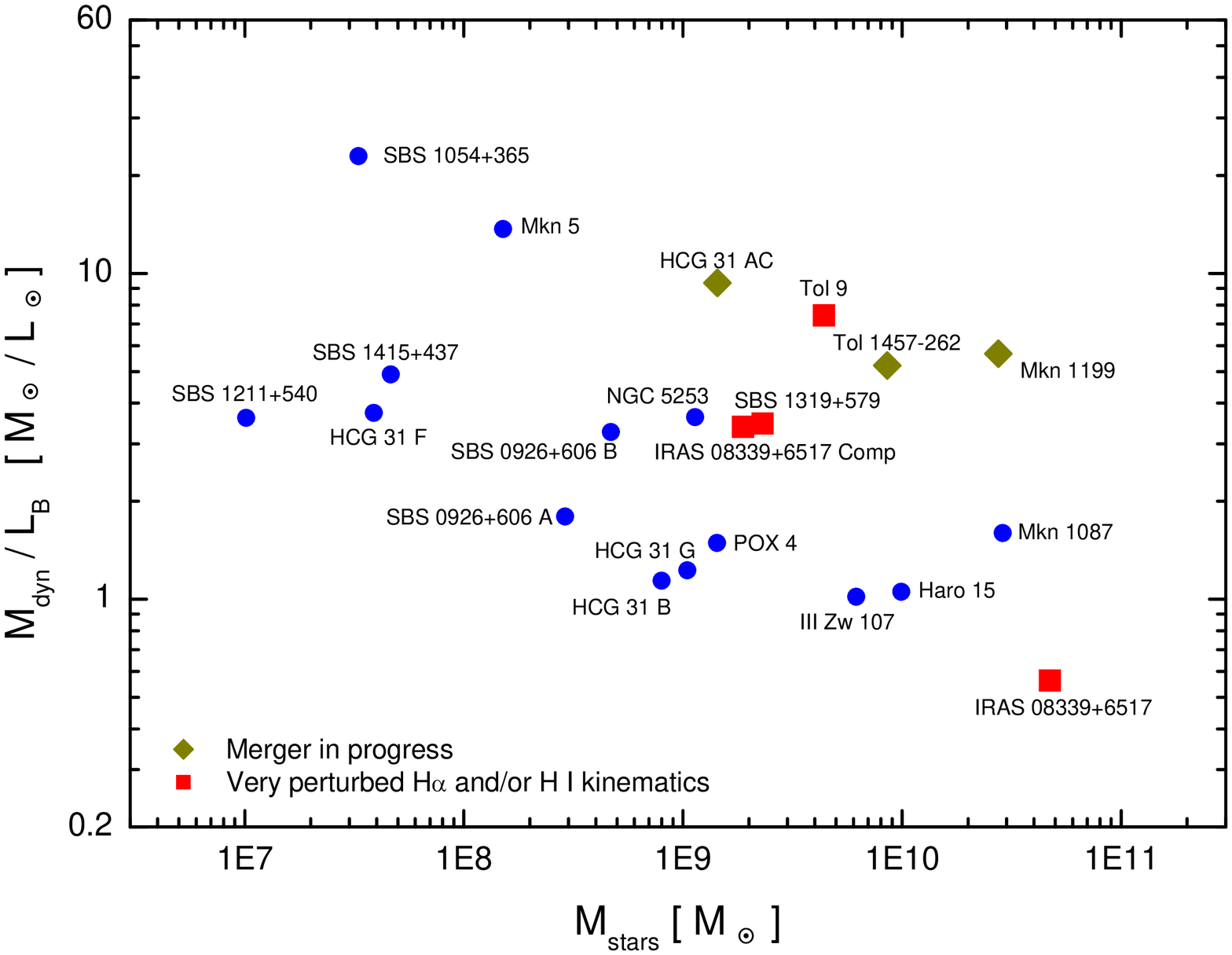} \\
\end{tabular}
\protect\caption[ ]{\footnotesize{Comparison between the stellar mass and some mass-to-light ratios for our sample galaxies.}}
\label{mlrat}
\end{figure*}

\begin{figure}[t]
\includegraphics[angle=270,width=\linewidth]{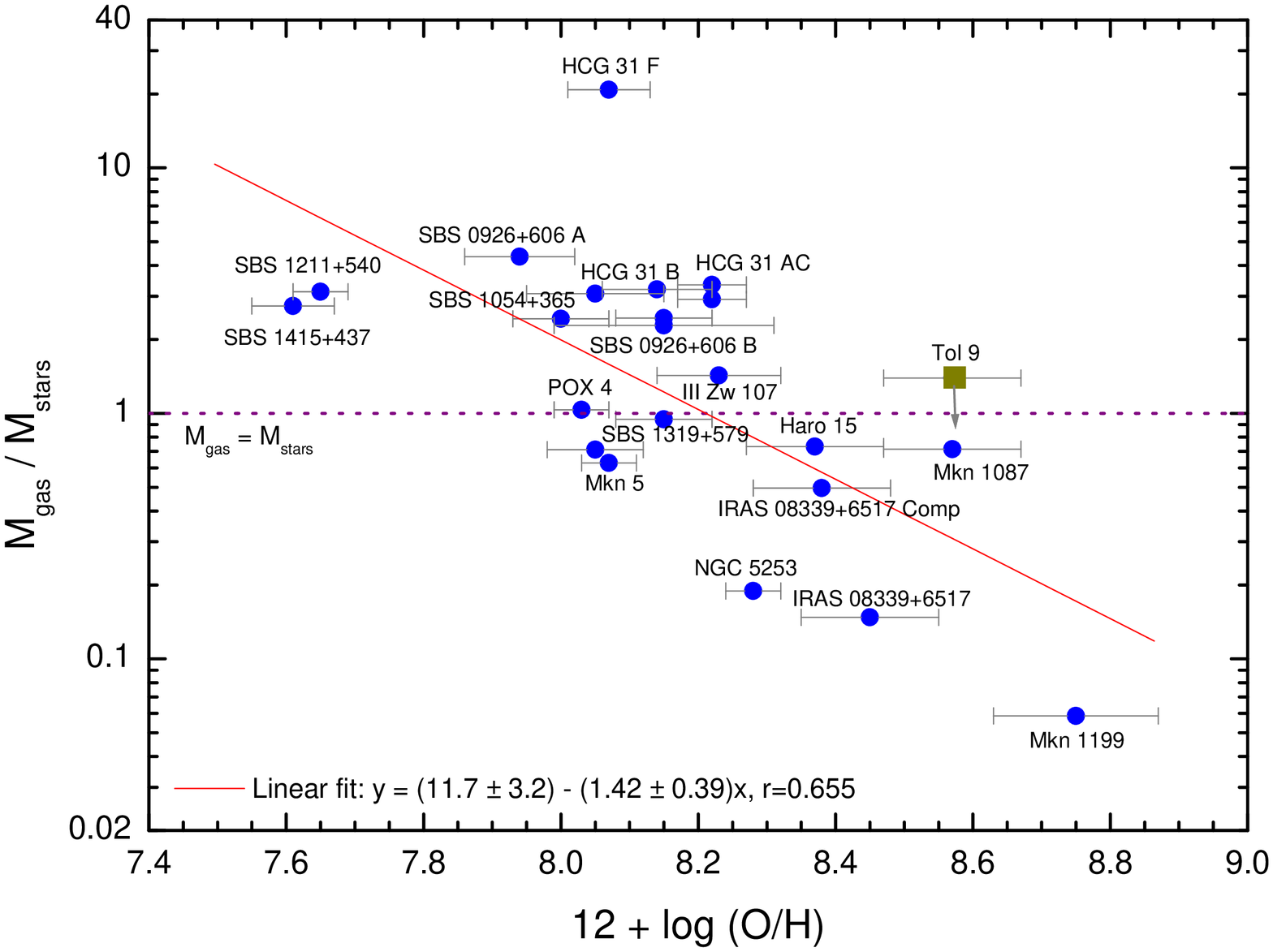}
\protect\caption[ ]{\footnotesize{Comparison of the $M_{gas}/M_{stars}$ ratio with the oxygen abundance for our sample galaxies. The dotted horizontal  
pink line indicates the position of $M_{gas}$=$M_{stars}$. The red continuous line is a linear fit to the data, neglecting Tol~9, for which the  
neutral gas mass corresponds to this object and some dwarf surrounding galaxies (see text and LS08).}}
\label{mgms_abox}
\end{figure}

Table~\ref{mlratios} compiles all the mass-to-$B$ luminosity ratios derived in this work. Some interesting relations are plotted in  
Figure~\ref{mlrat}, that compares some mass-to-light ratios with the stellar mass derived from the $H$-band luminosity.

\begin{figure*}[t]
\begin{tabular}{cc}
\includegraphics[angle=270,width=0.45\linewidth]{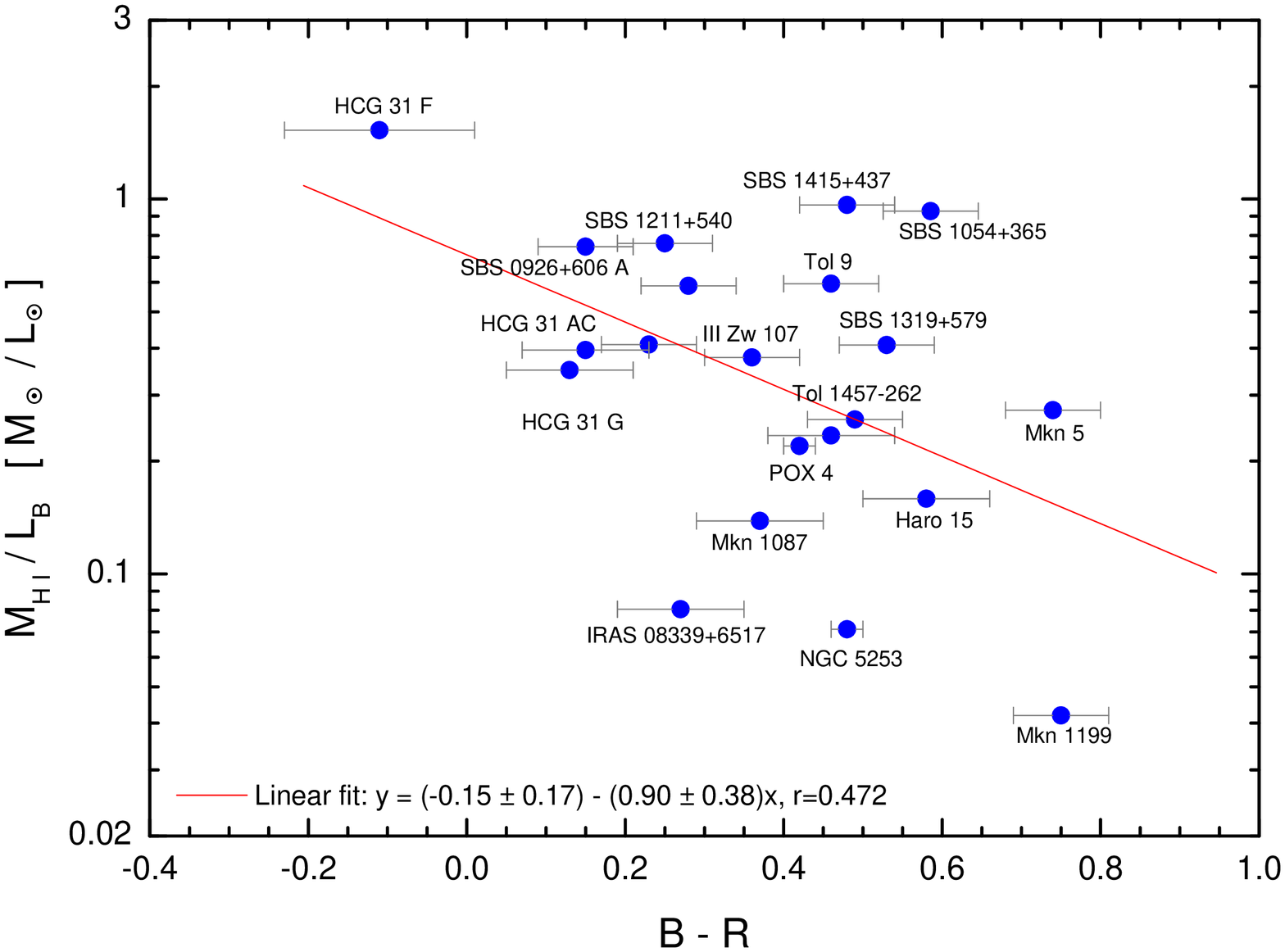} &
\includegraphics[angle=270,width=0.45\linewidth]{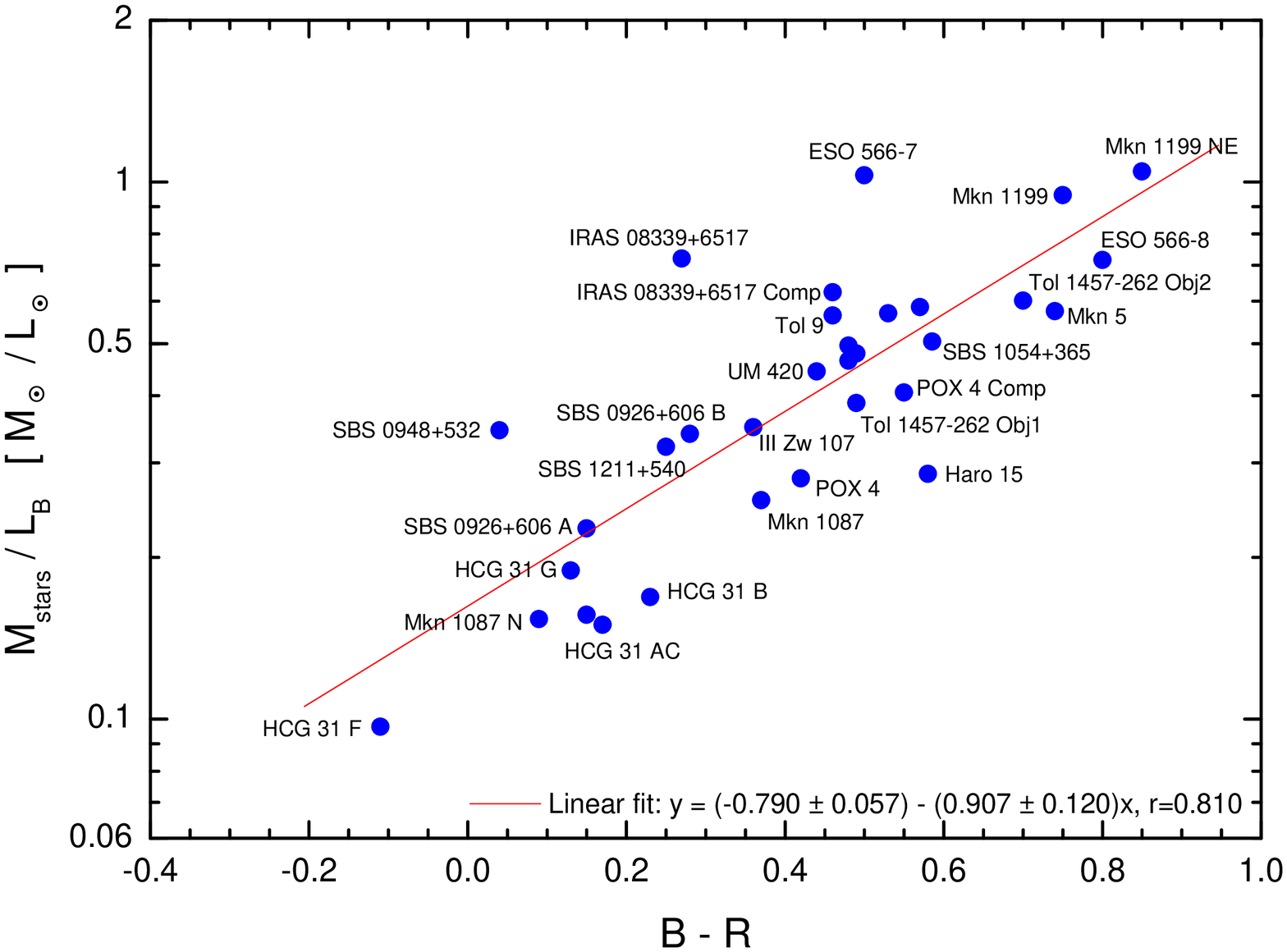} \\
\end{tabular}
\protect\caption[ ]{\footnotesize{Relation between the \MHil\ ratio (left panel) and the $M_{stars}/L_B$ ratio (right panel) with the $B-R$ color for  
our sample galaxies. Linear fits are plotted with a continuous red line. Some objects have been labeled.}}
\label{mlcolor}
\end{figure*}

The \HI\ mass-to-light ratio of a galaxy is a distance-independent quantity that compares the \HI\ mass with the luminosity in the $B$-band. This  
property correlates with many galaxy parameter, as the galaxy type, galaxy color or galaxy mass \citep{RH94}.
%(Robert \& Haynes 1994). 
Indeed, the comparison of the \MHil\ ratio with the stellar mass in our sample galaxy clearly indicates that less massive galaxies have a higher mass  
fraction of neutral gas. 
The majority of the galaxies have a \HI-mass-to-light ratio between 0.1 and 1.0, in agreement with previous estimations in star-forming dwarf galaxies  
\citep{Salzer02,HKP05}.
%(Salzer et al. 2002; Huchtmeier et al. 2005). 
%The median \MHil\ in type Sm/Im galaxies is 0.78 \Mol, 
%with an inner quartile range of 0.44 -- 1.32 \Mol.
We note some peculiar objects in this diagram. The \HI-mass of Tol~9 has been overestimated because the \HI\ cloud in which it is embedded includes  
several dwarf galaxies \citep{LS08b,LS+10b}. On the other hand, two galaxies (Mkn~5 and NGC~5253) are very \HI-deficient. 
In particular, NGC~5253 is very far from the typical position of the galaxies, showing a \MHil\ of $\sim$0.051 \Mol. The \MHil\ ratio of Mkn~1199 is  
also slightly low, even for a massive galaxy. As we already suggested (Sect.~3.4.3 of Paper~II), Mkn~1199 may has lost part of its neutral gas in the  
interaction process with its NE companion. Neglecting the contribution of Tol~9, Mkn~5 and NGC~5253, a linear fit provides the empirical relation
\begin{eqnarray}  
\log \frac{M_{\rm H\,I}}{L_B} = (2.32\pm0.41) - (0.309\pm0.045) \log M_{stars}, 
\end{eqnarray} 
that has a correlation coefficient of $r=0.858$.

We do not find any high \MHil\ ratio ($>1$\Mol) in our sample galaxy, except in the case of the \TDG\ candidate HCG~31~F, that has 1.53 \Mol. High  
\HI\ mass-to-light ratios has been reported in some few galaxies. The detailed analysis of the gas-rich low surface brightness dwarf irregular galaxy  
ESO~215-G009 performed by \citet*{Warren04} confirmed an extremely high \MHil\ of $22\pm4$ \Mol\ in this galaxy, for which the \HI\ disk extends  
$6.4\pm0.4$ times the Holmberg radius. They concluded that ESO~215-G009, that is very isolated (no neighbors identified out to 1 Mpc), has a low \SFR\  
that probably remained unchanged throughout the galaxy´s existence.  In a subsequent paper \citep*{Warren06} these authors suggested that high \MHil\  
galaxies are not lacking the baryons to create stars, but are underluminous as they lack either the internal or external stimulation for more  
extensive star formation. 

\citet*{Warren07} derived an empirical upper envelope for \MHil\ as a function of the absolute $B$-magnitude, that accounts for the maximum amount of  
atomic hydrogen gas a galaxy of a particular luminosity can retain in the Universe today. All our sample galaxies satisfy this empirical relation. 

The derived \MHiil\ ratios for our galaxy sample lie in the range $10^{-3}$--$10^{-4}$ \Mol.
Although the scatter of our data is high (a tentative linear fit gives a very low correlation coefficient), they indicate that the \MHiil\ ratio  
slightly decreases with increasing stellar mass, suggesting that the ionized gas to stars ratio is higher in dwarf galaxies. SBS~0948+532 is away from  
the rest of the objects because of its very high \Ha\ flux (see Sect.~3.11 in Paper~I). On the other hand, the ionizing star cluster mass-to-light  
ratio, \Mstarl, seems to be rather constant with the stellar mass, showing an average value of $\sim$0.0022 \Mol. Two galaxies, SBS~1319+579 and the  
companion object of IRAS~08339+6517 lie apart from this tendency, as we also saw in Fig.~\ref{masasmb}.

The \Mdynl\ ratio seems to slightly decrease with the stellar mass. However, the analysis of this diagram is difficult, because interactions notably  
modify the estimation of the dynamical mass. Usually, perturbed kinematics yield to higher \Mdyn\ (HGC~31~AC, Mkn~1199, Tol~1457-262, Tol~9), but  
sometimes the existence of tidal tails with a rather constant velocity give a lower \Mdyn\ than the real one (IRAS~08339+6517). SBS~1319+579 (a  
probable merging of two dwarf objects) and IRAS~08339+6517~Comp (in interaction with the main galaxy of the system) also show somewhat high \Mdynl\  
ratios. Hence, we may suggest that galaxies Mkn~5 and SBS~1054+365, that lie far from the rest of the objects, have a perturbed kinematics, being  
\Mdyn\ overestimated in both cases.  

%On the other hand, the $M_{stars}/L_B$ ratio seems to increase with increasing dynamical mass. This fact indicates that 
Figure~\ref{mgms_abox} compares the ratio between the gas and the stellar masses with the oxygen abundance. Clearly, $M_{gas}/M_{stars}$ decreases  
with increasing metallicity, indicating that the importance of the stellar component to the total mass is higher in more massive galaxies. We should  
expect this result, which is qualitatively the inverse behavior of the \MHil\ ratio with the stellar mass. A tentative linear fit to the data  
(excluding Tol~9) is plotted in Fig.~\ref{mgms_abox}) with a continuous red line, and provides
\begin{eqnarray}  
\log \frac{M_{gas}}{M_{stars}} = (11.7\pm3.2) - (1.42\pm0.39)x, 
\end{eqnarray} 
being x=\abox, and has a correlation coefficient of $r=0.655$. Following this analysis, we should expect that galaxies with \abox$\sim$8.2--8.3 have  
relatively equal gas and stellar masses. HCG~31~F is the galaxy with highest $M_{gas}/M_{stars}$ ratio (and lowest $M_{stars}/L_B$ ratio), indicating  
its low content of evolved stars. This result agrees with our suggestion \citep{LSER04a} that this \TDG\ has created stars mainly using the neutral  
gas from the long arm-like \HI\ structure \citep{VM05} found between members AC and G of the HCG~31 galaxy system.

Finally, we compare some mass-to-light ratios with the colours of the galaxies. \citet{Amorin09} reported 
%a correlation between the \HI\ gas mass and the host luminosity of \BCG s, 
%showing that galaxies with larger gas reservoirs are expected to be 
%optically more luminous and also more extended. These authors also indicated 
that the underlying component (host) of blue compact galaxies is redder with decreasing \MHil.
We do not find any correlation between the  \MHil\ ratio and the optical colors of the underlying component, but we did not perform a detailed  
analysis of the structural parameters of the host underlying the starburst as \citet{Amorin09} did. The comparison of the global $B-R$ colour and the  
neutral gas mass-to-light ratio is shown in the left panel of Fig~\ref{mlcolor}, and has a huge scatter. We should remember that Mkn~1199, Mkn~5 and  
NGC~5253 seem to be \HI\ deficient, and that IRAS~08339+6517 is a luminous blue compact galaxy \citep*{LSEGR06}, and hence their real positions in  
this diagram is uncertain. Although a tentative fit to the data suggests that galaxies with redder $B-V$ colours have lower \MHil\ ratios, the huge  
scatter does not allow us to confirm such tendency.

However, we do observe a clear relation between the stellar-to-light ratio and the global $B-R$ colour of the galaxies (right panel of  
Fig.~\ref{mlcolor}). This tendency seems to be also a consequence of the building of the galaxies, as more massive galaxies have experienced more  
star-formation events than less massive objects, and hence tend to show redder stellar populations than dwarf galaxies. This result also agrees quite  
well with the observed tendencies that \MHil\ ratio decreases with the stellar mass and that $M_{gas}/M_{stars}$ ratio decreases with increasing   
metallicity.

%Besides the scatter, Figure~\ref{mlcolor} shows that redder galaxies tend 
%to have lower \MHil\ ratios. A linear fit, with $r=0.469$, is plot in 
%this figure with a red continuous line.  The slope of that fit, $-$0.88, 
%is similar to that found by  Amor\'{\i}n et al. (2009) analyzing the hosts 
%of \BCDG s, $-0.81$. Galaxies IRAS~08339+6517, Mkn~1199 and NGC~5253 show 
%relatively low \MHil\ ratios for their $B-R$ colors, indicating again that 
%they are neutral gas deficient.

%\section{Warm dust mass-extintion relation}
\section{Dust properties in star-forming galaxies}
   
\begin{figure}[t!]
\includegraphics[angle=270,width=\linewidth]{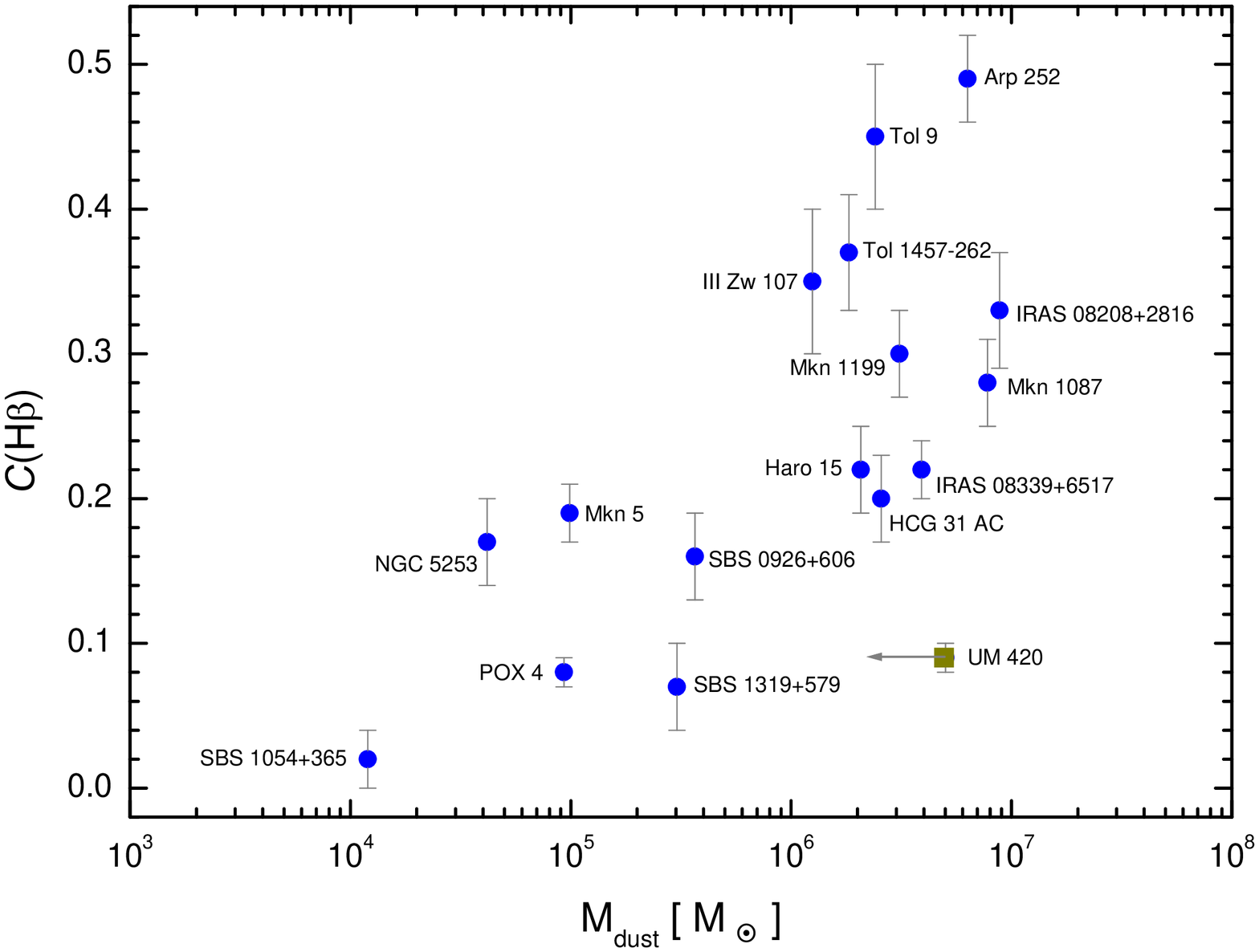}
\protect\caption[ ]{\footnotesize{Reddening coefficient, \CHb, vs. warm dust mass, \Mdust. We note that the \Mdust\ value in UM~420 is overestimated  
because of the \FIR\ contribution of UGC~01809.}}
\label{mpolvo}
\end{figure}

Our data set allows us to investigate the properties and effects of the dust content in low-metallicity star-forming galaxies.  
%We also compare the total masses with other properties. 
Figure~\ref{mpolvo} plots the reddening coefficient, \CHb\ --obtained using our optical spectra--, as a function of the warm dust mass, \Mdust\  
--derived from \FIR\ data--. Neglecting the data for UM~420 (as we said before, \Mdust\ is overestimated because of the \FIR\ contribution of the  
foreground galaxy UGC~01809), we see a clear correlation between both quantities: galaxies with higher amount of warm dust (and hence, as we saw in  
Figure~\ref{masasmb}, higher luminosity) show higher extinction. This conclusion is in agreement with other results previously found in this work, as  
the correlation between \CHb\ and the oxygen abundance discussed in Paper~IV. 
More important, this result indicates that most of the dust is internal to the galaxy and not in the line of sight.
%, as is very probably happening for UM~420.
Detailed analysis of the dust distribution within nearby galaxies (i.e. Mu\~noz-Mateos et al. 2009) found clear relationships between the dust content  
and general properties of nearby spiral galaxies, such as galaxy type, luminosity, and metallicity. Here we confirm in an independent way, such as the  
extinction was derived from our optical spectra, that the dust content and therefore the extinction in dwarf galaxies depends on their metallicities  
and luminosities, and very probably also on their star-formation histories. A proper estimation of the amount of dust within such objects is needed to  
perform appropriate statistical analysis involving larger galaxy samples.

%\subsection{Dust-to-gas ratio}

\begin{figure}[t]
\includegraphics[angle=270,width=\linewidth]{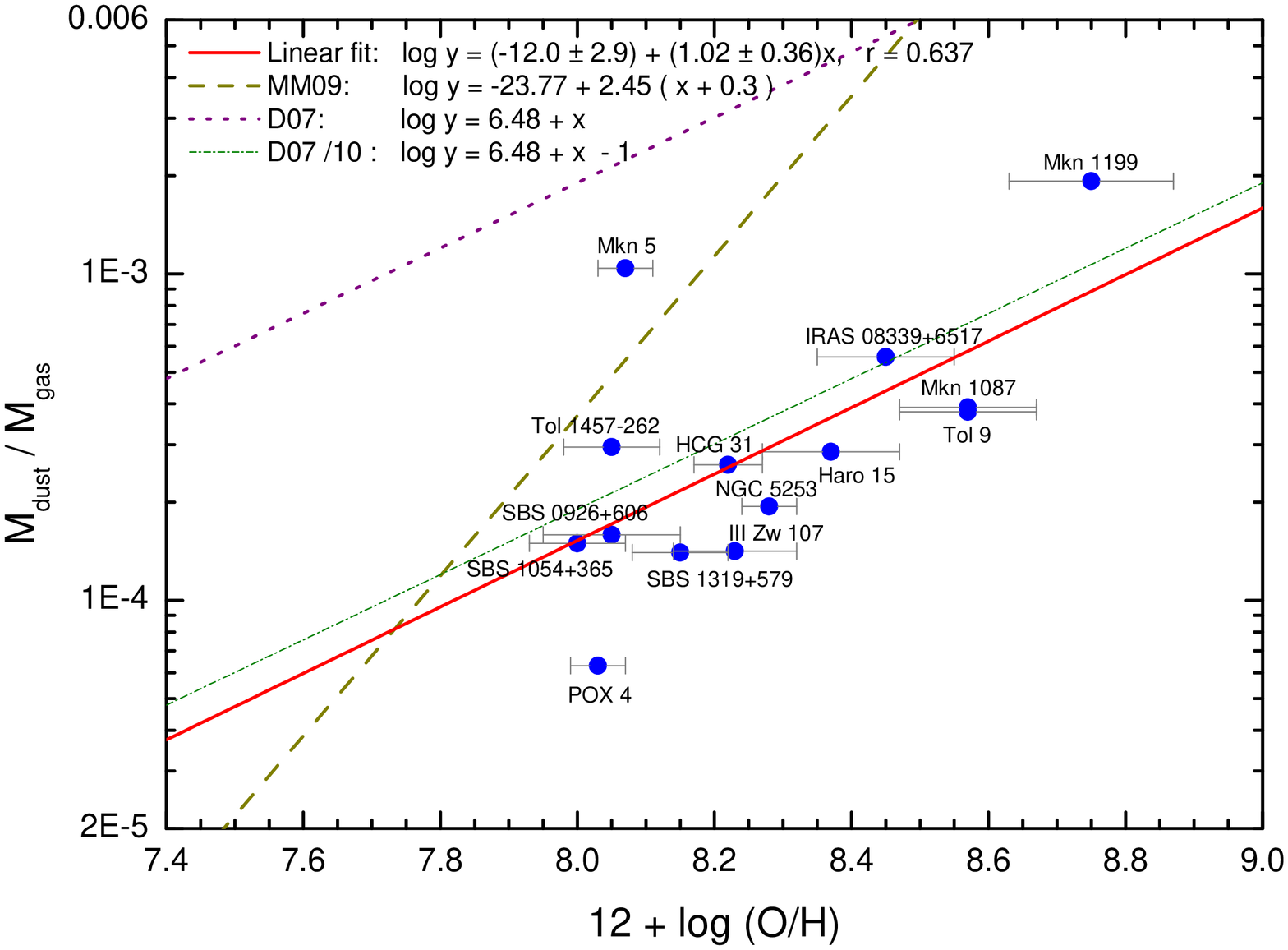}
\protect\caption[ ]{\footnotesize{Dust-to-gas ratio, $M_{dust}/M_{gas}$, vs. the oxygen abundance for our sample galaxies. The red continuous line
is a fit to our data (neglecting Mkn~5). The dashed yellow line indicates the relation found by \citet{MunosMateos09} analyzing the
radial dust-to-gas profiles for a larger sample of spiral galaxies. This relation has been corrected by 0.3 dex because it was derived assuming the  
\citet{KK04} calibration to compute the oxygen abundances, that overestimates in 0.2--0.4 dex the oxygen abundance provided by the direct \Te\ method.  
The dotted pink line is the relation provided by \citet{Draine07} in their analysis of a sample of spiral and irregular galaxies. The dashed-dotted  
green line indicates the \citet{Draine07} relation divided by a factor 10.}}
\label{dgr}
\end{figure}

We now investigate the dust-to-gas ratio, $M_{dust}/M_{gas}$, of our sample galaxies, a very important quantity when studying the chemical enrichment  
of the ISM, as it accounts for the amount of metals locked up onto dust grains through the stellar yields. The existence of a correlation between the  
$M_{dust}/M_{gas}$ and the oxygen abundance has been reported in many studies (i.e., Lisenfeld \& Ferrara 1998; James et al. 2002; Draine et al. 2007;  
Mu\~noz-Mateos et al. 2009). 
Figure~\ref{dgr} shows the dust-to-gas ratio as a function of the oxygen abundance.
%We investigate the relation between the dust-to-gas ratio and the oxygen abundance 
%in our sample galaxies in Figure~\ref{dgr}. 
The gas mass was computed assuming only the \HI\ and the \HeI\ gas, 
%but also heavier elements, $M_{gas}$=1.36\MHi. However, we do not consider 
but not the contribution of the molecular gas, that as we already explained is not important in dwarf low-metallicity objects.
The derived dust-to-gas ratio of each galaxy for which we have both \HI\ and \FIR\ data are compiled in the last column of Table~\ref{mlratios}. From  
Figure~\ref{dgr}, it is evident that objects with higher metallicities tend to have higher $M_{dust}/M_{gas}$ ratios, such as the amount of dust  
increases while the neutral gas is consumed as the galaxies experience new star-formation phenomena.  The linear fit to our data (continuous red line  
in Fig.~\ref{dgr}) provides this tentative relation 
\begin{eqnarray}  
\log (M_{dust}/M_{gas}) = (-12.0 \pm 2.9) + (1.02 \pm 0.36)x, 
\end{eqnarray}
with x=\abox\ and correlation coefficient of $r=0.637$. 
Mkn~5 lies far the the majority of the points, but as we already commented this object seem to be very deficient in \HI, so we did not include this  
point in the fit. 
%Maybe that is also the situation of Mkn~1199.
Although with higher uncertainties, we also observe that more massive objects also tend to have higher $M_{dust}/M_{gas}$ ratios. 

\citet{Draine07} provided a relation between $M_{dust}/M_{gas}$ and the oxygen abundance --computed following the \citet{PT05} calibration-- in a  
sample of spiral and irregular galaxies, that we may re-write as
\begin{eqnarray}  
\log (M_{dust}/M_{gas}) = 6.48 + x, 
\end{eqnarray} 
being x=\abox, and plotted in Figure~\ref{dgr} with a dotted pink line. The factor 6.48 was derived from  $\log [ (M_{dust}/M_{gas})_{MW} / 1.32]+  
x_{MW}$ assuming $(M_{dust}/M_{gas})_{MW}=0.010$ and $x_{MW}=8.6$ for the Milky Way. As we see, this relation lies away from our data, although the  
slope is the same in both cases. 
\citet{Draine07} also found that the global dust-to-gas ratio of all their galaxies with \abox$<$8.1 falls below this equation, sometimes by a factor  
larger than 10. The dashed-dotted green line in Figure~\ref{dgr} plots the \citet{Draine07} relation divided by a factor 10. As we can see, this  
relation agrees better with our data. 
\citet{Draine07} also remarked that many of the low-metallicity galaxies have large \HI\ envelopes mainly composed by un-enriched material. However,  
the metallicity is derived from the brightest \HII regions, that usually show the highest metallicities (most recent star formation) within the  
system. Consequently, the derived $M_{dust}/M_{gas}$ ratio should correspond to lower oxygen abundances.

%The recent study performed by 
Recently, \citet{MunosMateos09} analyzed the 
radial dust-to-gas profiles for a larger sample of spiral galaxies
and found a steeper relation between  $M_{dust}/M_{gas}$ and the metallicity,
%dashed yellow line in Figure~\ref{dgr})
%found that in nearby spiral galaxies both quantities are related as
%\begin{eqnarray}  
%\log (M_{dust}/M_{gas}) = -23.77 + 2.45x, 
%\end{eqnarray} 
%being x=\abox. They 
that they explained because the outskirts of spiral galaxies seem to have a much lower $M_{dust}/M_{gas}$ than the central regions 
%found that the outskirts of the spiral galaxies seem to show 
%a steeper slope than the central regions 
(see their figure~16).
These authors suggested a link with the behavior found in dwarf galaxies because in the external regions of the spiral galaxies the neutral gas has  
not yet undergone star-formation activity. However, this trend could also be a consequence of the radial decrease of the star-formation efficiency  
found in nearby spirals (i.e., Thilker et al. 2007; Leroy et al. 2008).

We may check these hypotheses comparing our data with the relation provided by \citet{MunosMateos09}. 
However, we should modify slightly their equation, because these authors used the \citet{KK04} method to derive the metallicities, and this  
calibration overestimates the oxygen abundances provided by the direct \Te\ method in 0.2--0.4~dex (see Paper~IV). The modified relation that is  
plotted with a  dashed yellow line in Figure~\ref{dgr} is
\begin{eqnarray}  
\log (M_{dust}/M_{gas}) = -23.77 + 2.45 (x+0.3), 
\end{eqnarray} 
being x=\abox. 
%This relation is plotted with a dashed yellow line in Figure~\ref{dgr}.
We see that except for Mkn~5 all our data have lower  $M_{dust}/M_{gas}$ ratios that those predicted following this relation. Hence, we suggest that the low  
$M_{dust}/M_{gas}$ ratios found in dwarf low-metallicity galaxies is a consequence of the large reserves of un-enriched neutral gas in these systems,  
while the low $M_{dust}/M_{gas}$ ratios reported by  \citet{MunosMateos09} in the external regions of spiral galaxies are due to the decreasing of the  
star-formation efficiency in those areas. Indeed, the outskirts of spiral galaxies seem to show lower \HI\ surface densities --$\log (M_{\rm  
H}/area)\sim$0.4--0.8 \Mo\ pc$^{-2}$ for external regions of M~51 \citep{K07} and NGC~1512 \citep{KoribalskiLS09}-- than the neutral gas envelopes of  
dwarf galaxies --$\log (M_{\rm H}/area)\sim$1.4--2.0 \Mo\ pc$^{-2}$ following Fig.~\ref{sfra}--, that is translated in a decreasing of the  
star-formation activity following the Schmidt-Kennicutt law. 
However, more high-quality data and a detailed analysis of the dust and the \HI\ distribution in nearby dwarf galaxies should be performed to confirm  
all these issues.

\section{Comparison with the closed-box model}

\begin{figure*}[t]
\begin{tabular}{cc}
\includegraphics[angle=270,width=0.45\linewidth]{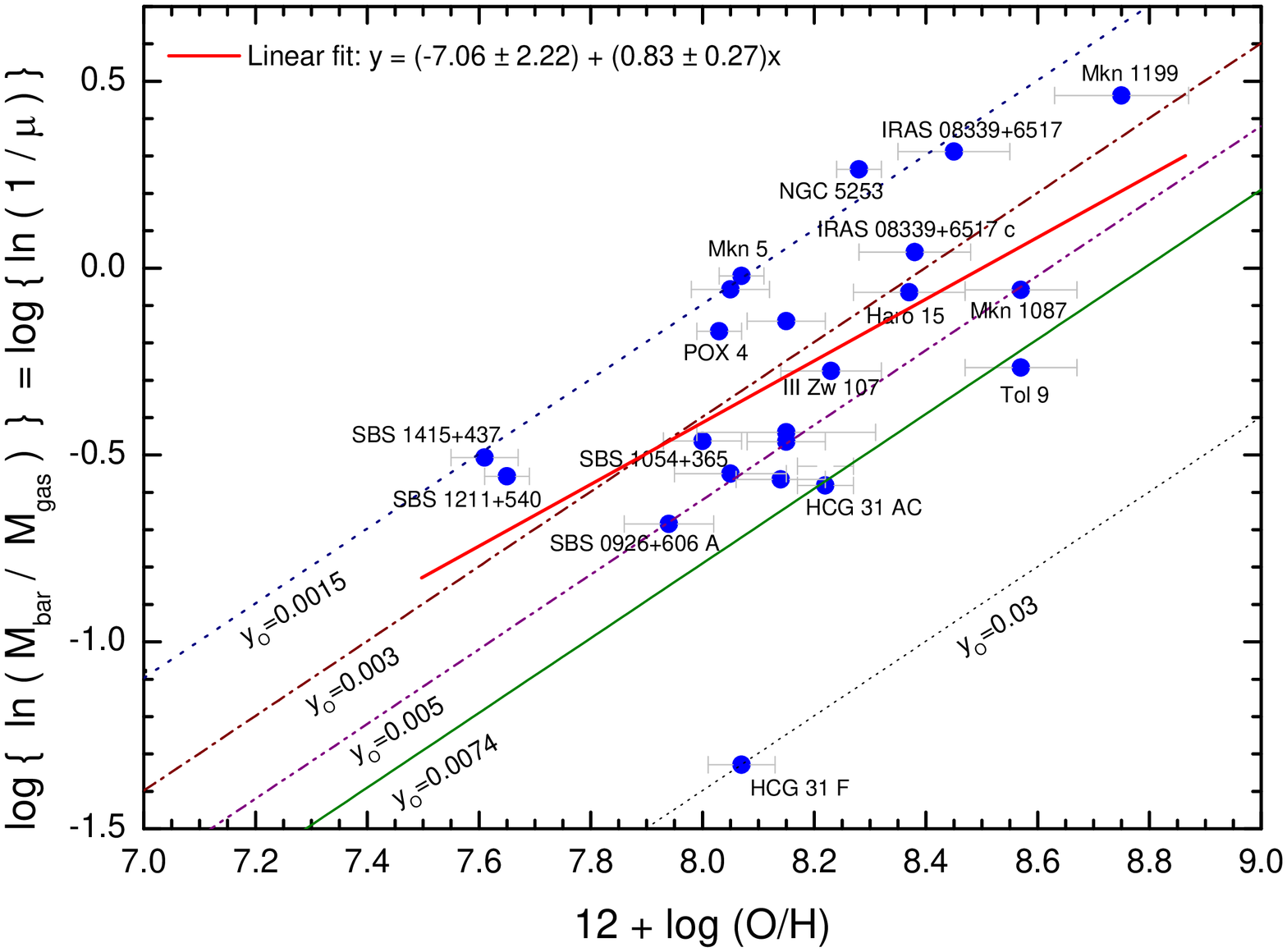} &
\includegraphics[angle=270,width=0.45\linewidth]{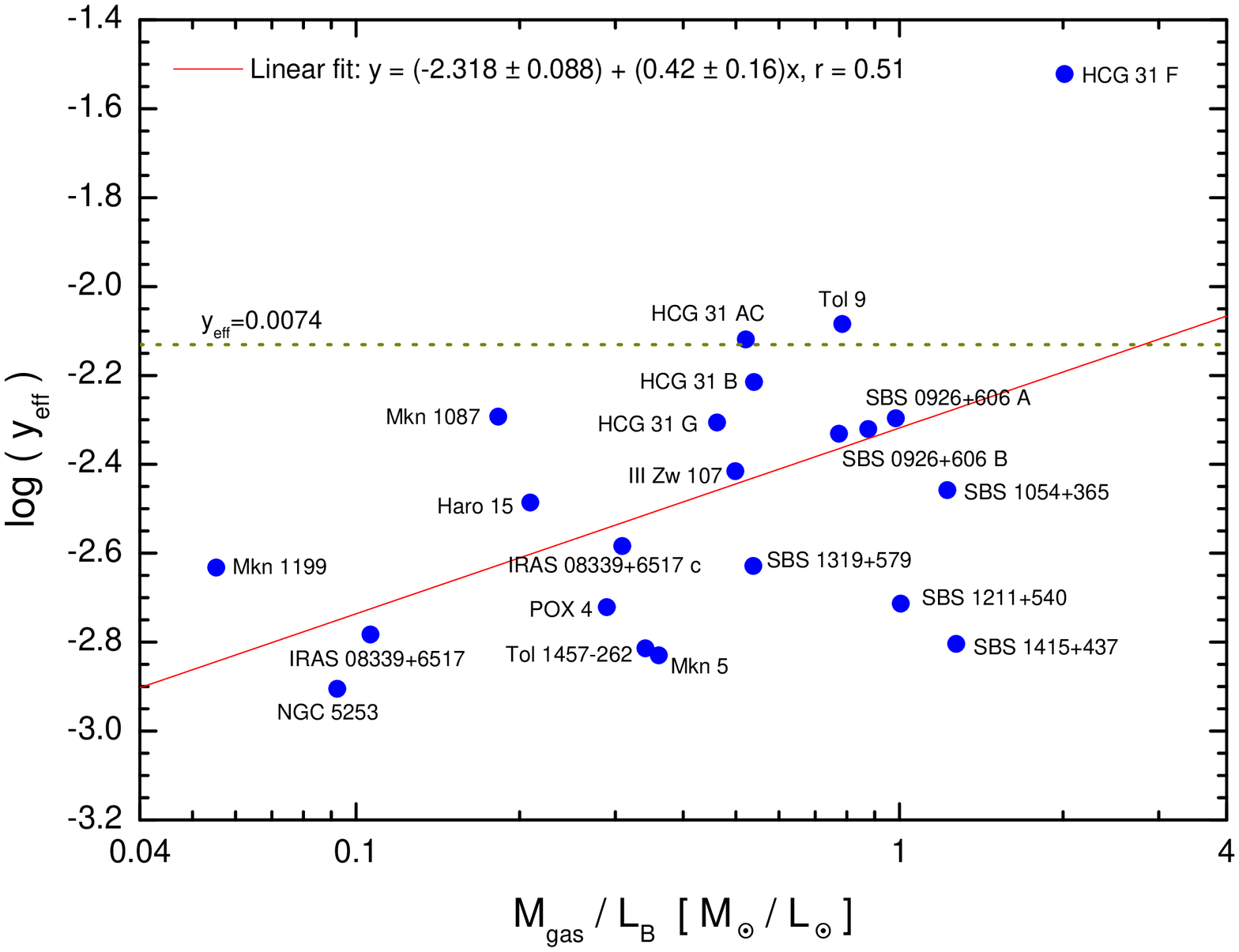} \\
\end{tabular}
\protect\caption[ ]{\footnotesize{(Left panel) Comparison of the observed oxygen abundance and that predicted by simple closed-box chemical evolution  
models with instantaneous recycling and constant star-formation rates. The continuous green line indicates the expected trend if the galaxies are  
closed boxes with an oxygen yield $y_{\rm O}$=0.0074 \citep{MM02,vZH06}. Closed-box models with  $y_{\rm O}$=0.0015, 0.003, 0.005, and 0.03 are also  
plotted. A linear fit to our data is shown with a continuous red line, and indicates an oxygen yield of $y_{\rm O}$=0.003--0.005.
(Right panel) Effective yield plotted as a function of the $M_{gas}/L_B$ ratio. The closed-box yield is plotted by a dotted yellow line. A linear fit  
to the data is shown with a continuous red line. }}
\label{closebox}
\end{figure*}

To study the environment effects of the gas content and the chemical enrinchment in galaxies, it is common to compare with the so-called  
\emph{closed-box chemical evolution} model \citep{Schmidt63,SearleSargent72,Edmunds90}. 
According to this model, a galaxy consists initially of gas with no stars and no metals. The stellar IMF is assumed to be constant on time. Stars that  
end their life as supernovae are assumed to enrich the ISM with metals immediately. Throughout its life, the galaxy experiences instantaneous  
recycling and the products of stellar nucleosynthesis are neither diluted by infalling pristine gas nor lost via outflow of enriched gas. Hence, the  
metallicity at any given time is only determined by the fraction of baryons which remains in gaseous form.
The model can be written as 
\begin{eqnarray}
\label{yield} {\rm Z_O} = y_{\rm O}\ln(1/\mu),
\end{eqnarray}
where Z$_{\rm O}$ is the oxygen mass fraction, $y_{\rm O}$ is the yield by mass and $\mu$ is the ratio of the gas mass to the baryonic mass,  
$\mu=M_{gas}/M_{bar}$. The gas mass corresponds to the hydrogen atomic gas with a correction for neutral helium, but it does not include molecular gas  
($M_{gas}$=1.32\MHi) that, as we already said, can be neglected in low-metallicity galaxies.
%The assumption of neglecting the molecular gas is valid in low-mass galaxies, 
%because both the difficulty of detect CO and the uncertainties of the correspondence 
%between CO and H$_2$ in low-metallicity objects (i.e., Taylor et al. 1998; Wilson 1995).

Left panel of Figure~\ref{closebox} compares the observed oxygen abundances to those predicted by closed-box models, that are plotted with lines with  
different $y_{\rm O}$. The green continuous line indicates the model with $y_{\rm O}$=0.0074, that is the theoretical yield of oxygen expected for  
stars with rotation following \citet{MM02} models \citep{vZH06}. As we can see, the majority of the galaxies show oxygen abundances lower than the  
expected by the closed-box models. The yield of oxygen that best fits our data (the \emph{effective yield}) is  $y_{\rm O}$=0.003--0.005, in agreement  
with previous results found in the literature (i.e., Lee et al. 2003; van Zee \& Haynes 06; Lee, Zucker \& Grebel 2007). Hence, the sample galaxies  
are generally not well reproduced by the simple closed box model, and therefore inflow of pristine gas or outflow of enriched gas have played an  
important role in their chemical evolution.

Interestingly, there is a galaxy that appears far away from the predictions given by the theoretical closed box model, but in the opposite direction  
to the rest of the galaxies. Indeed, HCG~31~F shows a much \emph{higher} oxygen abundance than that expected following the closed box model. The  
explanation of this behavior is that this object is a \TDG\ that was very probably formed from the material stripped from HCG~31~AC during the fly-by  
encounter between member G and the A+C complex \citep{LSER04a}. The \TDG\ has acretted a large fraction of the pre-enriched \HI\ gas available in the  
arm-like structure and now hosts a very intense star-formation activity.

Finally, the comparison of the effective yield derived in each object with some global galaxy parameters (dynamical and baryonic mass, absolute  
magnitude, gas mass-to-luminosity ratio, and surface star formation rate) does not show any clear trend. This result is almost the same than that  
observed by \citet{vZH06} in the analysis of a sample of isolated dwarf irregular galaxies. The difference is that these authors reported a strong  
correlation with the gas mass-to-luminosity ratio, that they explained in the sense that gas-rich galaxies are more likely to be closed boxes. But, as  
we see in the right panel of Figure~\ref{closebox}, such tight correlation is not satisfied by our data, although we do observe the trend that  
galaxies with higher $M_{gas}$/$L_B$ ratios have lower effective yields (a linear fit to our data is shown with a continuous red line). Therefore, for  
intense star-forming and gas-rich galaxies, the closed box model is also not valid. We then conclude that environment effects are playing a crucial  
role in the evolution of these galaxies.

\section{Quantification of the interaction features}

\begin{table*}[t!]
\centering
  \caption{\footnotesize{Interaction features in our WR galaxy sample. }}
  \label{interacciones}
  \tiny
  \begin{tabular}{l  cc  cc cc  cc cc c}
  \tableline
   \noalign{\smallskip}
 & \multicolumn{4}{c}{Morphological features} & &\multicolumn{2}{c}{Kinematics feat.} & Differ. in & $M/L_B$  & Other   & INTERACTION \\
 \cline{2-5}  \cline{7-8}  \noalign{\smallskip}  
Galaxy & Plume & Tail & Merger & \TDG s & &  \HII gas & \HI\ gas  &   abundances & ratios & features  & DEGREE \\      
\tableline
\noalign{\smallskip}  
HCG 31~AC$^a$  &  X  & X  &  X & X  & &  X   &  X       &   X  &    X    &    X   &  VERY HIGH \\
Mkn 1087       &  -- & X  & -- & X  & &  X   & \nodata  &   X  &   --    &    --  &  HIGH \\
Haro 15        &  ?  & -- & X  & ?  & &  X   & \nodata  &   X  &   --    &    --  &  VERY HIGH \\
Mkn 1199$^a$   &  X  & -- & X  & -- & &  X   & \nodata  &   X  &    X    &    X   &  VERY HIGH \\
Mkn 5          &  -- & -- & -- & -- & &  --  & \nodata  &   ?  &    X    &    X   &  LOW \\
IRAS 08208+2816&  -- & X  & ?  & ?  & &  X   & \nodata  &   X  &   --    &    --  &  VERY HIGH \\
IRAS 08339+6517$^a$&X& -- & -- & ?  & &  X   &  X       &  --  &   --    &    --  &  HIGH \\
POX 4          &  X  & -- & -- & ?  & &  X   &  X       &  --  &   --    &    X   &  HIGH \\
UM 420         &  -- & X  & ?  & -- & &  ?   & \nodata  &  --  &   --    &    --  &  VERY HIGH \\
SBS 0926+606~A &  X  & -- & X  & -- & &  X   & \nodata  &  --  &   --    &    --  &  VERY HIGH \\
SBS 0926+606~B &  X  & X  &    & ?  & &  X   & \nodata  &  --  &   --    &    --  &  HIGH \\
SBS 0948+532   &  -- & X  & -- & -- & &  ?   & \nodata  &  --  &    X    &    --  &  PROBABLE \\
SBS 1054+365   &  -- & -- & -- & ?  & &  X   & \nodata  &  --  &    X    &    X   &  LOW \\
SBS 1211+540   &  X  & -- & ?  & -- & &  X   & \nodata  &  --  &   --    &    --  &  PROBABLE \\
SBS 1319+579   &  -- & -- & ?  & -- & &  X   & \nodata  &   ?  &    X    &     X  &  PROBABLE \\
SBS 1415+437   &  -- & -- & -- & -- & &  ?   & \nodata  &  --  &   --    &    --  &  LOW \\
III Zw 107     &  X  & -- & ?  & -- & &  X   & \nodata  &   ?  &    X    &   --   &  HIGH \\
Tol 9          &  X  & -- & -- & -- & &  X   &  X       &  --  &    X    &    X   &  HIGH \\
Tol 1457-262   &  X  & -- & X  & X  & &  X   & \nodata  &   X  &   --    &   --   &  VERY HIGH \\
Arp 252$^a$    &  X  & X  & -- & X  & &  X   & \nodata  &   X  &   --    &   --   &  VERY HIGH \\
NGC 5253       &  -- & -- & -- & -- & &  X   &  X       &   X  &    X    &    X   &  PROBABLE$^b$ \\  
\noalign{\smallskip}
\tableline
  \end{tabular}
  \begin{flushleft}
  $^a$ Some interaction features in this galaxy were previously reported by other authors.\\ 
  $^b$ The chemical differences and the kinematics features can be explained by other reasons. See \citet{LSEGRPR07} and \citet{LS+10a}.
  \end{flushleft}
\end{table*}

Throughout this paper series we have compiled new evidences of the interaction-induced star-formation activity in starburst galaxies, in particular in  
dwarf galaxies. 
%The starburst phenomenon in dwarf galaxies has been always a challenge 
%as the wave-density theory (REF) cannot be applied because of the low mass of these objects. 
Alternative mechanisms, as the Stochastic Self-Propagating Star Formation \citep*{GerolaSS80} model (that assumes statistical fluctuations of \SFR),  
or the ideas of the cyclic gas re-processing of the ISM \citep{DaviesPhillipps88} or gas compression by shocks due to the mass lost by galactic winds  
followed by the cooling of the ISM \citep{Th91,Hi00}, fail to explain some observational characteristics and the triggering mechanism of dwarf  
starburst galaxies (i.e., Garc\'{\i}a-Lorenzo et al. 2008; Cair\'os et al. 2009).
%As we said in the introduction, alternative methods, as the self-propagating 
%star formation model (Gerola et al. 1980) or gas compression by shocks due 
%to the mass lost by galactic winds \citep{Th91,Hi00}, fail to explain the triggering 
%of the starbursts in dwarf galaxies. 
In the previous section, we demonstrated that the closed-box model is not valid to explain the chemical evolution experienced by our sample of  
galaxies, emphasizing the idea that environment effects are needed to understand their observed properties. Indeed, the interaction/merger scenario  
explains, in a natural way, the starburst activity in these objects just as a consequence of the evolution of the galaxies throughout the cosmic time  
following hierarchical formation models \citep{KW93,K97,Springel05}. These models predict that most galaxies have formed by merging of small clouds of  
protogalactic gas and that galaxy interactions between dwarf objects were very common at high redshifts.

However, the interaction features in dwarf objects are, in many cases, not evident because of the lack of deep and high-resolution images and spectra  
\citep{ME00} and detailed multi-wavelength analyses. Now it is well known that interactions in dwarf galaxies are not usual with nearby giant galaxies  
\citep{CA93,TT95,TellesMaddox00}  but with low surface brightness galaxies \citep{WLM96,N01,Pustilnik01}, or \HI\ clouds (e.g., Taylor et al. 1993,  
1995, 1996; Thuan et al. 1999; van Zee, Salzer \& Skillman 2001; Begum et al. 2006; Ekta, Chengalur \& Pustilnik 2006; Hutchmeier et al. 2008;  
L\'opez-S\'anchez \& Esteban 2008). \citet{ME00} suggested, for the first time, that interactions with or between dwarf objects could be the main star  
formation triggering mechanism in dwarf galaxies. Later, \citet{Ostlin01} and \citet{BO02} suggested that a merger between two galaxies with different  
metallicities  
%(one gas-rich and one gaspoor,
or infall of intergalactic clouds could very probably explain the starburst activity in the most luminous \BCDG s.
Since then, studies focused on individual objects have also shown that interactions do
play a decisive role in the evolution of these systems
\citep{JohnsonIWK04,BravoAlfaro04,BravoAlfaro06,Cumming08,GLCCMIK08,JamesTsamisBW09,James09}.

Our exhaustive multi-wavelength analysis of starburst galaxies combining broad-band optical/\NIR\ and \Ha\ photometry, optical spectroscopy, and  
X-ray, $UV$, $FIR$, 21-cm \HI\ line, and 1.4~GHz radio-continuum data compiled from the literature allows us to perform a quantitative analysis of the  
interaction features detected in each object.
%In this section, we quantify the interaction features detected in each 
%object of our WR galaxy sample using all available multi-wavelength data.
A summary of the results found in each individual system of our WR galaxy sample is presented in Appendix~A.
To quantify the interaction features, we compile in Table~\ref{interacciones} some interaction indicators 
%(low, probable, high and very high) 
classified in several categories, that we describe below. 
\begin{enumerate}
\item Morphological features, as the detection of faint plumes or bridges (Haro~15, IRAS~08339+6517, SBS~0926+606~B, SBS~1211+540, III~Zw~107, Tol~9,  
Tol~1457-262, Arp~252), prominent tails (HCG~31, Mkn~1087, IRAS~08208+2816, UM~420, SBS~0948+532, Arp~252), disturbed morphology (HCG~31, POX~4,  
Tol~1457-262), \TDG s candidates (HCG~31, Mkn~1087, SBS~0926+606~B, Arp~252) or mergers.% between galaxies. 
\item Kinematical features detected in the analysis of the ionized gas (see Paper~II) and the neutral gas (only in those systems for which  
interferometer \HI\ maps are available). The kinematical evidences found in the ionized gas of our sample galaxy includes: presence of objects with  
velocities decoupled from the main rotation pattern (Mkn~1087, Haro~15), sinusoidal velocity patterns that suggest a merging process (HCG~31~AC,  
Mkn~1199, IRAS~08208+2816, SBS~0926+606~A, III~Zw~107, \emph{Object~1} in Tol 1457-262), reversals in the velocity distribution
(Tol~9, Arp~252), indications of tidal streaming (HCG~31, IRAS~08208+2816, SBS~1319+579, Tol~9), or the presence of TDG candidates 
(HCG~31~F1 and F2, Mkn~1087, IRAS~08339+6517, POX~4, Tol 1457-262). 
\item Chemical abundance differences within several star-forming regions within the same system: Mkn~1087, Haro~15, and Mkn~1199 are clearly  
interacting with dwarf galaxies with lower O/H and N/O ratios. NGC~5253, IRAS~08208+2816, and Tol~1457-262 contain zones of different chemical  
composition. In the case of NGC~5253, this is produced by localized pollution of massive stars, but in the cases of IRAS~08208+2816 and Tol~1457-262  
the different chemical compositions seem to be caused by the regions corresponding to different galaxies in interaction. 
\item Furthermore, our multiwavelength analysis has provided us further indications of galaxies that do not follow their expected behavior. The  
analysis of the mass-to-light ratios indicates: very low \MHil\ in SBS~1319+579 and NGC~5253, high \MHiil\ in SBS~0948+532, low \Mstarl\ in  
SBS~1319+579, 
%and IRAS~08339+6517~Comp, 
and high \Mdynl\ in HCG~31~AC, Mkn~1199, Mkn~5, SBS~1054+365, Tol~9, and Tol~1457-262. Other evidences are: low \HI\ mass content from single-dish  
data (Mkn~1199, Mkn~5), very extended \HI\ emission embedding several nearby galaxies as HCG~31 (VM03) and Tol~9 (LS08, LS+10b), high \Mkep\  (Mkn~5,  
IRAS~08339+65~Comp, SBS~1211+540, SBS~1319+579), or important deviations of the star-formation law (Mkn~1199, SBS~1319+579 and NGC~5253). These  
features may have been produced by interactions (loss of \HI\ mass, enhancing of the star-formation activity, perturbed dynamics) but they are just  
indirect evidences that should be confirmed by new deep observations (i.e., \HI\ maps).    
\end{enumerate}

A question mark in Table~\ref{interacciones} indicates that the available data do not allow us to confirm this indicator. Last column in  
Table~\ref{interacciones} compiles the interaction degree that each system is experiencing after considering all positive interaction indicators. We  
have divided the interaction degree in four classes: low (no clear signs of interactions), probable (there are some interaction indicators, but  
deepest data are needed to confirm it), high (we found clear evidences of interactions, but we do not see merger features) and very high (in the cases  
of finding clear merger features). 

Evident mergers between independent objects with relatively similar masses (major mergers) have been detected in HCG~31~AC, SBS~0926+606~A,  
IRAS~08208+2816 and Tol~1457+262. The galaxy pair Arp~252, that is composed by ESO~566-7 and ESO~566-8, seems also to be experiencing the first stages  
of a major merger. Minor mergers are found in Haro~15 and Mkn~1199. UM~420 seems also to be experiencing a merger, as deep 3D optical spectroscopy  
(James et al. 2009) suggests. All these galaxies have a very high interaction degree. 

Mkn~1087, IRAS~08339+6517, POX~4, SBS~0926+606~B, III~Zw~107 and Tol~9 are experiencing clear interactions with nearby dwarf objects. In the case of  
POX~4, we still have to investigate \citep{LS+10b} if its dwarf companion galaxy is a \TDG\ candidate and the interaction was with a nearby diffuse  
\HI\ cloud, or if this object actually is an independent dwarf galaxy that crossed the main body of POX~4 \citep{ME97}. 
These six galaxies have a high interaction degree.

On the other hand, we find probable evidences of interaction in SBS~0948+532 (enhanced star formation activity, long optical tail, probable disturbed  
kinematics of the ionized gas), SBS~1211+540 (diffuse optical plumes, minor merger indications), SBS~1319+579 (peculiar \MHil, \MHiil, and \Mstarl\  
ratios, perturbed kinematics suggesting merging or tidal stream phenomena). The \HI\ data of NGC~5253 shows disturbed morphology and kinematics  
\citep{LS08a}, suggesting that this \BCDG\ has disrupted or accreted recently a dwarf gas-rich companion \citep{KS08,LS+10a}.

Only Mkn~5, SBS~1054+364, and SBS~1415+437 do not show interaction evidences in our exhaustive multi-wavelength study. However, Mkn~5 seems to be  
\HI-deficient and seems to possess a perturbed neutral gas kinematics because of its relatively high \Mdynl\ ratio. SBS~1054+364 also show a high   
\Mdynl\ ratio that may suggest perturbed \HI\ kinematics. Furthermore, the chemical abundance of this galaxy is very high in comparison to that  
expected from its baryonic mass. On the other hand, SBS~1415+437 shows a relatively low oxygen abundance for its baryonic mass.

Considering all indicators, we find that 13 up to 20 systems (68\% of our WR galaxy sample) are classified with a high or very high interaction  
degree. We note, however, that four of these objects  (HCG~31, Mkn~1199, IRAS~08339+6517 and Arp~252) show well known interaction evidences, but our  
analysis reinforces the evidences and improve our knowledge of these systems.  Only three galaxies (Mkn~5, SBS~1054+364, and SBS~1415+437) do not show  
interaction features, but they show considerable divergences of some properties when comparing with similar objects. Hence, it is evident that {\bf  
the majority of the analyzed galaxies} (17 up to 20) are interacting or merging with or between dwarf objects. Our analysis therefore demonstrates the  
importance of the low-luminosity galaxies, \HI\ clouds and dwarf objects in the evolution of the galaxies. 
%as hierarchical formation models predict (i.e. Kauffmann \& White 1993; Springer 2003). 
Interactions with dwarf galaxies also may initiate star-formation events in normal spiral galaxies, such it occurs in the external arms of Mkn~1199,  
in Haro~15, surrounding Mkn~1087, or in the impressive galaxy pair NGC~1512/1510 \citep{KoribalskiLS09}. Definetively, interactions between dwarf  
galaxies is one of the main triggering mechanism of the star-formation activity in starburst galaxies, but these dwarf objects are only detected when  
deep optical images and spectroscopy and complementary \textsc{H\,i} observations are obtained. 
%Therefore, BCDGs seem not to be real isolated systems.

\section{Conclusions}

We have presented a comprehensive analysis of a sample of 20 starburst galaxies that show the presence of a substantial population of very young  
massive stars, most of them classified as Wolf-Rayet galaxies. In this paper, the last of the series, we analyze the global properties of our galaxy  
sample using all multiwavelength data, that include X-ray, \FUV, \FIR, and radio (both \HI\ spectral line at 21~cm and 1.4~GHz radio-continuum)  
results. Our main conclusions are the following:
\begin{enumerate}
\item We compared the values of the \SFR\ derived from several indicators that consider fluxes at different wavelengths. The results agree well within  
the experimental errors and with our \Ha-based values, that were obtained after correcting for reddening and [\ion{N}{ii}] contribution. However, we  
consider that the new \Ha-based calibration provided by \citet{Calzetti07} should be preferred over the well-known and extensively used \citet{K98}  
calibration. Additionally, we checked that the \FUV-based \SFR\ very often shows similar results to those obtained using the emission of the ionized  
gas, providing a powerful tool to analyze independently the star-formation activity in both global and local scales. 
\item  We checked that the \SFR/$L_B$ ratio decreases with increasing metallicity. We derived empirical relationships between the $U$-band, $B$-band,  
and X-ray luminosities and the \SFR, that can be only used in starburst galaxies and as a first estimation of the real \SFR\ value.
\item All objects except one in our galaxy sample satisfy the \FIR-radio correlation, indicating that they are pure star-forming systems. Only the  
galaxy ESO~566-8 lies away from the \FIR-radio correlation because it seems to host some kind of nuclear activity. The non-thermal-to-thermal ratio  
seems to increases with increasing luminosity, suggesting that the cosmic-ray confinement is more efficient in massive galaxies than in dwarf objects.
\item We provided empirical relationships between the ionized gas mass, neutral gas mass, dust mass, stellar mass, and dynamical mass with the  
$B$-luminosity. Although all mass estimations increase with increasing luminosity, we find important deviations to the general trend in some objects,  
that seem to be consequence of peculiarities in these galaxies. The comparison between the dynamical mass (derived from the kinematics of the neutral  
gas) with the Keplerian mass (obtained from the kinematics of the ionized gas) and the stellar mass (from the $H$-band luminosity) provides further  
clues about systems in which the dynamics seem to be highly perturbed. We remark the importance of this study, because it is not common to find in the  
literature a comprehensive and detailed analysis of a sample of galaxies for which the total (dynamical or stellar) mass, the reddening-corrected  
luminosity in optical and \NIR\ filters, and the \Te-based oxygen abundance, have been derived in a coherent way. 
\item We investigated some mass-metallicity relations and compared with previous results found in the literature. As pointed out by \citet{KE08}, the  
choice of the metallicity calibration has a strong effect in the derived $M-Z$ relation. The tightness of the $M_{dyn}-Z$ calibration indicates that  
that the dark matter content also increases with metallicity. The scatter in the $M_{stars}-Z$ and $M_{bar}-Z$ relations are consequence of both the  
nature (dwarf galaxies, \TDG\ candidates, mergers) and the star-formation histories experienced in each galaxy. 
%We conclude that a similar analysis to that performed here should be done 
%in a larger sample of starburst galaxies to understand how metallicity and 
%mass are related in these kind of galaxies.
\item We found that our sample galaxies agree well with the Schmidt-Kennicutt scaling law of star-formation derived by \citet{K07}, that considers  
individual star-forming regions within M~51. Some important deviation are found in NGC~5253 and Mkn~1199, that are very \HI-deficient, and in  
SBS~1319+579, where the star-formation activity seems to be supressed. 
\item The study of the mass-to-light ratios reinforces some of the results found in our analysis. We found that the neutral-gas-mass-to-luminosity  
ratio clearly decreases with increasing mass, as it seems to happen with the ionized-gas-mass-to-luminosity ratio. The  
ionizing-cluster-mass-to-luminosity ratio, however, seems to be constant with metallicity. The fact that we do not find any dwarf galaxy high \MHil\  
ratio indicates that they have not experienced a lonely life. The analysis of the $M_{gas}/M_{stars}$ ratio suggests that this kind of galaxies have  
equal amount of neutral and stellar masses for metallicities \abox$\sim$8.2--8.3.  The stellar-mass-to-luminosity ratio clearly increases with the  
$B-R$ colour.
\item We found that the reddening coefficient derived from the Balmer decrement clearly increases the the warm dust mass, indicating that the  
extinction is mainly internal to the galaxy and not in the line-of-sight. We confirmed that the dust-to-gas ratio increases with the metallicity, and  
suggested that the low $M_{dust}/M_{gas}$ ratios in dwarf low-metallicity galaxies is consequence of the large reserves of un-enriched neutral gas.  
However, the low $M_{dust}/M_{gas}$ ratios observed at the outskirts of spiral galaaxies seem to be a result of the decreasing of the star-formation  
efficiency in these regions.
\item The comparison of our data with the closed-box model clearly indicates that environment effects have played and important role in the evolution  
of the analyzed galaxies. The main effective yield we derived for our data agrees quite well with results found in the literature, in particular with  
results found in other starburst or irregular dwarf galaxies.  
\end{enumerate}  
Considering all available data, we quantified how many galaxies are experiencing interaction or merger processes. We found that 17 up to 20 objects  
are clearly interacting or merging with low-luminosity dwarf objects or \HI\ clouds, and all the remnant three galaxies (Mkn~5, SBS~1054+364, and  
SBS~1415+437) show considerable divergences of some properties when comparing with similar objects. However, the interacting/merging features are only  
detected when deep optical spectroscopy and a detailed multi-wavelength analysis, remarking analysis of the kinematics and distribution of the neutral  
gas, are obtained. We therefore conclude that interactions do play a fundamental role in the triggering mechanism of the strong star-formation  
activity observed in dwarf starburst galaxies. This observational result completely agrees with the hierarchical model of galaxy formation that  
considers that large galactic structures were built up from the accretion of dwarf galaxies. 
%New deep \HI\ observations of a sample of Blue Compact Dwarf Galaxies 
%carried out at the \emph{Australia Telescope Compact Array} (LS08a; LS09; vEKLS10) 
%also confirm this result, and will be soon analyzed in detail elsewhere (LS10a,LS10b).                     

\begin{acknowledgements}

Based on observations made with NOT (Nordic Optical Telescope), INT (Isaac Newton Telescope) and WHT (William Herschel Telescope) operated on the  
island of La Palma jointly by Denmark, Finland, Iceland, Norway and Sweden (NOT) or 
the Isaac Newton Group (INT, WHT) in the Spanish Observatorio del Roque de Los Muchachos of the Instituto de Astrof\'{\i}sica de Canarias. 
Based on observations made at the Centro Astron\'omico Hispano Alem\'an (CAHA) at Calar Alto, 
operated by the Max-Planck Institut f\"ur Astronomie and the Instituto de Astrof\'{\i}sica de Andaluc\'{\i}a (CSIC). Based on observations made with the ATCA
(Australia Telescope Compact Array), which is funded by the Commonwealth of Australia for operation as a National Facility managed by CSIRO. 

\'A.R. L-S thanks C\'esar Esteban (his formal PhD supervisor) for all the help and very valuable explanations, talks and discussions along these  
years. He also acknow\-ledges Jorge Garc\'{\i}a-Rojas, Sergio Sim\'on-D\'{\i}az and Jos\'e Caballero for their help and friendship during his PhD,  
extending this acknowledge to all people at Instituto de Astrof\'{\i}sica de Canarias (Spain). 
\'A.R. L-S. \emph{deeply} thanks the Universidad de La Laguna (Tenerife, Spain) for force him to translate his PhD thesis from English to Spanish; he 
had to translate it from Spanish to English to complete this publication. 
This was the main reason of the delay of the publication of this research, because the main results shown here were already included in the PhD 
dissertation (in Spanish) which the author finished in 2006 \citep{LS06}. 
The author is indebted to the people at the CSIRO Astronomy and Space Science / Australia 
Telescope National Facility, especially B\"arbel Koribalski, for their support and friendship while translating his PhD.  The author also thanks B\"arbel Koribalski (CSIRO/ATNF) for  
her help analyzing HIPASS data and all the talk and discussions about radio-astronomy.

%The authors are very grateful to A\&A language editor, Claire Halliday, 
%for her kind revision of the manuscript. 
This work has been partially funded by the Spanish Ministerio de Ciencia y Tecnolog\'{\i}a (MCyT) under projects AYA2004-07466 and AYA2007-63030. 
This research has made use of the NASA/IPAC Extragalactic Database (NED) which is operated by the Jet Propulsion Laboratory, California Institute of  
Technology, under contract with the National Aeronautics and Space Administration. 
The  Galaxy  Evolution  Explorer (GALEX) 
%(Martin  et  al.  2005)  
is  a  NASA Small  Explorer,  launched  in  April 2003.  We  gratefully  acknowledge NASA's  support  for  construction, 
operation, and science analysis for the GALEX mission. 
The Infrared Astronomical Satellite (IRAS) mission was a collaborative effort by the United States (NASA), the Netherlands (NIVR), and the United  
Kingdom (SERC).
This research has made extensive use of the SAO/NASA Astrophysics Data System Bibliographic Services (ADS).

\end{acknowledgements}

%\listofobjects

\clearpage
\appendix
\normalsize

\section{Final summary of individual galaxies}

In this Appendix we compile the main results found in our multiwavelength analysis of each individual system within our sample of Wolf-Rayet galaxies.  
In Papers~I and II we described the optical/\NIR\ broad-band and \Ha\ photometry, and the intermediate-resolution optical spectroscopy analysis,  
respectively, of 16 up to 20 galaxies studied in this work. The analysis of the additional 4 systems were presented in previous papers: NGC~1741  
(member AC within the HCG~31 group) in \citet{LSER04a}, Mkn~1087 and their surrounding galaxies in \citet{LSER04b}, the luminous blue compact galaxy  
IRAS 08339+6517 in \citet{LSEGR06}, and NGC~5253 in \citet{LSEGRPR07}. 
Paper~III compiled the localization of the WR-rich star clusters within the galaxies and the analysis of their massive stellar populations. Paper~IV  
compiles the global analysis of colours, and the physical properties and chemical abundances of the ionized gas. This paper, the last of the series,  
completes our analysis via a multi-wavelength analysis involving X-ray, $UV$, $FIR$, and radio data in both the 21-cm \HI\ line and the 1.4~GHz  
radio-continuum.

	\begin{itemize}

    \item {\bf NGC 1741} hosts a very strong star-formation event, that is very probably a consequence of the merging of two spiral galaxies. NGC~1741  
is the main member (AC) of the galaxy group HCG~31, and it is interacting with other galaxies in the group, including Mkn~1090 (HCG~31~G). We detect  
both the blue and red WR bumps in its brightest region, as well as the nebular \ion{He}{ii} $\lambda$4686 line. HCG~31~AC seems to have a slightly  
higher N/O ratio. Some dwarf objects (members E, F1, F2 and H) are tidal dwarf galaxy (\TDG) candidates. See \citet{LSER04a} for details. 
    \item {\bf Mkn 1087} is a Luminous Compact Blue Galaxy (\LCBG) in interaction with the nearby galaxy KPG~103a and with a dwarf surrounding galaxy  
(N companion). Deep optical images show long stellar tails connecting the main body of the galaxy with diffuse objects, some of them hosting  
star-formation activity, and several \TDG\ candidates. Although WR features were previously reported by other authors, we do not detect any. See  
\citet{LSER04b} for details.
	\item {\bf Haro 15} probably is a medium-size Sc spiral in interaction with two nearby dwarf objects. Knot~A shows a very high star-formation  
activity and WR features; its disturbed kinematics suggests that it is experiencing a minor merger with Haro~15. Knot~B is an independent object  
because of its morphology, decoupled kinematics and chemical abundances. 
%. The main galaxy also seems to have WR features.
 	\item {\bf Mkn 1199} is a system composed by a Sb-Sc spiral and a dwarf galaxy, both in clear interaction, as they may be at the first stages of a  
minor merger. The interaction has triggered the star formation activity in some areas of the main galaxy. Both the blue and red \WRBUMP s are detected  
in the central region of Mkn~1199, that has solar metallicity. It seems that a substantial fraction of the \HI\ gas has been expelled to the  
intergalactic medium because of its low \MHil\ ratio.
    \item {\bf Mkn 5} is a \BCDG\ with a strong star-forming burst located in the external part of the galaxy. The blue \WRBUMP\ is detected in this  
starbursting region, that also possesses an important underlying old stellar component. We do not find any evidence of interaction, but the amount of  
\HI\ gas of the galaxy is very low compared with that expected for a \BCDG. Furthermore, its dynamical mass is higher than that expected for an dwarf  
galaxy with similar properties. Both results suggest that Mkn~5 has lost its neutral gas in some moment of its past and still has a disturbed \HI\  
kinematics.
    \item {\bf IRAS 08208+2816} is a Luminous Infrared Galaxy (\LIRG) showing two long tails with a very high star-formation activity. The kinematics  
of the ionized gas clearly indicate merger features and the existence of two long tidal tails with \TDG\ candidates. The chemical abundances of the  
brightest knots also seem to be different. We detect both the blue and red \WRBUMP s in the central region, that possesses a high N/O ratio.  
   \item {\bf IRAS 08339+6517} is a luminous infrared galaxy (\LIRG) and a luminous blue compact galaxy (\LCBG) in clear interaction with a nearby  
dwarf galaxy. The majority of the \HI\ gas of the system has been expelled to the intergalactic medium because of this interaction (Cannon et al.  
2004). Our deep optical images reveal a faint stellar plume coincident with the \HI\ tail, and a disturbed morphology in the outskirts of the galaxy.   
A particular bright knot may be a TDG candidate of the remnant of a previous minor merger. We detect weak WR features in its central burst and  
quantified the star formation history of the galaxy \citep{LSEGR06}. 
    \item {\bf POX 4} is a morphology-disturbed low-metallicity \BCDG\ showing strong star-forming bursts throughout all the galaxy. It seems to be in  
interaction with a nearby dwarf object that may have passed through the main body of the galaxy, being the origin of its ring-like morphology  
\citep{ME97} and kinematics. However, this objects may also be a \TDG\ candidate originated by the interaction with a nearby and diffuse \HI\ cloud  
\citep{LS+10b}. The \ion{He}{ii} $\lambda$4686 emission line is clearly detected in its brightest region, as well as both the blue and the red \WRBUMP  
s.
    \item {\bf UM 420}: is a blue compact galaxy, but not a dwarf object, hosting intense star-formation activity. Besides it is located at 237~Mpc,  
we observe a central region and two kind of bright \Ha\ tails pointing towards different directions. Its kinematics is also perturbed. It has a very  
low metallicity for an object with its absolute optical/\NIR\ luminosities, suggesting that it is a merging of two independent galaxies. We detect the  
\ion{He}{ii} $\lambda$4686 emission line but no the blue \WRBUMP\ in its brightest region. We found a probable N/O enrichment in the central region.  
Its colors and properties are somewhat contaminated by the spiral disk of the foreground galaxy UGC~1809, located at 97 Mpc.
	    \item {\bf SBS 0926+606}: is a galaxy pair with high star formation activity. Member A is a \BCDG\ that shows a double nucleus; both its  
morphology and kinematics strongly suggests that it is a galaxy merger. We do not detect WR features in this galaxy but only the  \ion{He}{ii}  
$\lambda$4686 emission line. On the other hand, member B (another \BCDG) hosts less star-formation activity, but it also shows hints of interactions,  
remarkably a long diffuse optical tail that shows a \TDG\ candidate. SBS 0926+606~B has a huge emission in $UV$; the \SFR\ derived from the $FUV$  
luminosity is more than one order of magnitude higher than the \Ha-based \SFR. The system still hosts a huge amount of neutral gas.  
	\item {\bf SBS 0948+532}: is a very compact and blue object that hosts a very high star formation activity. Its \MHiil\ ratio is very high in  
comparison with objects with similar properties. Although usually classified as \BCDG s, its total $B$-luminosity indicates that it is not a dwarf  
object. We detect a faint optical tail mainly composed by old stars and with a slightly disturbed kinematics. We observe the nebular and broad  
\ion{He}{ii} $\lambda$4686 lines.
    \item {\bf SBS 1054+365}: is a very nearby \BCDG\ showing several star-forming regions embedded in a elliptical envelope composed by old stars.  
The kinematics of the ionized gas seems to be slightly disturbed. The main starbursting region shows the nebular and broad \ion{He}{ii} $\lambda$4686  
lines. Although we do not detect any clear interaction feature, its dynamical mass is too high in comparison with that observed in similar objects,  
and its metallicity is too high for a dwarf object. Further studies are needed to clarify its nature.
    \item {\bf SBS 1211+540}: is a very low-metallicity \BCDG. It is composed by two bright \Ha\ regions surrounded by a relatively old stellar  
component. This \BCDG\ seems to show a higher metallicity than expected for a dwarf object with its same properties. The detection of two faint  
optical tails and its disturbed kinematics suggest that this galaxy is experiencing its first stages of a merger process. Although reported  
previously, we do not detect any WR features.
	\item {\bf SBS 1319+579}: is a cometary-like \BCDG\ showing two chains of intense star-forming regions over an underlying low-luminosity component  
dominated by old stars. We detect a very faint blue WR feature in the brightest knot. The analysis of the kinematics of the ionized gas strongly  
suggests that it is composed by two objects in interaction, that it is happening edge-on. Although there is plenty of neutral gas, the star formation  
is not very efficient, showing very low \MHiil\ and \Mstarl\ ratios in comparison with similar objects. Furthermore, it does not satisfy the  
Schmidt-Kennicutt law of star formation and the \HI\ dynamics seem to be perturbed. We consider that the neutral gas has been expelled from the  
galaxy, but interferometric observations are needed to probe it. 
    \item {\bf SBS 1415+437}: is a very low-metallicity  \BCDG\ that hosts a very strong star-forming region in which the nebular \ion{He}{ii}  
$\lambda$4686 emission line is observed. It possesses an important old stellar population underlying the starburst. We do not detect any optical  
nearby companions and it does not show any evidence of interactions.
    \item {\bf III Zw 107}: is a \BCDG\ showing two strong star-forming bursts embedded in an irregular envelope. A diffuse prominent tail is detected  
in this object. The broad \ion{He}{ii} $\lambda$4686 line is found in the brightest knot, that shows a slightly higher N/O ratio. The neutral gas may  
have been expelled and/or dispersed. This galaxy is likely composed by two dwarf objects in process of interaction or merging.
	\item {\bf Tol 9}: is a \BCG\ that belongs to the Klemola 13 galaxy group. It is a elliptical-shaped galaxy with intense nebular emission and  
chemically evolved. We have detected morphological and kinematical pattern that suggest interaction features. Our deep \Ha\ image reveals an extended  
filamentary structure with two main features that are located almost perpendicular to the main optical axis of the galaxy. The probable origin of this  
structure is a galactic wind. We detect both the blue and red \WRBUMP s in the central region. The \HI\ morphology and kinematics is quite intriguing,  
because this galaxy and two surrounding dwarf objects are embedded  in the same \HI\ cloud (LS08b,LS+10b). 
    \item {\bf Tol 1457-262}: is a system composed by two bright objects and two dwarf galaxies, all showing nebular emission. We detect the nebular  
\ion{He}{ii} line in the brightest knots of the main object. The regions within this system show chemical differences and peculiar kinematics. The  
neutral gas content seems to be very high, and its dynamics highly perturbed, although detailed \HI\ map should be required to quantify this. We  
consider that this system is a galaxy group in which its members are in interaction.     
	\item {\bf Arp 252}: is a galaxy pair composed by two spiral galaxies, ESO~566-8 (A) and ESO~566-7 (B), in the first stages of a major merger.  
This object shows two long tails mainly composed by old stars but hosting some star-forming regions and \TDG\ candidates. ESO~566-8 shows the broad  
and nebular \ion{He}{ii} $\lambda$4686 emission line and the red \WRBUMP. Its N/O is quite high for a galaxy with its oxygen abundance. ESO~566-8 has  
a strong star-formation and may host some kind of nuclear activity, because the \FIR/radio relation is not satisfied in it. Although it was previously  
observed by other authors, we do not observe any WR feature in ESO~566-7.
    \item {\bf NGC 5253}: is a very nearby \BCDG\ showing many peculiarities with respect to objects of similar characteristics. We detect clear broad  
WR features in the central regions, indicating the presence of both WNL and WCE stars.  We confirmed the presence of a localized N enrichment in  
certain zones of the center of the galaxy and suggested a possible slight He overabundance in the same areas. We demonstrated that the enrichment  
pattern agrees with that expected for the pollution by the ejecta of WR stars. The amount of enriched material needed to produce the observed  
overabundance is consistent with the mass lost by the number of WR stars estimated in the starbursts (see L\'opez-S\'anchez et al. 2007 for details).
%\citep{LSEGRPR07}. 
%We detected, for the first time in a dwarf starburst galaxy, 
%faint \ion{C}{ii} and \ion{O}{ii} recombination lines \citet{LSEGRPR07}. 
Although the kinematics of the ionized gas is somewhat peculiar, stellar kinematics seem to be consequence of rotation.  Our optical study has not  
reveal any disturbed feature of a recent interaction process. However, its \HI\ morphology is disturbed and its kinematics is quite intriguing,  
because it does not show any sign of regular rotation. The origin of this anomaly is most likely the disruption/accretion of a dwarf gas-rich  
companion or the interaction with another galaxy in the M\,83 subgroup \citep{KS08,LS08a,LS+10a}. Furthermore, its \MHil\ and \Mdustl\ ratios are very  
low and it does not satisfy the Schmidt-Kennicutt law of star formation. 
	\end{itemize} 

%% The following command ends your manuscript. LaTeX will ignore any text

\end{document}